%% file: colour_factor_counting_draft_penult.tex
\documentclass[12pt,a4paper]{article}
\usepackage{jheplikemod,ifpdf,tocloft,rotating,amsmath,textcomp,mathrsfs,mathtools,braket,pgffor,rotating,pdflscape,graphicx,multirow,longtable}
\usepackage[group-separator={\hspace{0.07cm}}]{siunitx}
\usepackage{tikz,tikz-cd,pgfplots}
\usetikzlibrary{calc,decorations.markings,shapes.geometric}
\usepackage{diagbox}
\usepackage[backgroundcolor=green!5, linecolor=blue!60]{todonotes}
\input{main_headers.tex}
\graphicspath{{figs/}}

\title{\texorpdfstring{{\huge \mbox{The Many Colours of Amplitudes\hspace{-10pt}}}\\[-0pt]}{The Many Colours of Amplitudes}}
\author{\vspace{-24pt}Jacob~L.~Bourjaily,}\emailAdd{bourjaily@psu.edu}
\author{\hspace{-2pt}Michael~Plesser,}\emailAdd{mkp5771@psu.edu}
\author{\hspace{-2pt}Cristian~Vergu}\emailAdd{cfv5175@psu.edu}
\affiliation{Institute for Gravitation and the Cosmos, Department of Physics,\\Pennsylvania State University, University Park, PA 16802, USA}
\abstract{%
We study the colour-dependence of scattering amplitudes in Yang-Mills theory with arbitrary (but fixed) gauge group and various representations of charged matter. When the rank of the gauge theory is taken arbitrarily large compared to the number of particles involved in an amplitude, it is well known that the number of independent colour-structure tensors grows \emph{factorially} with multiplicity; however, for any fixed gauge group, this number grows at most \emph{exponentially} with multiplicity. We review how this counting arises in representation theory and survey its implications for a wide variety of specific gauge groups with various representations of charged matter, uncovering several surprising structures along the way.%
}
\preprint{}

\begin{document}

\maketitle\thispagestyle{empty}
\pagenumbering{roman}\clearpage

\setcounter{section}{0}

\pagenumbering{arabic}
\vspace{0pt}%
\section{Introduction and Overview}\label{sec:introduction}\vspace{0pt}
Yang-Mills is the unique (unitary, local, Lorentz-invariant) quantum field theory of massless \mbox{spin-one} particles~\cite{Benincasa:2007xk}. This follows from the rigidity of the $S$-matrix for three massless particles, which can depend on any non-dynamic quantum numbers $\{\r{a},\r{b},\ldots,\r{c}\}$ (`colours') used to distinguish particles only through some rank-three tensor $f^{\r{a\,b\,c}}$; Bose symmetry requires that this tensor be fully \emph{antisymmetric} in colour labels, and locality and unitarity require~\cite{Benincasa:2007xk} (via consistency among four-particle factorization channels) that these tensors satisfy the Jacobi relation:\\[-10pt]
\eq{\kappa_{\g{e}\,\g{f}}\left(f^{\r{a}\,\r{b}\,\g{e}}f^{\g{f}\,\r{c}\,\r{d}}{+}f^{\r{b}\,\r{c}\,\g{e}}f^{\g{f}\,\r{a}\,\r{d}}{+}f^{\r{c}\,\r{a}\,\g{e}}f^{\g{f}\,\r{b}\,\r{d}}\right)\,=0\,,\vspace{-0pt}\label{jacobi_relation}}
where $\kappa_{\g{e}\,\g{f}}$ is the (inverse of the)\footnote{Non-interacting `gluons' do not participate in this identity, and so the non-degeneracy of the Killing form (and the semi-simplicity of the algebra) are tautologically ensured.} \emph{Cartan-Killing form} which encodes the overlap between these particles' labels in the propagator. Due to (\ref{jacobi_relation}), the tensors $f^{a\,b\,c}$ form the \emph{structure constants} of some Lie algebra $\mathfrak{g}$; and the labels assigned to distinguishable gluons \emph{necessarily} furnish the \emph{adjoint representation}  `$\mathbf{ad}$' of $\mathfrak{g}$. 

Similarly, the interactions between gluons and any set of distinguishable spin-1/2 Fermions labelled by $\{\b{i},\b{j},\ldots,\b{k}\}$ (also called `colours') must involve a constant rank-three tensor $T^{\r{a}\,\b{i}}_{\phantom{a\,i\,}\b{j}}$, constrained by locality and unitarity to satisfy
\eq{\begin{split}&T^{\r{a}\,\b{i}}_{\phantom{\r{a}\,\b{i}\,}\g{j}}T^{\r{b}\,\g{j}}_{\phantom{b\,j\,}\b{k}}{-}T^{\r{b}\,\b{i}}_{\phantom{\r{b}\,\b{i}\,}\g{j}}T^{\r{a}\,\g{j}}_{\phantom{a\,j\,}\b{k}}\equivL\,\big[\underline{T}^{\r{a}},\underline{T}^{\r{b}}\big]\hspace{-8pt}\phantom{T}^{\b{i}}_{\phantom{i\,}\b{k}}\!=\!\kappa_{\g{c\,d}}f^{\r{a\,b\,}\g{c}}\,T^{\g{d}\,\b{i}}_{\phantom{d\,i}\,\b{k}}\equivL\, f^{\r{a\,b}}_{\phantom{a\,b\,}\g{c}}\,T^{\g{c}\,\b{i}}_{\phantom{g\,i\,}\b{k}}\equivL f^{\r{a\,b}}_{\phantom{a\,b\,}\g{c}}\,\underline{T}^{\g{c}}\,;\end{split}\label{representation_definition}}
that is, the `colour-charges' $T^{\r{a}\,\b{i}}_{\phantom{a\,i\,}\b{j}}\equivL\,\underline{T}^{\r{a}}$ of any matter must furnish a \emph{representation}, denoted `$\mathbf{R}$', of the Lie algebra $\mathfrak{g}$. The dimension of this representation is that of the span of the indices $\b{i},\b{j}$, the possible values of which are far from arbitrary: they must adhere to the restriction following from the theory of Lie algebra representations. As with gluons, which must come in distinguishable sets of possible colours according to the adjoint representation of some Lie algebra, any matter which interacts with gluons must come in sets according to some particular representation of that Lie algebra. Although these facts are well known, it is worth appreciating that they follow from locality and unitarity alone, independent of any Lagrangian description.

In this work, we will be interested in how scattering amplitudes involving colour-charged particles depend on the {colours} of particles involved. In general, the $S$-matrix will depend on both the kinematic data describing each of the particles (their masses, momenta, and spins/helicities) and also all their distinguishable quantum numbers (their `colours', species, etc.). These dependencies factorize separately in the Feynman rules, and so it is obvious that an amplitude involving coloured particles can be written as a sum over the product of colour-dependent tensors and kinematic-dependent functions:
\eq{\mathcal{A}\big(\{p,\epsilon,\r{c}_{{}}(\mathbf{R}_{})\}_{\text{in}},\{p_{},\epsilon_{},\r{c}_{}(\mathbf{R}_{})\}_{\text{out}}\big){=}\sum_{\Gamma}\Gamma\big[\{\r{c}_{}(\mathbf{R}_{})\}_{\text{in}},\{\r{c}_{}(\mathbf{R}_{})\}_{\text{out}}\big]A_{\Gamma}\big(\{p,\epsilon\}_{\text{in}},\{p_{},\epsilon_{}\}_{\text{out}}\big).\nonumber}
Each colour-dependent factor can be viewed as a tensor whose indices span the possible colours of all the various particles involved. The individual colour-tensors that arise from the Feynman rules for gauge theory are rarely independent: minimally, they satisfy the relations (\ref{jacobi_relation}) and (\ref{representation_definition}). It is natural to wonder: how many linearly independent colour-tensors exist for a given process? 

The decomposition of amplitudes into such tensors has a rich history of study \mbox{(see \emph{e.g.}~\cite{Mangano:1988kk,Mangano:1990by,birdtracks,DelDuca:1999rs,Dixon:2013uaa,Melia:2015ika,Johansson:2015oia,Ochirov:2019mtf})}, and the interplay between colour- and kinematic-dependence of amplitudes has led to remarkable new insights connecting gauge theories and gravitational theory (see \emph{e.g.}~\cite{BCJ,Bern:2010ue,Bern:2019prr,Adamo:2022dcm,Bern:2023zkg,Bourjaily:2023ycy,Bourjaily:2023uln}). Most of this work has been done in the context of perturbation theory, either making no assumptions about the particular gauge theory algebra $\mathfrak{g}$ beyond generic identities such as (\ref{jacobi_relation}) or upon restricting to the particular case of $\mathfrak{su}_{N_c}$ gauge theories (especially in the limit of large-$N_c$) (see \emph{e.g.}~\cite{Naculich:2023wyp}). The colour-structure of amplitudes for \emph{particular} gauge theories (beyond those relevant to the Standard Model) especially in the limit of large multiplicity of particles (relative to the number of distinguishable gluons) has been largely unexplored to date.

Obviously, the number of independent colour-tensors relevant to a given process depends on the gauge theory's Lie algebra and the representations and multiplicities of the particles involved. Our goal is to describe how representation theory can be used to assess this number, survey a wide range of particular cases, and describe some of the general features that arise.

\subsection{Representation-Theoretic Counting of Independent Colour-Tensors}

Although the question of how many independent colour-tensors actually arise from the Feynman expansion at a given order of perturbation theory appears rather difficult to answer in general, it turns out that representation theory provides a way to directly answer this question \emph{non-perturbatively}. As described in more detail in \mbox{section~\ref{sec:representation_theory_review}} below, the problem is straight-forward from the perspective of representation theory. 

Without loss of generality (via crossing), we may assume that all the particles are incoming; it is clear that the $S$-matrix for coloured particles must correspond to a mapping from the tensor product of all the particles' representations into the un-coloured vacuum, which itself must transform as the trivial representation, denoted `$\mathbf{1}$'. The number of such mappings counts the rank of linearly-independent tensors, and is therefore determined by the \emph{multiplicity} of the representation $\mathbf{1}$ in the decomposition of this tensor product into \emph{irreducible} representations:
\eq{\mathcal{C}\big[\{\mathbf{R}_i\}\big]\equivR\mathrm{rank}\Big[\Gamma\big[\{\r{c}_i(\mathbf{R}_i)\},\{\}\big]\Big]={m}\Big(\!\bigotimes_i\mathbf{R}_i\!\to\!\mathbf{1}\Big)\,,\vspace{-4pt}}
where ${m}\big(\mathbf{T}\!\to\!\mathbf{R}\big)$ counts the \emph{multiplicity} of the \emph{irreducible} representation $\mathbf{R}$ in the decomposition of the \emph{arbitrary} representation $\mathbf{T}$: 
\eq{\mathbf{T}\simeq\mathbf{R}^{\oplus\b{m}(\mathbf{T}\to\mathbf{R})}\oplus\cdots\,.}
For example, consider the case of scattering $\r{n}$ particles whose colours are all grouped into the adjoint representation, denoted `$\mathbf{ad}$' of $\mathfrak{g}$. Then the number of independent colour-tensors $\mathcal{C}\big[\mathbf{ad}^{\r{n}}\big]\!\equivL\,\mathcal{C}_{\mathfrak{g}}^{\r{n}}$ would be given by ${m}(\mathbf{ad}^{\otimes\r{n}}\!\!\to\!\mathbf{1})$. To be clear, this is not necessarily the same as the number of independent tensors that actually \emph{arise} via the Feynman expansion: for example, this number also counts those colour-factor tensors which could arise only through anomalies. Similarly, for the scattering of $\r{n}$ gluons with $\b{q}$ matter lines transforming under the representation $\mathbf{R}$, the number of independent colour-tensors would be given by 
\eq{\mathcal{C}\big[\mathbf{ad}^{\r{n}},(\mathbf{R}\bar{\mathbf{R}})^{\b{q}}\big]\equivL\,\,\mathcal{C}_{\mathfrak{g}}^{\r{n}\,\b{q}}(\mathbf{R})=m\big(\mathbf{ad}^{\otimes\r{n}}\!\otimes\!(\mathbf{R}\!\otimes\!\bar{\mathbf{R}})^{\otimes\b{q}}\!\!\to\!\mathbf{1}\big)\,.}

As reviewed in \mbox{section~\ref{sec:representation_theory_review}} below, the theory of Lie algebras and their representations readily allows us to determine this number for any particular case of interest. A survey of results uncovers a number of surprising features and some general patterns in the limits of large rank or multiplicity. 

\begin{figure}[b]\vspace{-30pt}$$\fig{-180pt}{1}{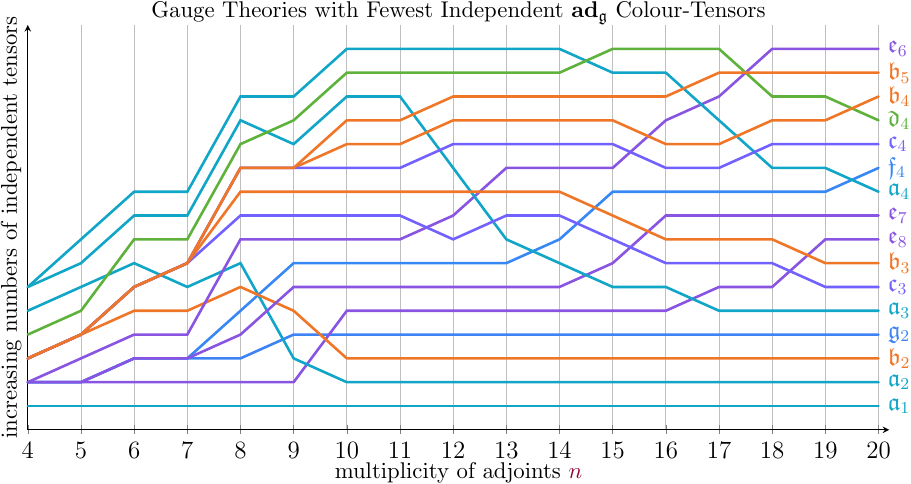}\vspace{-18pt}$$\caption{Gauge theories with fewest numbers of independent colour-tensors required for low-multiplicity scattering. The scale of the vertical axis is arbitrary: this plot merely shows the relative ordering of the Lie algebras with the fewest numbers of independent colour-tensors for each multiplicity $\r{n}$. \label{gauge_theories_with_fewest_tensors}}\vspace{-40pt}\end{figure}

Consider again the case of scattering $\r{n}$ adjoint-coloured (particles such as gluons). Although it is broadly true that the number of independent colour-structures grows with the number of distinguishable gluons, many exceptions arise at low multiplicity. For example, for fewer than $\r{10}$ particles, $\mathfrak{e}_\b{8}$ gauge theory admits \emph{fewer} independent colour-tensors than \emph{any other gauge group} besides $\mathfrak{a}_1\!\!\sim\!\mathfrak{su}_\b{2}$. In \mbox{Figure~\ref{gauge_theories_with_fewest_tensors}}, we list the gauge theories with the fewest independent colour-tensors for $\r{n}\!\leq\!20$ adjoint-charged particles. 

\newpage
\subsection{Organization and Outline}
This paper is organized as follows. In section~\ref{sec:representation_theory_review} we review in some generality the necessary ingredients from the theory of simple Lie algebras and their representations.  In section~\ref{sec:gluonic_amplitudes} we discuss the counting of colour-structures for processes involving only particles charged under the adjoint representation for all simple Lie algebras. In particular, we investigate the limit of large number of particles and, when applicable, the large-rank limit.  Here we encounter a few surprises, such as an excess for orthogonal groups in even dimensions, which we explain in section~\ref{subsec:dtensors}. Surprisingly, the number of invariants for $\mathfrak{e}_8$ at low multiplicity of gluons turns out to be remarkably small when compared to other algebras of much smaller dimension and rank.  

In section~\ref{sec:adjoints_and_matter} we discuss scattering of gluons (or other adjoints) with charged matter in various representations and the saturation of the number of independent colour-structures when the rank increases.  It turns out that in some cases, such as for orthogonal groups and matter charged in the spinor representation, the number of independent colour-structures increases with the rank, which forbids any coherent notion of a large-rank limit to exist. In section~\ref{sec:open_problems} we conclude with a discussion of some open problems and future work.\\

We have also included a number of appendices which detail many results and the methods used to obtain them. In Appendix~\ref{computing_multiplicities_appendix} we describe structurally how counting the number of independent tensors can be recursively constructed from answering this question for arbitrary triples of representations, and in Appendix~\ref{racahspeiser-Appendix} we discuss the Racah-Speiser method for decomposing tensor products to compute these numbers concretely. In Appendix~\ref{appendix_enumerating_ranks} we present tables for the counting of adjoints scattering for a range of lie algebras and multiplicities, and in Appendix~\ref{asymptotic_appendix} we describe the asymptotics of this counting as the number of adjoint-charged particles grows arbitrarily large.\\

Finally, we have included with our submission an ancillary file (accessible from the right-panel of this work's abstract page on \texttt{arXiv}) `\texttt{adjoint\_tensor\_ranks.tex}' which enumerates the numbers of independent colour-tensors for adjoint scattering for various multiplicities for all rank $k\!\leq\!8$ gauge theories. This data is formatted as a list of entries of the form $\{\mathfrak{g},\{n_1,n_2,\ldots\}\}$ where the first integer listed denotes the number of colours for $1$-particle scattering, and so on.

\newpage
\vspace{0pt}%
\section{Review: Lie Algebras and Their Representations}\label{sec:representation_theory_review}\vspace{0pt}

We begin with a cursory review of the essential ideas we require from the study of Lie algebras and their representations mostly in order to establish notation and ideas that will prove useful to us in what follows. This is of course a vast subject in mathematics; we refer the reader to \emph{e.g.}~\cite{Slansky:1981yr,MR1153249, MR1424041} for more thorough discussions and background.

\subsection{Generators, Structure Constants, and (Adjoint) Representations}\label{subsec:generators}\vspace{0pt}

Physicists often encounter the theory of Lie algebras and their representations through particular examples (often motivated by Lie \emph{groups}). In this spirit, let us start with the view that one has some particular representation of some particular Lie algebra on hand. Such a \emph{representation}, denoted `$\mathbf{R}$' consists of a collection of $\b{k}\!\times\!\b{k}$ matrices $\{\underline{T}_{\mathbf{R}}^{\r{a}}\}$ indexed by $\r{a}$ which commute into themselves according to (\ref{representation_definition}):
\eq{\big[\underline{T}_{\mathbf{R}}^{\r{a}},\underline{T}_{\mathbf{R}}^{\r{b}}\big]=f^{\r{a}\,\r{b}}_{\phantom{a\,b\,}\g{c}}\,\underline{T}_{\mathbf{R}}^{\g{c}}\,.\label{matrix_version_of_representation}}
These matrices are called the \emph{generators} of the representation, their size $\b{k}$ is the representation's \emph{dimension}; and they come in a collection indexed by $\r{a}$ which spans the \emph{algebra}'s dimension. As $\b{k}\!\times\!\b{k}$ matrices, they may be viewed as encoding particular linear maps that {act} upon a vector space (or `module') of dimension $\b{k}$. 

For many Lie algebras, there is a \emph{defining} or `\emph{fundamental}'\footnote{This is the physicists' language; more commonly, any representation whose Dynkin labels sum to 1 would be called fundamental. We use the more restrictive meaning here for reasons discussed below in \mbox{section~\ref{subsec:dtensors}}.} representation `$\mathbf{F}$'. For example, the space of antisymmetric $\b{k}\!\times\!\b{k}$ matrices is spanned by the set ${{T}_{\mathbf{F}}}^{\hspace{-5pt}\r{(ij)}}_{\hspace{-7pt}\phantom{(ij)}\b{r\,s}}\equivR\delta^{\r{i}}_{\b{r}}\delta^{\r{j}}_{\b{s}}{-}\delta^{\r{i}}_{\b{s}}\delta^{\r{j}}_{\b{r}}$ indexed by $(\r{ij})\!\in\!\binom{[\b{k}]}{2}$ furnish a $\b{k}$-dimensional representation of the Lie algebra $\mathfrak{so}(\b{k})$---the \emph{fundamental} representation. In this case, it is not hard to see that these matrices satisfy (\ref{matrix_version_of_representation}); the structure constants are given (in terms of the metric $\eta$)
\eq{f^{\r{(a\,b)}\,\r{(c\,d)}}_{\phantom{(i\,j)\,(j\,k)\,}\r{(e\,f)}}={-}\frac{1}{2}\eta^{\r{a\,c}}\delta_{\r{e}}^{\r{b}}\delta_{\r{f}}^{\r{d}}{+}\text{symmetries}\,}
where the omitted terms can be obtained by anti-symmetrizing in $\{\r{(a\,b)},\r{(c\,d)},\r{(e\,f)}\}$. Although in this case the structure constants are fully antisymmetric, this need not be true in general: from the definition in (\ref{matrix_version_of_representation}), it is only necessary that $f^{\r{a\,b}}_{\phantom{a\,b\,}\g{c}}$ be antisymmetric in its first two indices. The fully anti-symmetric tensor $f^{\r{a\,b\,c}}$ is related to the those appearing in (\ref{matrix_version_of_representation}) via
\eq{f^{\r{a\,b\,c}}\equivR\,f^{\r{a\,b}}_{\phantom{a\,b}\,\g{d}}\,\kappa^{\r{c}\,\g{d}}}
where $\kappa^{\r{a}\,\r{b}}$ is the symmetric \emph{Cartan-Killing} form, given by\footnote{As defined this way, `$\kappa^{a b}$' depends on the representation $\mathbf{R}$. However, it turns out that the $\kappa$s for different irreducible representations are always proportional. It is more conventionally defined by the trace in the representation $\mathbf{ad}$ and dividing by twice the dual Coxeter number of the algebra.  For other representations, the proportionality factor is called the \emph{Dynkin index} of the representation.}
\eq{\kappa^{\r{a\,b}}\equivR\mathrm{tr}\big(\underline{T}_{\mathbf{R}}^{\r{a}}.\underline{T}_{\mathbf{R}}^{\r{b}}\big)\equivR\mathrm{tr}_{\mathbf{R}}(\r{a\,b})\equivR{{T}_{\mathbf{R}}}^{\hspace{-5pt}\r{a}\,\g{i}}_{\hspace{-5pt}\phantom{a\,i\,}\g{j}}{{T}_{\mathbf{R}}}^{\hspace{-5pt}\r{b}\,\g{j}}_{\hspace{-5pt}\phantom{a\,i\,}\g{i}}\,.}
A Lie algebra is \emph{semi-simple} iff $\kappa$ is non-degenerate, in which case it can be used as a metric to raise/lower adjoint indices by defining $\kappa_{\r{a\,b}}\equivR(\kappa^{\r{a\,b}})^{{-}1}$. Also, using our definitions here it is not hard to show that the antisymmetric $f^{\r{a\,b\,c}}$ is related to the generators via
\eq{f^{\r{a\,b\,c}}=\mathrm{tr}_{\mathbf{R}}(\r{a\,b\,c}){-}\mathrm{tr}_\mathbf{R}(\r{b\,a\,c})\,.\label{f_as_traces}}

(We apologize for what might appear to be an excessive degree of caution in our discussion above. Physicists often choose linear combinations of generators in order to diagonalize $\kappa_{\r{a\,b}}\!\mapsto\!\kappa_{\r{a'}\r{b'}}\!=\!\delta_{\r{a'b'}}$, in which case there would be no difference between raised/lowered indices. But such a choice frequently requires that generators involve algebraic numbers such as $\sqrt{2}$ or $i$, and such numbers \emph{dramatically} encumber most computer algebra software. For (mostly) this reason, we have been careful to avoid such (often implicit) choices in our discussions here.)\\

Regardless of what particular representation is used to define an algebra, there are \emph{infinitely many} \emph{other} sets of other matrices---different sets of generators furnishing \emph{different} representations---which all satisfy (\ref{matrix_version_of_representation}) \emph{with the same coefficients} $f^{\r{a\,b}\,}_{\phantom{a\,b}\,\r{c}}$. We will soon have more to say about the range of possible representations, but for now it is worth noting that given \emph{any} generators which satisfy (\ref{matrix_version_of_representation}), there is another important representation defined directly in terms of the coefficients $f^{\r{a\,b}}_{\phantom{a\,b}\,\r{c}}$: the `adjoint' representation denoted `$\mathbf{ad}$' and defined via: 
\eq{{T^{\r{a}}_\mathbf{ad}}^{\hspace{-2pt}\phantom{}\,\b{b}}_{\hspace{-0pt}\phantom{\b{b}}\,\b{c}}\equivR\,{-}f^{\r{a}\,\b{b}}_{\phantom{a\,b\,}\b{c}}\,.}
It is easy to see that these matrices satisfy exactly the same commutation relations with the same coefficients as in (\ref{matrix_version_of_representation}). Moreover, when applied to these matrices, the identity in (\ref{matrix_version_of_representation}) can be seen as nothing but the Jacobi relation (\ref{jacobi_relation}). Expressed the other way, given any rank-three tensor which satisfies the Jacobi relation, there exists a set of generators which commute into themselves according to (\ref{matrix_version_of_representation}) and thereby define a Lie algebra.

It is natural to wonder about the scope of possibilities for objects that satisfy (\ref{jacobi_relation}) or (\ref{matrix_version_of_representation}). Indeed, the classification of Lie algebras and their possible representations is one of the great achievements of mathematics (see refs.~\cite{MR1510529, MR1509040, MR1509054}). We will not have time to do justice to the vast body of research on this subject, but it will prove helpful later on to summarize a few basic facts and ideas, starting with the classification of Lie algebras. 

\newpage
\subsection{The Classification of Lie Algebras}\label{sec:lie-algebras}\vspace{0pt}
Suppose one is given a set of generators satisfying (\ref{matrix_version_of_representation}) for some set of coefficients $f^{\r{a\,b}}_{\phantom{a\,b}\,\r{c}}$. If the Cartan-Killing form is non-degenerate, then the algebra is said to be \emph{semi-simple}. (If not, there exists a subspace of generators---of dimension equal to the degeneracy of the Cartan-Killing form---which form a commuting sub-algebra; these can always be projected out to yield a semi-simple sub-algebra.) A general theorem is that any semi-simple Lie algebra can be written as a product of \emph{simple} Lie algebras, whose classification we now review. 

There are four infinite families of `classical' Lie algebras indexed by their \emph{rank} $\b{k}$ and five \emph{exceptional} algebras. These are listed in Table~\ref{classification_of_simple_lie_algebras_table}. Notice that the restriction in ranges of allowed ranks enforces non-redundancy and simplicity: although some authors may speak of `$\mathfrak{sp}_\b{2}$', it would be isomorphic to $\mathfrak{b}_\b{2}$; similarly, `$\mathfrak{d}_\b{3}$' would be isomorphic to $\mathfrak{a}_{\b{3}}$, and `$\mathfrak{d}_\b{2}$' would not in fact be simple: $\mathfrak{d}_2 \!\sim\!\mathfrak{a}_\b{1}\!\times\!\mathfrak{a}_\b{1}$. 

\begin{table}[t]\caption{The simple Lie algebras' ranks, dimensions, and \emph{fundamental} representations.}\label{classification_of_simple_lie_algebras_table}\vspace{-8pt}$$\begin{tabular}{|l@{$\;$}c@{$\;\;\;$}c@{$\;\;$}c|}\hline
\text{algebra}&\text{ranks}&\text{dim}($\mathbf{ad}$)&\text{$\mathrm{dim}(\mathbf{F})$}\\\hline\hline
$\mathfrak{a}_\b{k}\!\sim\!\mathfrak{su}_{\b{k}{+}1}$&$\b{k}\geq1$&$\b{k}(\b{k}+2)$&${\b{k}{+}1}$\\
$\mathfrak{b}_\b{k}\!\sim\!\mathfrak{so}_{2\b{k}{+}1}$&$\b{k}\geq2$&$\b{k}(2\b{k}+1)$&${2\b{k}{+}1}$\\
$\mathfrak{c}_\b{k}\!\sim\!\mathfrak{sp}_{\b{k}}$&$\hspace{0pt}\b{k}\geq3$&$\b{k}(2\b{k}+1)$&${2\b{k}}$\\
$\mathfrak{d}_\b{k}\!\sim\!\mathfrak{so}_{2\b{k}}$&$\hspace{0pt}\b{k}\geq4$&${\b{k}(2\b{k}{-}1)}$&${2\b{k}}$\\\hline
\multirow{3}{*}{$\mathfrak{e}_\b{k}\hspace{36pt}\left\{\rule{0pt}{25pt}\right.$\hspace{-40pt}}&$\b{k}=6$&${78}$&${27}$\\
&$\b{k}=7$&$133$&${56}$\\
&$\b{k}=8$&$248$&${248}$\\
$\mathfrak{f}_\b{k}$&$\b{k}=4$&${52}$&${26}$\\
$\mathfrak{g}_\b{k}$&$\b{k}=2$&${14}$&${7}$\\\hline
\end{tabular}\vspace{-20pt}$$\end{table}
Among the possible Lie algebras for gauge theory, a great deal of attention has been paid to the case of $\mathfrak{su}_{N_c}\!\sim\!\mathfrak{a}_{N_c{-}1}$. Beyond the obvious phenomenological motivations, there are important simplicities afforded by the `large-$N_c$' limit of such gauge theories. By comparison, relatively little attention has been paid to gauge theories involving other Lie algebras (at least insofar as \emph{particular} gauge theories are considered).\footnote{There are important outliers, however---most notably, $\mathfrak{d}_\b{5},\mathfrak{e}_\b{6},$ and $\mathfrak{e}_\b{8}$, which have played important roles in unified model building and string theory.}

\subsection{The Classification of Lie Algebras' Representations}\label{sec:labelling_irreps}\vspace{0pt}

Any interactions between massless spin-one particles (`gluons') and matter must be dictated by some \emph{representation} of the Lie algebra encoded by the gluons' self-interactions. The scope of distinct representations of a given Lie algebra begins---as with the classification of Lie algebras---with a mention of \emph{reducibility}, as it is a general theorem that {any} finite-dimensional representation of a Lie algebra may be expressed as a sum over \emph{irreducible representations} with various multiplicities. And so, let us first review the classification of the \emph{irreducible} representations of simple Lie algebras. 

\subsubsection{Irreducible Representations Simple Lie Algebras}\label{subsubsec:irreps}

When considering a particular representation of some Lie algebra, physicist often use the size of its generators (equivalently, the \emph{dimension} of the vector space on which these matrices are taken to act) as a convenient label. This works well in the case of $\mathfrak{a}_\b{1}\!\sim\!\mathfrak{su}_2$, for which all irreducible representations are uniquely identified by a single positive integer $\mathbf{2j{+}1}$, where $j\!\in\!\text{\textonehalf}\mathbb{Z}$ is called the `spin' of the representation. In this language, $\mathbf{ad}({\mathfrak{a}_\b{1}})$ would be denoted `$\mathbf{3}$', a spin-1/2 representation a `$\mathbf{2}$', and so-on.

This works reasonably well and is often sufficient, but there are many cases where inequivalent representations of a Lie algebra have the same dimension. For example, there are $9$ inequivalent $\mathbf{672}$-dimensional irreducible representations of $\mathfrak{d}_\b{4}$, and distinguishing between them can require considerable notational complications. 

Nevertheless, the simplicity of the case of $\mathfrak{a}_\b{1}$ is not entirely spurious. Physicists first learn to identify representations of $\mathfrak{su}_2$ by their `spin', which we can think of as the largest possible eigenvalue of spin in the `$z$-direction'. What is so special about the `$z$-direction'? In the standard presentation of the generators of $\mathfrak{su}_2$ due to Pauli \cite{Pauli:1927qhd},
\eq{\underline{T}^{\r{x}}\equivR\!\frac{1}{2}\left(\begin{array}{@{}cc@{}}0&1\\1&0\end{array}\right),\quad\underline{T}^{\r{y}}\equivR\!\frac{1}{2}\left(\begin{array}{@{}cc@{}}0&\fwbox{15pt}{{-}i\phantom{{-}}}\\\fwbox{15pt}{i}&0\end{array}\right),\quad \underline{T}^{\r{z}}\equivR\!\frac{1}{2}\left(\begin{array}{@{}cc@{}}\fwbox{15pt}{1}&0\\0&\fwbox{15pt}{{-}1\phantom{{-}}}\end{array}\right)\,.}
The generator $\underline{T}^{\r{z}}$ is diagonal, and can be viewed as the generator of the 1-dimensional \emph{Cartan} sub-algebra of $\mathfrak{su}_2$. Its eigenvalues are $\pm\text{\textonehalf}$. For any other representation, one may consider the eigenvalues of the generator $\underline{T}^{\r{z}}_\mathbf{R}$ and thereby construct a `\emph{lattice}' (in this case, simply a list) of its eigenvalues or `weights'. The weights of an irreducible representation of $\mathfrak{su}_2$ are $\{j,j{-}1,\ldots,1{-}j,-j\}$ for some $j\!\in\!\text{\textonehalf}\mathbb{Z}_{\geq0}$. As most physicists learn early on, we may label representations by their `highest-weight' `${j}$' or by their dimension `$(\mathbf{2j{+}1})$'.\\

Thinking of lattices of weights encoding eigenvalues (of some special generators) and labelling representations by their `highest' weights turns out to be quite general and powerful. For every representation, we can describe a set of weights corresponding to the eigenvalues of the generators of the Cartan sub-algebra (generated by the maximal subspace of mutually commuting generators). Moreover, \emph{irreducible} representations are uniquely characterized by their \emph{highest-weights}---those vectors from which an associated set of weights may be obtained by iteratively subtracting some number of so-called positive simple roots (corresponding to the rows of the Cartan matrix). 

Labelling irreducible representations by their highest-weights is closely related to a simple but deep theorem: the irreducible representations of any rank-$\b{k}$ Lie algebra $\mathfrak{g}$ are uniquely characterized by a $\b{k}$-tuple of non-negative integers called its `Dynkin label' and every such $\b{k}$-tuple labels an irreducible representation of $\mathfrak{g}$. Given the Dynkin label for any irreducible representation of a Lie algebra, one can readily determine the representation's dimension and construct its system of weights. Both these labels and their associated weights prove extremely useful computationally---especially in the decomposition of tensor products into irreducible representations. We refer the reader to more thorough treatments elsewhere (\emph{e.g.}~\cite{MR1510529, MR1509040, MR1509054}).

\subsubsection{Conjugate Representations of Irreducible Representations}\label{subsubsec:conjugates}
An important tool we will use is \emph{Schur's lemma}.  To state it, we first introduce the notion of intertwiners.  Given representations $\mathbf{R}_i$ acting on complex vector spaces $V_i$, then a linear map $f \colon V_1 \to V_2$ which commutes with the action of the representations (in the sense that $f(\mathbf{R}_1(T)) = \mathbf{R}_2(f(T))$ for all $T \in \mathfrak{g}$) is called an \emph{intertwiner}.  The same definition can be made for group representations and the condition for Lie algebra representations can be obtained by expanding around the identity and using the linearity of $f$.

The intertwiners form a complex vector space, $\operatorname{Hom}(\mathbf{R}_1, \mathbf{R}_2)$.  Schur's lemma in this language is the statement that if $\mathbf{R}_1$ and $\mathbf{R}_2$ are irreducible and isomorphic, then the space $\operatorname{Hom}(\mathbf{R}_1, \mathbf{R}_2)$ is one dimensional, generated by the identity intertwiner.  Otherwise, $\operatorname{Hom}(\mathbf{R}_1, \mathbf{R}_2)$ is the trivial vector space containing only the zero intertwiner.

We will use Schur's lemma together with the following properties
\eq{\begin{split}  \operatorname{Hom}(\oplus_i \mathbf{R}_i, \mathbf{R}) = \oplus_i \operatorname{Hom}(\mathbf{R}_i, \mathbf{R}), \\
  \operatorname{Hom}(\mathbf{R}, \oplus_i \mathbf{R}_i) = \oplus_i \operatorname{Hom}(\mathbf{R}, \mathbf{R}_i), \\
  \operatorname{Hom}(\mathbf{A}, \mathbf{B}) = \operatorname{Hom}(\mathbf{A} \otimes \bar{\mathbf{B}}, \mathbf{1}),
\end{split}}
where $\bar{\mathbf{R}}$ is the \emph{dual} representation of a given representation $\mathbf{R}$.\footnote{In the mathematics literature, this is often denoted `$\mathbf{R}^*$'.}

For $f\!\in\!\operatorname{Hom}(\mathbf{A},\mathbf{B})$ we have $f\!\colon\!\!\mathbf{A}\!\to\!\mathbf{B}$, an intertwiner.  Hence for $a\!\in\!\mathbf{A}$ we obtain $f(a)\!\in\!\mathbf{B}$.  Similarly, for $g\!\in\!\operatorname{Hom}(\mathbf{A}\!\otimes\!\bar{\mathbf{B}}, \mathbf{1})$, we have $g\!\colon\!\!\mathbf{A}\!\otimes\!\bar{\mathbf{B}}\!\to\!\mathbb{C}$, where we identify the trivial representation $\mathbf{1}$ with the base field $\mathbb{C}$.  In this case, for $a\!\in\!\mathbf{A}$ and $\bar{b} \in \bar{\mathbf{B}}$, we obtain $g(a\!\otimes\!\bar{b})\!\in\!\mathbb{C}$.  We can establish a correspondence between $f$ and $g$ by taking $g(a\!\otimes\!\bar{b})\!=\!\bar{b}(f(a))$.  One can check that this is well-defined, that is it yields the same answer for $\alpha (a\!\otimes\bar{b}){=}(\alpha a)\!\otimes\!\bar{b}{=}a\!\otimes\!(\alpha \bar{b})$ for $\alpha\!\in\!\mathbb{C}$ and also for $a_1\!\otimes\!b{+}a_2\!\otimes\!b{=}(a_1{+}a_2)\!\otimes\!b$, etc.

If $\mathbf{B}{=}\bar{\mathbf{C}}$ we have, by the bi-dual theorem, since all these spaces are finite-dimensional, that $\bar{\mathbf{B}}{=}\mathbf{C}$.  Taking $c\!\in\!\mathbf{C}$, we can define $f$ in terms of $g$ via \mbox{$f(a)(c){=}g(a\!\otimes\!c)$}.  It can be checked as above that this is well-defined.

The dual (sometimes called contragredient) representation is defined as follows.  Consider a vector space $V$, then its dual space $\bar{V}$ is the set (actually a vector space as well) of linear functionals on $V$.  In other words, $f\!\in\!\bar{V}$ if $f(v)\!\in\!\mathbb{C}$ for all $v\!\in\!V$ and $f(v_1{+}v_2){=}f(v_1){+}f(v_2)$ and $f(\alpha v){=}\alpha f(v)$.

Next, consider a left action of a group $G$ on $v$.  In other words, we have a map $G\!\times\!V\!\to\!V$ such that $(g_1 g_2) v{=}g_1 (g_2 v)$ and $g (v_1{+}v_2){=}g v_1{+}g v_2$ (that is, $V$ is a right module for $G$).  From this we can build a dual action on $\bar{V}$ by taking
\eq{(g f)(v) = f(g^{-1} v),\qquad\forall f\!\in\!\bar{V}, \quad v\!\in\!V, \quad g\!\in\!G.}
The $g^{-1}$ can be understood by requiring that $(g_1 g_2) f{=}g_1 (g_2 f)$.  Alternatively, we could take $G$ to be a \emph{left} action on $\bar{V}$ and in that case we can take
\eq{
(f g)(v) = f(g v),\qquad\forall f\!\in\bar{V}, \quad v\!\in\!V, \quad g\!\in\!G.}

Let us look at this in a given basis.  We take $e_i$ to be a basis of $V$ and $e^i$ to be the dual basis in $\bar{V}$.  Therefore $e^i(e_j) = \delta_j^i$.  Then we can decompose $f{=}f_i e^i$, $v{=}v^i e_i$.  Taking $g^{-1} e_i{=}(g^{-1})_i{}^j e_j$ and $g e^i = (\bar{g})^i{}_j e^j$ and plugging into the definition of the definition of the dual action, we find
\eq{(\bar{g})^i{}_j = (g^{-1})_j{}^i.}
This derivation was done for Lie groups; the analogous statement for Lie algebras can be obtained by expanding around identity, as usual.

Using Schur's lemma we can say that the multiplicity of the trivial representation in $\mathbf{R}_1\!\otimes \cdots \otimes\!\mathbf{R}_p$ is given by
\eq{\operatorname{dim}\big(\!\operatorname{Hom}(\mathbf{R}_1\!\otimes \cdots \otimes\!\mathbf{R}_p, \mathbf{1})\big).}
This is the same as the number of invariant tensors or the number of independent colour-structures. The same argument holds if we take some of the particles to be incoming and others to be outgoing.  In that case the number of independent colour-structures is
\eq{\begin{split}
  &\operatorname{dim}\big(\!\operatorname{Hom}(\mathbf{R}_1\!\otimes\!\cdots \otimes\!\mathbf{R}_p,~ \mathbf{R}_{p + 1}\!\otimes\!\cdots\!\otimes\!\mathbf{R}_{p + q})\big)\\
  =&\operatorname{dim}\big(\!\operatorname{Hom}(\mathbf{R}_1\!\otimes\!\cdots \otimes\!\mathbf{R}_p\!\otimes\!\bar{\mathbf{R}}_{p + 1}\!\otimes\!\cdots\!\otimes\!\bar{\mathbf{R}}_{p + q},~ \mathbf{1})\big).
\end{split}}

A consequence of Schur's Lemma is that for any irreducible representation $\mathbf{R}$ of simple Lie algebra $\mathfrak{g}$, there is a unique \emph{dual} or `conjugate' representation $\bar{\mathbf{R}}$ such that 
\eq{\mathbf{R}\!\otimes\!\bar{\mathbf{R}}=\mathbf{1}\!\oplus\!\ldots\,.}

Representations for which $\bar{\mathbf{R}}\!\simeq\!\mathbf{R}$ are called \emph{real} (or pseudoreal\footnote{The distinction between real and pseudoreal does not matter for us here; but if $\mathbf{R}\!\otimes_{\text{sym}}\!\mathbf{R}{=}\mathbf{1}\!\oplus\ldots$ then the representation is called real and if  $\mathbf{R}\!\otimes_{\text{anti-sym}}\!\mathbf{R}{=}\mathbf{1}\!\oplus\ldots$ then it would be called pseudoreal.}), and those for which $\bar{\mathbf{R}}\!\not\simeq\!\mathbf{R}$ are called \emph{complex}. It is not hard to see that the only algebras which admit complex representations are $\mathfrak{a}_{k>1}$, $\mathfrak{d}_{2k{+}1}$ and $\mathfrak{e}_6$; examples of such include the fundamental representations of $\mathfrak{a}_{k>1}$ and $\mathfrak{e}_6$, or the spinor representations of $\mathfrak{d}_{2k{+}1}$ for which spinor representations satisfy $\bar{\mathbf{S}_{\pm}}\!\simeq\!\mathbf{S}_{\mp}$. 

The conservation of charge along the propagator for a charged matter particle implies that if it is charged under the representation $\mathbf{R}$, its antiparticle must be charged under the conjugate representation $\bar{\mathbf{R}}$. Moreover, because the adjoint representation is real for all simple Lie algebras, in order for there to be non-vanishing interactions between incoming particles are charged under the representations $\mathbf{R}$ and $\bar{\mathbf{R}}$ to exist, we must have that $\mathbf{ad}\!\otimes\!\mathbf{R}\!\otimes\!\bar{\mathbf{R}}\simeq\mathbf{1}\oplus\ldots$.  The invariant tensor corresponding to this interaction is $\mathbf{R}(T^a)^i{}_j$ where $a$ is an index that runs over a basis of $\mathfrak{g}$, $j$ runs over a basis of the representation space of $\mathbf{R}$ and $i$ over a basis of the representation space of $\bar{\mathbf{R}}$.  This construction works except when $\mathbf{R}{=}\mathbf{1}$.

Using the duality transformation and the reality of $\mathbf{ad}$ we can move $\mathbf{ad}$ to the right hand side to obtain $\mathbf{R}\!\otimes\!\bar{\mathbf{R}}\simeq\mathbf{ad}\!\oplus\ldots$.

\subsection{Decomposition of Tensor Products into Irreducible Representations}\label{tensor_product_decompositions}
As outlined above, the number of independent colour-tensors for a given process involving $\r{n}$ gluons and some numbers of matter charged under various representations is computed by the multiplicity of the trivial representation $\mathbf{1}$ in the tensor product
\eq{\mathbf{ad}^{\otimes\r{n}}\!\bigotimes_{i}(\mathbf{R}_i\!\otimes\!\bar{\mathbf{R}}_i)^{\otimes\b{q_i}}\,.}
We will not review in much detail how this number can be computed, but we mention that it is relatively straightforward in terms of the weights of the various representations involved. Roughly speaking, if $\operatorname{wt}(\mathbf{a})$ and $\operatorname{wt}(\mathbf{b})$ are the sets of weights (with multiplicities) for representations $\mathbf{a},\mathbf{b}$, then the set of weights for $\mathbf{a}\!\otimes\!\mathbf{b}$ is given by
\eq{\operatorname{wt}(\mathbf{a}\!\otimes\!\mathbf{b})=\left\{w_a{+}w_b \mid w_a\in\operatorname{wt}(\mathbf{a}),\; w_b\in\operatorname{wt}(\mathbf{b})\right\}\,.}
These weights can be written as a union of the weights of irreducible representations by successively removing those encoded by their highest weights. The standard procedure for doing this is the Racah-Speiser algorithm.

We give a description of this algorithm in Appendix~\ref{racahspeiser-Appendix} where we present also a few worked out examples.

\newpage
\vspace{0pt}%
\section{Counting Colours for \texorpdfstring{$\r{n}$}{n} Adjoint-Charged Particles}\label{sec:gluonic_amplitudes}\vspace{0pt}
Let us begin our survey of colour-structures in the case of $\r{n}$ particles (such as gluons) charged under the adjoint representation $\mathbf{ad}$ of some Lie algebra. We've seen that the number of independent colour-tensors would be given by  
\eq{\mathcal{C}\big[\mathbf{ad}(\mathfrak{g})^{\r{n}}\big]\equivL\,\mathcal{C}_{\mathfrak{g}}^{\r{n}}=m\!\left(\mathbf{ad}^{\otimes\r{n}}\!\!\to\!\mathbf{1}\right)\,.}
This multiplicity is easy to study in any particular case using the tools of representation theory as outlined above. 

Consider for example the simplest possible gauge theory: the case of $\mathfrak{a}_1\!\sim\!\mathfrak{su}_2$. Recall that for $\mathfrak{a}_1$, irreducible representations are uniquely characterized by their dimension, suggesting that we describe $\mathbf{ad}(\mathfrak{a}_1)$ as the `$\mathbf{3}$' representation; tensor products involving the irreducible representation $\mathbf{d}\!\neq\!\mathbf{1}$ always take the form: 
\eq{\mathbf{3}\!\otimes\!\mathbf{d}\simeq\mathbf{(d{-}1)}\!\oplus\!\mathbf{d}\!\oplus\!\mathbf{(d{+}1)}\,.}
From this, we may recursively compute the decomposition of all $\mathbf{3}^{\otimes\r{n}}$: 
\eq{\fwbox{0pt}{\begin{array}{r@{}c@{}ll}
\mathbf{3}^{\hspace{-1pt}\otimes\r{1}}&{=}&\mathbf{3}\\
\mathbf{3}^{\hspace{-1pt}\otimes\r{2}}&{=}&\mathbf{1}^{\hspace{-1pt}\oplus\b{1}}\hspace{-1pt}\!\oplus\!\mathbf{3}\hspace{-1pt}\!\oplus\!\mathbf{5}\\
\mathbf{3}^{\hspace{-1pt}\otimes\r{3}}&{=}&\mathbf{1}^{\hspace{-1pt}\oplus\b{1}}\hspace{-1pt}\!\oplus\!\mathbf{3}^{\hspace{-1pt}\oplus3}\hspace{-1pt}\!\oplus\!\mathbf{5}^{\hspace{-1pt}\oplus2}\hspace{-1pt}\!\oplus\!\mathbf{7}\\
\mathbf{3}^{\hspace{-1pt}\otimes\r{4}}&{=}&\mathbf{1}^{\hspace{-1pt}\oplus\b{3}}\hspace{-1pt}\!\oplus\!\mathbf{3}^{\hspace{-1pt}\oplus6}\hspace{-1pt}\!\oplus\!\mathbf{5}^{\hspace{-1pt}\oplus6}\hspace{-1pt}\!\oplus\!\mathbf{7}^{\hspace{-1pt}\oplus3}\hspace{-1pt}\!\oplus\!\mathbf{9}\\
\mathbf{3}^{\hspace{-1pt}\otimes\r{5}}&{=}&\mathbf{1}^{\hspace{-1pt}\oplus\b{6}}\hspace{-1pt}\!\oplus\!\mathbf{3}^{\hspace{-1pt}\oplus15}\hspace{-1pt}\!\oplus\!\mathbf{5}^{\hspace{-1pt}\oplus15}\hspace{-1pt}\!\oplus\!\mathbf{7}^{\hspace{-1pt}\oplus10}\hspace{-1pt}\!\oplus\!\mathbf{9}^{\hspace{-1pt}\oplus4}\hspace{-1pt}\!\oplus\!\mathbf{11}\\
\mathbf{3}^{\hspace{-1pt}\otimes\r{6}}&{=}&\mathbf{1}^{\hspace{-1pt}\oplus\b{15}}\hspace{-1pt}\!\oplus\!\mathbf{3}^{\hspace{-1pt}\oplus36}\hspace{-1pt}\!\oplus\!\mathbf{5}^{\hspace{-1pt}\oplus40}\hspace{-1pt}\!\oplus\!\mathbf{7}^{\hspace{-1pt}\oplus29}\hspace{-1pt}\!\oplus\!\mathbf{9}^{\hspace{-1pt}\oplus15}\hspace{-1pt}\!\oplus\!\mathbf{11}^{\hspace{-1pt}\oplus5}\hspace{-1pt}\!\oplus\!\mathbf{13}\\
\mathbf{3}^{\hspace{-1pt}\otimes\r{7}}&{=}&\mathbf{1}^{\hspace{-1pt}\oplus\b{36}}\hspace{-1pt}\!\oplus\!\mathbf{3}^{\hspace{-1pt}\oplus91}\hspace{-1pt}\!\oplus\!\mathbf{5}^{\hspace{-1pt}\oplus105}\hspace{-1pt}\!\oplus\!\mathbf{7}^{\hspace{-1pt}\oplus84}\hspace{-1pt}\!\oplus\!\mathbf{9}^{\hspace{-1pt}\oplus49}\hspace{-1pt}\!\oplus\!\mathbf{11}^{\hspace{-1pt}\oplus21}\hspace{-1pt}\!\oplus\!\mathbf{13}^{\hspace{-1pt}\oplus6}\hspace{-1pt}\!\oplus\!\mathbf{15}\\
\mathbf{3}^{\hspace{-1pt}\otimes\r{8}}&{=}&\mathbf{1}^{\hspace{-1pt}\oplus\b{91}}\hspace{-1pt}\!\oplus\!\mathbf{3}^{\hspace{-1pt}\oplus232}\hspace{-1pt}\!\oplus\!\mathbf{5}^{\hspace{-1pt}\oplus280}\hspace{-1pt}\!\oplus\!\mathbf{7}^{\hspace{-1pt}\oplus238}\hspace{-1pt}\!\oplus\!\mathbf{9}^{\hspace{-1pt}\oplus154}\hspace{-1pt}\!\oplus\!\mathbf{11}^{\hspace{-1pt}\oplus76}\hspace{-1pt}\!\oplus\!\mathbf{13}^{\hspace{-1pt}\oplus28}\hspace{-1pt}\!\oplus\!\mathbf{15}^{\hspace{-1pt}\oplus7}\hspace{-1pt}\!\oplus\!\mathbf{17}\\
\mathbf{3}^{\hspace{-1pt}\otimes\r{9}}&{=}&\mathbf{1}^{\hspace{-1pt}\oplus\b{232}}\hspace{-1pt}\!\oplus\!\mathbf{3}^{\hspace{-1pt}\oplus603}\hspace{-1pt}\!\oplus\!\mathbf{5}^{\hspace{-1pt}\oplus750}\hspace{-1pt}\!\oplus\!\mathbf{7}^{\hspace{-1pt}\oplus672}\hspace{-1pt}\!\oplus\!\mathbf{9}^{\hspace{-1pt}\oplus468}\hspace{-1pt}\!\oplus\!\mathbf{11}^{\hspace{-1pt}\oplus258}\hspace{-1pt}\!\oplus\!\mathbf{13}^{\hspace{-1pt}\oplus111}\hspace{-1pt}\!\oplus\!\mathbf{15}^{\hspace{-1pt}\oplus36}\hspace{-1pt}\!\oplus\!\mathbf{17}^{\hspace{-1pt}\oplus8}\hspace{-1pt}\!\oplus\!\mathbf{19}\,
\end{array}}}
and so on. Thus, the number of independent colour-tensors required for scatting $\r{n}$ adjoints in $\mathfrak{a}_1$ gauge theory would be given by the sequence $\{0,1,1,3,6,15,36,91,232,\ldots\}$. This sequence is recognizable as the \emph{Riordan numbers}, and is given in the Online Encyclopedia of Integer Sequences (`OEIS') as sequence \href{https://oeis.org/A005043}{[A005043]} \cite{oeis-riordan}. In particular, we have that 
%
\eq{\mathcal{C}^{\r{n}}_{\mathfrak{a}_1}=\sum_{r=0}^{\r{n}}({-}1)^r\binom{\r{n}}{r}\binom{r}{\lfloor r/2\rfloor}\,,}
and satisfies the recurrence relation
\eq{\mathcal{C}^{\r{n}}_{\mathfrak{a}_1}{=}\frac{\r{n}{-}1}{\r{n}{+}1}\big(2\,\mathcal{C}^{\r{n{-}1}}_{\mathfrak{a}_1}{+}3\,\mathcal{C}^{\r{n{-}2}}_{\mathfrak{a}_1}\big)\quad\text{with}\quad \mathcal{C}^{\r{0}}_{\mathfrak{a}_1}{=}1,\mathcal{C}^{\r{1}}_{\mathfrak{a}_1}{=}0\,.}

What is especially interesting is that for $\r{n}\!>\!5$-particle scattering, this number is \emph{fewer} than the $(\r{n}{-}2)!$ of independent colour-tensors for gluon scattering \emph{at tree-level} described by del Duca, Dixon and Maltoni (see ref.~\cite{Duca1999NewCD}); and yet, these numbers are valid to \emph{all-orders of perturbation}. Thus, the all-orders colour-structure of $\mathfrak{a}_1$ gauge theory will be considerably simpler than even that of \emph{tree-level} scattering in a algebra-agnostic approach. 

To be fair, the basis of tree- or loop-level colour-tensors described by DDM and others were derived using only generic identities among colour-tensors such as Jacobi (\ref{jacobi_relation}), and therefore should represent a \emph{spanning set} of tensors which lose their independence for particular Lie algebras. (For example, of the 24 colour-tensors appearing in the DDM basis at 6-particles, only 14 are independent for $\mathfrak{a}_1$ gauge theory.)

Although analyses such as that of ref.~\cite{Duca1999NewCD} have been done order-by-order in perturbation theory (and typically include the contributions exclusively arising via Feynman diagrams), they suggest that perhaps we may discover certain universal behavior among the number of independent colour-tensors in the limit of large numbers of independent gluons. This turns out to be the case.

\subsection{\emph{Saturation}: Counting Colours in the Limit of Large-Ranks}\label{subsec:large_rank_limit}\vspace{0pt}
For the classical Lie algebras it is interesting to consider how the number of independent colour-tensors involving adjoints grows with the rank of the algebra. The number of distinguishable gluons, $\mathrm{dim}(\mathfrak{g})$, grows quadratically with the rank; intuitively, when the number of distinguishable gluons vastly exceeds the number involved in a scattering process, the details of so many other states should cease to affect the answer. 

\begin{figure}$$\fig{0pt}{1}{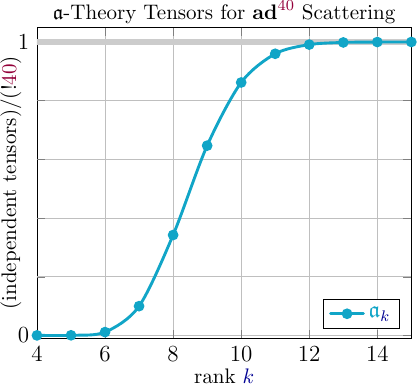}\;\;\;\fig{0pt}{1}{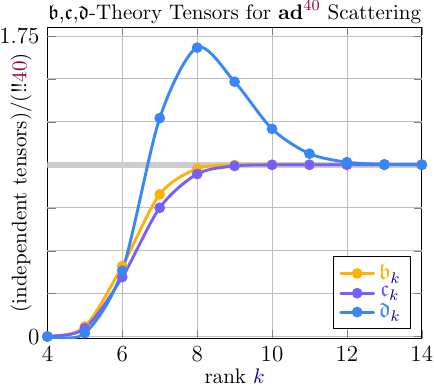}\vspace{-20pt}$$\caption{The numbers of independent colour-tensors for $\r{40}$-adjoint scattering for gluons defined by each of the classical Lie algebras $\mathfrak{a}_{\b{k}},\mathfrak{b}_{\b{k}},\mathfrak{c}_{\b{k}},\mathfrak{d}_{\b{k}}$ as a function of rank $\b{k}$.\label{saturation_figure}}\end{figure}

We can check this intuition by direct computation. Consider the scattering of $\r{n}{=}{40}$ adjoint-charged particles in each of the classical Lie algebras, for which the number of independent colour-tensors (relative to the their asymptotic values) are plotted in \mbox{Figure~\ref{saturation_figure}} as a function of rank. It is easy to see that these numbers \emph{saturate} in the large-rank limit, and to different values in the case of $\mathfrak{a}$-theory than the other algebras. Let us begin there. 

\subsubsection{Saturation in the Case of $\mathfrak{a}$-Type Gauge Theory}
Empirically, it is easily seen that the number of independent colour-tensors grows uniformly as a function of $\b{k}$ until some saturation number is reached for each of the classical Lie algebras---beyond which all algebras of higher rank share the same number of independent tensors. In the case of $\mathfrak{a}_{\b{k}}$-theory, we find these saturation numbers to be reached when $\b{k}\!\geq\!\r{n}{-}1$; for $\r{n}\!\in\!\{1,\ldots,10\}$, we find these numbers to be given by $\{0,1,2,9,44,265,1\,854,14\,833,133\,496,1\,334\,961\}$. This sequence is easily identified in the OEIS \cite{oeis-derangements} as the number of \emph{derangements}, denoted `$!\r{n}$':
\eq{!\r{n}\,\equivR\,\r{n}!\sum_{r{=}0}^{\r{n}}\frac{({-}1)^r}{r!}\,.\label{derangements_explicit}}
More values are given in \mbox{Table~\ref{derangements_table}}. Somewhat surprisingly, there exists an even simpler closed formula for $!\r{n}$ given by:
\eq{!\r{n}=\mathrm{Round}[\r{n}!/e]\,\label{derangements_round_formula}}
(Here, `Round' means what you think it means: the integer nearest to the argument.) 

In the study of integer sequences, an important concept is that of a generating or exponential generating function. Specifically, for any sequence $a(\r{n})$ defined for $\r{n}\!\in\!\mathbb{Z}_{\geq0}$, we may associate with it the \emph{Exponential Generating Function} as follows:
\eq{a(\r{n})\,\,\Leftrightarrow\,\,\underset{a(\r{n})}{\text{EGF}}(\b{x})\equivR\sum_{\r{n}{=}0}^{\infty}\frac{a(\r{n})}{\r{n}!}\b{x}^\r{n}\,.}
It will be useful to note that $!\r{n}$ is encoded by the exponential generating function:
\eq{!\r{n}\,\,\Leftrightarrow\,\,\underset{!\r{n}}{\text{EGF}}(\b{x})\equivR\sum_{\r{n}{=}0}^{\infty}\frac{(!\r{n})}{\r{n}!}\b{x}^{\r{n}}=\frac{e^{{-}\b{x}}}{1{-}\b{x}}\label{derangements-egf}\,.}

At any rate, we have found that for any fixed multiplicity $\r{n}$, $\displaystyle\lim_{\b{k}\to\infty}C^{\r{n}}_{\mathfrak{a}_{\b{k}}}=\,!\r{n}$: 
\eq{\mathcal{C}^{\r{n}}_{\mathfrak{a}_{\b{k}}}=\,\,!\r{n}\qquad\forall\,\,\b{k}\geq\r{n}{-}1\,.\label{saturation_adj_a}}
This number is easy to understand for most physicists (if difficult to justify). Combinatorially, the number of derangements $!\r{n}$ is equal to the number of permutations $\sigma\!\in\!\mathfrak{S}([\r{n}])$ which are free of fixed-points; this number is equivalent to the number of `multi-trace' tensors that any physicist would write down immediately as a (possibly degenerate) basis of tensors.

To see this, we simply note that any permutation can be represented by a collection of \emph{cycles}, and that any fixed point would be correspond to a length-1 cycle.\footnote{Tracelessness of the fundamental generators follows from semi-simplicity of the Lie algebra, so we may discard any tensor involving the trace over a single generator of any representation.} For $\r{n}{=}4$, we have that $!\r{4}=9$; this is easy to see directly:
\renewcommand{\arraystretch}{0.25}\eq{\begin{array}{c}\begin{array}{lcccccc}\fwboxR{0pt}{\text{permutation:}}&\left(\!\begin{array}{@{$\,$}c@{$\;$}c@{$\;$}c@{$\;$}c@{$\,$}}1&2&3&4\\
\downarrow&\downarrow&\downarrow&\downarrow\\[2pt]
2&3&1&4\end{array}\!\right)&\left(\!\begin{array}{@{$\,$}c@{$\;$}c@{$\;$}c@{$\;$}c@{$\,$}}{\color{black}1}&{\color{black}2}&{\color{black}3}&{\color{black}4}\\
{\color{black}\downarrow}&{\color{black}\downarrow}&{\color{black}\downarrow}&\downarrow\\[2pt]
{\color{black}2}&{\color{black}4}&{\color{black}1}&{\color{black}3}\end{array}\!\right)
&\left(\!\begin{array}{@{$\,$}c@{$\;$}c@{$\;$}c@{$\;$}c@{$\,$}}1&2&3&4\\
\downarrow&\downarrow&\downarrow&\downarrow\\[2pt]
3&4&2&1\end{array}\!\right)
&\left(\!\begin{array}{@{$\,$}c@{$\;$}c@{$\;$}c@{$\;$}c@{$\,$}}1&2&3&4\\
\downarrow&\downarrow&\downarrow&\downarrow\\[2pt]
3&1&4&2\end{array}\!\right)
&\left(\!\begin{array}{@{$\,$}c@{$\;$}c@{$\;$}c@{$\;$}c@{$\,$}}1&2&3&4\\
\downarrow&\downarrow&\downarrow&\downarrow\\[2pt]
4&3&1&2\end{array}\!\right)
&\left(\!\begin{array}{@{$\,$}c@{$\;$}c@{$\;$}c@{$\;$}c@{$\,$}}1&2&3&4\\
\downarrow&\downarrow&\downarrow&\downarrow\\[2pt]
4&1&2&3\end{array}\!\right)
\\[14pt]
\fwboxR{0pt}{\text{cycles:}}&(1\,2\,3\,4)&(1\,2\,4\,3)&(1\,3\,2\,4)&(1\,3\,4\,2)&(1\,4\,2\,3)&(1\,4\,3\,2)\end{array}\\[24pt]
\begin{array}{lccc}
&\left(\!\begin{array}{@{$\,$}c@{$\;$}c@{$\;$}c@{$\;$}c@{$\,$}}\r{1}&\r{2}&\b{3}&\b{4}\\
\r{\downarrow}&\r{\downarrow}&\b{\downarrow}&\b{\downarrow}\\[2pt]
\r{2}&\r{1}&\b{4}&\b{2}\end{array}\!\right)&\left(\!\begin{array}{@{$\,$}c@{$\;$}c@{$\;$}c@{$\;$}c@{$\,$}}\r{1}&\b{2}&\r{3}&\b{4}\\
\r{\downarrow}&\b{\downarrow}&\r{\downarrow}&\b{\downarrow}\\[2pt]
\r{3}&\b{4}&\r{1}&\b{2}\end{array}\!\right)
&\left(\!\begin{array}{@{$\,$}c@{$\;$}c@{$\;$}c@{$\;$}c@{$\,$}}\r{1}&\b{2}&\b{3}&\r{4}\\
\r{\downarrow}&\b{\downarrow}&\b{\downarrow}&\r{\downarrow}\\[2pt]
\r{4}&\b{3}&\b{2}&\r{1}\end{array}\!\right)
\\[14pt]
&(\r{1\,2})(\b{3\,4})&(\r{1\,3})(\b{2\,4})&(\r{1\,4})(\b{2\,3})\end{array}\end{array}\hspace{-40pt}\renewcommand{\arraystretch}{1}}
This association leads to a set of colour-tensors for the scattering of $\r{4}$ adjoint-charged particles of the form \renewcommand{\arraystretch}{1}
\eq{\left\{\begin{array}{c}\mathrm{tr}_\mathbf{R}(1\,2\,3\,4),\mathrm{tr}_\mathbf{R}(1\,2\,4\,3),\mathrm{tr}_\mathbf{R}(1\,3\,2\,4),\mathrm{tr}_\mathbf{R}(1\,3\,4\,2),\mathrm{tr}_\mathbf{R}(1\,4\,2\,3),\mathrm{tr}_\mathbf{R}(1\,4\,3\,2),\\
\mathrm{tr}_\mathbf{R}(1\,2)\mathrm{tr}_\mathbf{R}(3\,4),\mathrm{tr}_\mathbf{R}(1\,3)\mathrm{tr}_\mathbf{R}(2\,4),\mathrm{tr}_\mathbf{R}(1\,4)\mathrm{tr}_\mathbf{R}(2\,3)\end{array}\right\}}
for any (complex) representation $\mathbf{R}$. (As discussed below, when $\mathbf{R}\simeq\bar{\mathbf{R}}$, these traces enjoy a dihedral/reflection symmetry among their arguments, leading to a shorter list of only 6 distinct tensors.) 

The reason these numbers may seem somewhat unsurprising to many physicists is that it is common to decompose colour-tensors of $\mathfrak{a}_{\b{k}}$ gauge theory into sums of products of traces over \emph{fundamental}-representation generators via (\ref{f_as_traces}), then to use the `Fierz identity' to eliminate all reference to internal adjoint-labels. The Fierz identity for $\mathfrak{a}_\b{k}$ tensors is often expressed in a form that requires a specific choice of basis for the fundamental generators (and their scaling); but it can be stated without reference to any such basis as follows: 
\eq{\begin{split}\kappa_{\r{a\,b}}\mathrm{tr}_{\mathbf{F}}(\r{a},\g{x\rule[-0.5pt]{10pt}{0.5pt}})\mathrm{tr}_{\mathbf{F}}(\r{b},\g{y\rule[-0.5pt]{10pt}{0.5pt}})&=\mathrm{tr}_\mathbf{F}(\g{x},\g{y}){-}\frac{1}{\b{k}{+}1}\mathrm{tr}_\mathbf{F}(\g{x})\mathrm{tr}_\mathbf{F}(\g{y})\,\\
\kappa_{\r{a\,b}}\mathrm{tr}_{\mathbf{F}}(\r{a},\g{x\rule[-0.5pt]{10pt}{0.5pt}},\r{b},\g{y\rule[-0.5pt]{10pt}{0.5pt}})&=\mathrm{tr}_\mathbf{F}(\g{x})\mathrm{tr}_\mathbf{F}(\g{y}){-}\frac{1}{\b{k}{+}1}\mathrm{tr}_\mathbf{F}(\g{x},\g{y})
\end{split}\label{fierz}}
where $\g{x\rule[-0.5pt]{10pt}{0.5pt}}$,$\g{y\rule[-0.5pt]{10pt}{0.5pt}}$ denote any ordered sequence of indices (including an empty range---for which it should be understood that $\mathrm{tr}_\mathbf{F}(){=}\mathrm{dim}(\mathbf{F}){=}\b{k}{+}1$). Such trace tensors are especially useful in the context of the large-$\b{k}$ limit, as many of the multi-trace tensors will be strongly suppressed by orders of $\b{k}$, leading to a notion of planarity \cite{tHooft:1973jz}. 

Although it may have been easy to have guessed the fact that $\lim_{\b{k}\to\infty}\mathcal{C}_{\mathfrak{a}_{\b{k}}}^{\r{n}}$ saturates to the number of multi-trace tensors, it should be surprising---and in \emph{both} directions. On the one hand, these numbers would seem to be too small: the Fierz decomposition can be applied only to the tensors arising from Feynman diagrams \emph{involving adjoint- or \emph{internal} fundamental-coloured particles}. In the case of $\mathfrak{a}_k$ the fact that this reasoning should extend to internal loops of particles charged under \emph{arbitrary} representations is due to the fact that all irreducible representations of $\mathfrak{a}_{\b{k}}$ can be generated through tensor products of $\mathbf{F}$---a fact not universal to other algebras' fundamental representations (as we have defined them in Table~\ref{classification_of_simple_lie_algebras_table}).

On the other hand, it would appear that this number \emph{over counts} the number of actual tensors that would arise from the Feynman expansion. For $\r{n}{=}3$ (and $\b{k}\!\geq2$---as required to reach saturation), $!\r{3}{=}2$, reflecting the fact that, for $\mathfrak{a}_{k\geq2}$ the tensors
\eq{\mathrm{tr}_\mathbf{F}(\r{a\,b\,c})\quad\text{and}\quad\mathrm{tr}_\mathbf{F}(\r{c\,b\,a})\label{two_three_particle_tensors}}
are generally distinct---that is, linearly-independent. Recall that the structure constants $f^{\r{a\,b\,c}}$ are defined as the difference between these tensors, (\ref{f_as_traces}); the sum defines a distinct and generally non-vanishing rank-3 tensor via:
\eq{d^{\r{a\,b\,c}}\equivR\,\mathrm{tr}_{\mathbf{R}}(\r{a\,b\,c}){+}\mathrm{tr}_{\mathbf{R}}(\r{c\,b\,a})\,.}
(This tensor vanishes for \emph{all} other Lie algebras and representations.) Recall that Bose symmetry required that the constants appearing in gluon interactions must involve only the fully \emph{antisymmetric} $f^{\r{a\,b\,c}}$---and \emph{not} this symmetric tensor. How then can we understand a \emph{need} for such new tensors in our counting? 

The argument leading to the uniqueness of Yang-Mills theory applied only to the interactions of \emph{massless} spin-one particles. For the scattering of such particles, the appearance of a new, symmetric tensor for the scattering of gluons would be equivalent to the appearance of an anomaly in the gauge theory---in conflict with our premise of masslessness for the particles. This is certainly not excluded on general principles: it is of course possible for anomalies to arise in $\mathfrak{a}_k$ gauge theory (if there exists `anomalous' contents of charged matter which can contribute to loops).

If we insisted on only \emph{non-anomalous} gauge theory, we would expect $!\r{n}$ to overcount the number of independent tensors that would arise for the scattering of gluons. To see this, consider \emph{pure} gauge theory---where all Feynman diagrams involve only gluons and all colour-tensors are constructed directly from the antisymmetric tensor  $f^{\r{a\,b\,c}}$ defining the theory; as such, only the differences between reflected traces can arise via (\ref{f_as_traces}). 

In general, traces over the generators of conjugate representations are related by 
\eq{\mathrm{tr}_\mathbf{R}(\r{a\,b\cdots\,c\,d})=({-}1)^{|\r{a\,b\cdots c\,d}|}\mathrm{tr}_\mathbf{\bar{R}}(\r{d}\,\r{c}\cdots\r{b}\r{a})\,.\label{reflected_trace_relations}}
Because the fundamental representation of $\mathfrak{a}_{k\geq2}$ is \emph{complex}, reality of $f^{\r{a\,b\,c}}$ requires that only dihedrally-symmetrized combinations of trace tensors would be required. (We will soon return to the question of how many dihedrally-symmetric traces exist.)

All that being said, there is in fact another context in which even for \emph{non-anomalous} gauge theory the complete basis of $!\r{n}$ tensors may be relevant: the scattering of gluons and \emph{adjoint-charged matter}. The argument leading to the uniqueness of the tensor $f^{\r{a\,b\,c}}$ for gluon scattering is predicated on Bose symmetry, which would not apply to the interactions of gluons and charged particles of other spins. And so, in the more general context of scattering $\r{n}$ `adjoint-charged' particles (of arbitrary spin), we expect that $!\r{n}$ reflects the right number of independent tensors that would be relevant in the large-rank limit. 

Of course, the number of independent colour-dependent tensors which can exist {non-perturbatively} is not necessarily the same as the number which \emph{actually appear} at some order of perturbation theory. Even within the context of pure gauge theory, it remains an important, open question as to how many independent tensors arise order-by-order in perturbation theory; it is a question we must leave to future work.

\begin{table}[b!]\vspace{-10pt}$$\fwbox{0pt}{\begin{array}{|r@{$\,$}|@{$$}r|r|r|r|r|r|r|r|r|r|r@{$\,$}|}\multicolumn{1}{c@{$\,$}|@{$$}}{\,\,\,\r{n}\fwboxL{0pt}{}}&\multicolumn{1}{c}{\r{3}}&\multicolumn{1}{c}{\r{4}}&\multicolumn{1}{c}{\r{5}}&\multicolumn{1}{c}{\r{6}}&\multicolumn{1}{c}{\r{7}}&\multicolumn{1}{c}{\r{8}}&\multicolumn{1}{c}{\r{9}}&\multicolumn{1}{c}{\r{10}}&\multicolumn{1}{c}{\r{11}}&\multicolumn{1}{c}{\r{12}}&\multicolumn{1}{c}{\r{13}}\\\hline\hline
!\r{n}&2&9&44&265&1,\!854&14,\!833&133,\!496&1,\!334,\!961&14,\!684,\!570&176,\!214,\!841&2,\!290,\!792,\!932\\\cline{1-12}
!!\r{n}&1&6&22&130&822&6,\!202&52,\!552&499,\!194&5,\!238,\!370&60,\!222,\!844&752,\!587,\!764\\\cline{1-12}
\end{array}}\vspace{-10pt}$$\caption{Numbers of  \emph{derangements} $!\r{n}$ and \emph{dihedral} derangements $!!\r{n}$.\label{derangements_table}}\vspace{-30pt}\end{table}

\subsubsection{Saturation in the Cases of $\mathfrak{b},\mathfrak{c},\mathfrak{d}$-Type Gauge Theory}
By comparison to the case of the adjoints of $\mathfrak{a}$, we expect that $!\r{n}$ should substantially over-count the number of tensors, as \emph{all} representations are real and therefore dihedrally-related trace tensors cannot be independent. Indeed, from (\ref{reflected_trace_relations}) we expect that only dihedrally-related traces (over any representation's generators) should suffice. Also for $\mathfrak{d}$-type theories, the fundamental representation at least is real---and so only dihedrally-related traces over fundamental generators would be independent. 

We choose to denote this number of `\emph{dihedral} derangements' (the number of multi-trace symbols equivalent under a reversal of their indices) by `$!!\r{n}$'. Counting these for the cases $\r{n}\!\in\!\{1,\ldots,10,\ldots\}$, we find these to be given by the sequence $\{0, 1,1,6,22,130,822,6\,202,52\,552,499\,194,\ldots\}$, which can be identified in the OEIS as the sequence \href{https://oeis.org/A002137}{[A002137]} \cite{oeis-dihedral_derangements}, which we list in \mbox{Table~\ref{derangements_table}}. Although we know of no closed formula for $!!\r{n}$ analogous to (\ref{derangements_explicit}) or (\ref{derangements_round_formula}), the sequence $!!\r{n}$ corresponds to the exponential generating function: 
\eq{!!\r{n}\,\,\Leftrightarrow\,\,\underset{!!\r{n}}{\text{EGF}}(\b{x})\equivR\sum_{\r{n}{=}0}^{\infty}\frac{(!!\r{n})}{\r{n}!}\b{x}^{\r{n}}=\frac{e^{\b{x}^2/4{-}\b{x}/2}}{\sqrt{1{-}\b{x}}}\label{dihedral_derangements-egf}\,.}

This number exactly agrees with the empirical numbers of tensors computed using the decomposition of tensor products into irreducible representations for each of the $\mathfrak{b},\mathfrak{c},\mathfrak{d}$-type gauge theories in the limit of large rank (see \mbox{Tables~\ref{pure_glue_counting_table_b}, \ref{pure_glue_counting_table_d}, \ref{pure_glue_counting_table_d}}). Specifically, we find that saturation occurs for
\eq{\begin{split}\mathcal{C}^{\r{n}}_{\mathfrak{b}_\b{k}}&=\,\,!!\r{n}\qquad\forall\,\, \b{k}\geq\frac{1}{2}(\r{n}{-}1)\,;\\
\mathcal{C}^{\r{n}}_{\mathfrak{c}_\b{k}}&=\,\,!!\r{n}\qquad\forall\,\,\b{k}\geq\frac{1}{2}(\r{n}{-}1)\,;\\
\mathcal{C}^{\r{n}}_{\mathfrak{d}_\b{k}}&=\,\,!!\r{n}\qquad\forall\,\, \b{k}\geq(\r{n}{+}1)\,.\\
\end{split}\label{saturation_condition_bcd}}

One obvious feature of how saturation is reached as $\b{k}$ is increased (see \mbox{Figure~\ref{saturation_figure}}) is that while the limit is approached monotonically for $\mathfrak{b},\mathfrak{c}$ theories, there is more structure for the $\mathfrak{d}$-type algebras: fixing the multiplicity $\r{n}$, we see that $\mathcal{C}^{\r{n}}_{\mathfrak{d}_\b{k}}\!<\,!!\r{n}$ for low-rank algebras, then \emph{exceeds} this `limit' for some range of ranks, before reaching saturation \emph{from above} when $\b{k}\!>\!\r{n}$. Consulting \mbox{Table~\ref{pure_glue_counting_table_d}}, it is curious to note that when $\b{k}{=}\r{n}$, $\mathcal{C}^{\r{n}{=}\b{k}}_{\mathfrak{d}_\b{k}}{=}(!!\r{n}){+}1$.

Notice that $!!\r{n}\!<\,!\r{n}$ for all $\r{n}\!\geq\!3$; and thus, the number of independent colour-tensors for adjoint scattering for \emph{all} classical Lie algebras in the limit of large rank is consistent with a saturation basis consisting of the products of the traces over the generators of the algebra's \emph{fundamental} representation. Below saturation for $\mathfrak{a},\mathfrak{b},\mathfrak{c}$-theories and for all the exceptional algebras, any difference between $\mathcal{C}^{\r{n}}_{\mathfrak{g}}$ and $!\r{n}$ or $!!\r{n}$ (as appropriate) can be understood as reflecting \emph{identities} among the tensors arising at low rank. In the case of $\mathfrak{d}$-theories, we need to explain the appearance of \emph{new} tensors relative to the basis valid in the large-rank limit. 

\subsection{Origins of Identities Among (and the Appearance of New) Tensors}\label{subsection_approaching_saturation}

The counting described above is consistent with the notion that a natural, large-rank {basis} of colour-structures for $\mathbf{ad}^{\otimes\r{n}}$ may consist of the products of traces over the generators of the fundamental representation in each case: $!\r{n}$ counts the number of multi-trace `symbols', and $!!\r{n}$ counts the number of multi-trace symbols where a dihedral symmetry is enforced (valid for multi-trace tensors involving the generators of any \emph{real} representation according to (\ref{reflected_trace_relations})). Indeed, we have checked that traces over the fundamental representation $\mathbf{F}$'s generators do indeed provide an independent set of colour-tensors at large-rank for each of the classical Lie algebras.

But why must we choose the \emph{fundamental} representation's generators and not some other? After all, the relationship between the structure constants and traces over the generators (\ref{f_as_traces}) applies to \emph{any} representation of the algebra (even the reducible ones)! In the case of $\mathfrak{a}_k$-theory, the traces of the generators of any \emph{real} representation (the adjoint, say) would enjoy dihedral symmetry, leading to $(!\r{n}{-}!!\r{n})$ relations among the multi-trace tensors; as such, they would not prove sufficient to span all independent tensors that exist; but we do expect that the traces of any complex representation should also work. 

Saturation for the $\mathfrak{b},\mathfrak{c},\mathfrak{d}$ series is given by the number of dihedrally distinct multi-trace tensors. As this is justified for any real representation, it is natural suppose that multi-traces involving \emph{any} choice of representation should suffice. 

Interestingly, although the fundamental\footnote{Recall that we are \emph{defining} the fundamental representation to be the \emph{defining} representation for each algebra. Equivalently, the representation of smallest dimension (with the exception of $\mathfrak{b}_2$).} representation of $\mathfrak{d}_k$ is real, complex representations do exist when $k$ is odd. It is highly non-trivial that no more than $!!\r{n}$ independent multi-trace tensors exist above saturation even when considering traces involving complex representations, as no simple symmetries such as (\ref{reflected_trace_relations}) are expected. But of course, the numbers $\mathcal{C}^{\r{n}}_{\mathfrak{d}_k}$ are computed entirely independently of how such tensors are represented, and so we can rest assured that such relations among these objects \emph{must} exist.

At least for the $\mathfrak{a},\mathfrak{b},\mathfrak{c}$-type algebras, we may detect the fact that $\mathcal{C}^{\r{n}}_\mathfrak{g}$ is strictly bounded above by its saturation value through new relations being satisfied among the tensors in this na\"ive basis. Indeed, this turns out to be a simple consequence of linear algebra in the case of their fundamental representations. 

\subsubsection{Explaining Identities Among Colour-Tensors: Cayley Hamilton}

Let us begin our analysis with the case of $\mathfrak{a}$-type theories. As described above,
\eq{\left.\begin{array}{lr}\mathcal{C}^{\r{n}}_{\mathfrak{a}_{\b{k}}}=\,(!\r{n})&\text{for }\b{k}>\r{n}{-}2\\\mathcal{C}^{\r{n}}_{\mathfrak{a}_{\b{k}}}<\,(!\r{n})&\text{for }\b{k}\leq\r{n}{-}2\end{array}\right.\,.}
Consider the first case \emph{below} saturation: $(\r{n{+}1})$-particle scattering for $\mathfrak{a}_{\b{k}{=}\r{n}{-}1}$ gauge theory. We should find the difference between $\mathcal{C}^{\r{n}}_{\mathfrak{a}_{n{-}1}}$ and $!\r{n}$ to be explained by some new linear relation(s) among the multi-trace tensors constructed out of any (complex) representation. Indeed, we can identify these easily enough for multi-traces over the fundamental representation: which would be of dimension $\r{n}$. 

For any $\r{n}\!\times\!\r{n}$ matrix ${A}$, the successive matrix-powers of ${A}$ cannot all be linearly independent. They are constrained by the \emph{Cayley-Hamilton} theorem, which states: 
\eq{{A}^\r{n}{+}c_{\b{1}}(A){A}^{\r{n}{-}\b{1}}{+}\ldots{+}c_{\b{{n}{-}1}}(A){A}^{1}{+}c_{\b{{n}}}(A)\mathbb{I}_{\r{n}\!\times\!\r{n}}=0\,,\label{cayley_hamilton}}
where the coefficients are given by 
\eq{c_{\b{m}}(A)\equivR\frac{({-}1)^{\b{m}}}{\b{m}!}\left|\begin{array}{@{}cccccc@{}}\mathrm{tr}({A})&1&0&\cdots&0\\
\mathrm{tr}({A}^2)&\mathrm{tr}({A})&2&\ddots&\vdots\\
\vdots&\ddots&\ddots&\ddots&0\\
\mathrm{tr}({A}^{\b{m}{-}1})&\ddots&\ddots&\mathrm{tr}({A})&\b{m}{-}1\\
\mathrm{tr}({A}^{\b{m}})&\mathrm{tr}({A}^{\b{m}{-}1})&\cdots&\mathrm{tr}({A}^{2})&\mathrm{tr}({A})
\end{array}\right|\,.}
This identity does not immediately suggest anything about traces built from ordered products of generators; but we may apply this identity to the matrix $A\!\mapsto\!\sum_{a{=}1}^{\r{n}}T_{\mathbf{F}}^{\r{\sigma(a)}}$, which would then involve the symmetrized products of subsets of generators chosen from the list indexed by $\sigma(a)$. 

Because we are interested in $(\r{n{+}1})$-particle scattering, we may contract this identity with any additional generator and take its trace, resulting in a new identity among the traces involving the $(\r{n{+}1})$ generators. 

Consider the case of $\mathfrak{a}_2$. We'd like to find an identity among the 4-particle multi-traces of the $\mathbf{3}$-dimensional generators of the fundamental representation. In this case, and using the fact that the generator matrices are traceless (true for any representation of any semi-simple Lie algebra), the identity (\ref{cayley_hamilton}) takes the form:
\eq{A^{\r{3}}{+}c_{\b{1}}(A)A^{\r{2}}{+}c_{\b{2}}(A)A^{\r{1}}{+}c_{\b{3}}(A)A^{\r{0}}=0\,,\label{eg_ch}}
with coefficients given by
\eq{\begin{split}c_\b{1}(A)&={-}\mathrm{tr}(A)=0\,;\\
c_\b{2}(A)&=\frac{1}{2}\mathrm{tr}(A^{1})^2{-}\frac{1}{2}\mathrm{tr}(A^{\r{2}})={-}\frac{1}{2}\mathrm{tr}(A^{\r{2}})\,;\\
c_\b{3}(A)&=\frac{1}{2}\mathrm{tr}(A^{\r{1}})\mathrm{tr}(A^{\r{2}}){-}\frac{1}{6}\mathrm{tr}(A^{\r{1}})^3{-}\frac{1}{3}\mathrm{tr}(A^{\r{3}})={-}\frac{1}{3}\mathrm{tr}(A^{\r{3}})\,.
\end{split}}
Relative to the general case, the only simplification arising here is due to the vanishing of $\mathrm{tr}(A)$.

Contracting the identity (\ref{eg_ch}) on the left with any \emph{other} (traceless) matrix $\b{B}$, and taking the trace of the result leads to the identity 
\eq{\begin{split}
0&=\mathrm{tr}\big[\b{B}.\!\left(A^{\r{3}}{+}c_{\b{1}}(A)A^{\r{2}}{+}c_{\b{2}}(A)A^{\r{1}}{+}c_{\b{3}}(A)A^{\r{0}}\right)\!\big]\\
&=\mathrm{tr}\big(\b{B}.A^{\r{3}}\big){+}c_{\b{1}}(A)\mathrm{tr}\big(\b{B}.A^{\r{2}}\big){+}c_{\b{2}}(A)\mathrm{tr}\big(\b{B}.A^{\r{1}}\big){+}c_{\b{3}}(A)\mathrm{tr}(\b{B})\\
&=\mathrm{tr}\big(\b{B}.A^{\r{3}}\big){+}c_{\b{2}}(A)\mathrm{tr}\big(\b{B}.A^{\r{1}}\big)\\
&=\mathrm{tr}\big(\b{B}.A^{\r{3}}\big){-}\frac{1}{2}\mathrm{tr}(A^\r{2})\mathrm{tr}\big(\b{B}.A^{\r{1}}\big)\,.
\end{split}}
Replacing $A$ with the sum over any subset of the $\mathrm{dim}(\mathfrak{a}_2){=}8$ generators and $\b{B}\!\mapsto T^{\b{a}}_{\mathbf{F}}$ any other will result in an identity
\eq{\hspace{-20pt}\sum_{\r{\sigma}\in\mathfrak{S}(\{\r{2},\r{3},\r{4}\})}\hspace{-10pt}\mathrm{tr}_{\mathbf{F}}(\b{1}\,\r{\sigma_1}\,\r{\sigma_2}\,\r{\sigma_3}){=}\mathrm{tr}_{\mathbf{F}}(\b{1}\,\r{\sigma_1})\mathrm{tr}_{\mathbf{F}}(\r{\sigma_2}\,\r{\sigma_3}){+}\mathrm{tr}_{\mathbf{F}}(\b{1}\,\r{\sigma_2})\mathrm{tr}_{\mathbf{F}}(\r{\sigma_1}\,\r{\sigma_3}){+}\mathrm{tr}_{\mathbf{F}}(\b{1}\,\r{\sigma_3})\mathrm{tr}_{\mathbf{F}}(\r{\sigma_1}\,\r{\sigma_2})\,.}

This generalizes in a nice way to all $\r{n}$: for $(\r{n{+}1})$-point scattering of adjoints in $\mathfrak{a}_{n{-}1}$ gauge theory, there exists a single identity:
\eq{\sum_{\substack{\text{distinct}\\\text{multi-traces}}}\hspace{-5pt}({-}1)^{|\mathrm{tr}_i|}\mathrm{tr}_i=0\,.}
This agrees with the empirical observation that $\mathcal{C}^{\r{n{+}1}}_{\mathfrak{a}_{n{-}1}}{=}\,\,!(\r{n{+}1}){+}1$ for all $\r{n}\geq2$ (see \mbox{Table~\ref{pure_glue_counting_table_a}}).

When the rank is lower and the size of the generators of the fundamental representation is substantially smaller than a multiplicity in question, there will be additional identities arising from replacing `$\b{B}$' in the argument above with products over more generators. Choosing different (non-symmetrized) sets of generators chosen for $\b{B}$ gives rise to many (not necessarily independent) identities. 

The same logic applies to the cases of $\mathfrak{b},\mathfrak{c}$-type gauge theory below saturation. The main distinction is that for these algebras saturation always occurs when the dimension of the fundamental representation is farther from the multiplicity than the marginal case encountered for $\mathfrak{a}$-series:
\eq{\left.\begin{array}{lr}\mathcal{C}^{\r{n}}_{\mathfrak{b}_{\b{k}},\mathfrak{c}_\b{k}}=\,(!!\r{n})&\text{for }\b{k}>\frac{1}{2}(\r{n}{+}1)\\\mathcal{C}^{\r{n}}_{\mathfrak{b}_{\b{k}},\mathfrak{c}_{\b{k}}}<\,(!!\r{n})&\text{for }\b{k}\leq\frac{1}{2}(\r{n}{+}1)\end{array}\right.\,.}
This explains a greater number of identities among the tensors away from saturation.

The above discussion helps to explain the decrease in the counting of independent tensors only because we considered multi-traces over the fundamental representation---which involves matrices small enough for Cayley-Hamilton's relation to apply. But what if we had chosen another representation? For example, although we have identified an identity among the $\r{n}{=}{6}$-particle multi-traces over the generators of the $\mathbf{5}$-dimensional representation of $\mathfrak{su}_5$, what if we had instead chosen a basis of multi-traces over the generators of the $\mathbf{10}$-dimensional representation? Of course, the number of independent colour-tensors has nothing to do with the basis in which we choose to represent them; but it is clear that the logic above seems strongly sensitive to the \emph{smallest}-dimensional representation that exists for a given algebra. It would be worthwhile to better understand how this story changes if other representations are used. 

\subsubsection{There be $\mathfrak{d}$ragons: Origins of non-Monotonicity for $\mathfrak{d}$-Theories}\label{subsec:dragons}

The $\mathfrak{d}_k$ series is remarkable because for \emph{low rank} there are \emph{extra} colour-tensors relative to the large-$k$ limit; as $k$ increases, these additional tensors must become dependent (or cease to exist). 

This should not be entirely surprising, however, as the $\mathfrak{d}$-type algebras (for odd rank) admit \emph{complex} representations; as such, even if multi-traces over (real) fundamental generators should saturate to $!!\r{n}$ (reflecting the dihedral symmetry of traces over any real representation), surely we'd expect fewer relations (and a larger basis) to be spanned by multi-traces over \emph{complex} representations such as spinors. This argument is a red-herring: for one thing, not all $\mathfrak{d}$ admit complex representations. Indeed, although no such simple pairwise relations as in (\ref{reflected_trace_relations}) exist for multi-traces over the spinors of $\mathfrak{d}_{2k{+}1}$, we have checked explicitly that they span a space of rank $!!\r{n}$ above the saturation limit. 

We can understand the origin of these `new' tensors looking just below the marginal limit. Consider, for example, $\r{4}$-adjoint scattering in $\mathfrak{d}_4$-type gauge theory. As described in \mbox{section~\ref{subsec:generators}}, the generators of the fundamental representation can be represented by the collection of matrices ${{T}_{\mathbf{F}}}^{\hspace{-5pt}\r{(ij)}}_{\hspace{-7pt}\phantom{(ij)}\b{r\,s}}\equivR\delta^{\r{i}}_{\b{r}}\delta^{\r{j}}_{\b{s}}{-}\delta^{\r{i}}_{\b{s}}\delta^{\r{j}}_{\b{r}}$ indexed by $(\r{ij})\!\in\!\binom{[\b{2k}]}{2}$. For $k{=}4$, we will have 28 generators of size $8\!\times\!8$. We can think of each as a two-form, and consider the antisymmetric product of any $\r{4}$ of them to obtain an $8$-form; dividing this by the volume form (that is, contracting the representation indices with a determinant) results in a completely symmetric tensor in the adjoint indices related to the \emph{Pfaffian} of the generators.

This object ceases to exist for $k\!>\!n$, and surely accounts for the \emph{one} excess that $\mathcal{C}^{\r{n}}_{\mathfrak{d}_{\b{k}{=}\r{n}}}{=}(!!\r{n}){+}1$. Moreover, such a construction requires that the representation be even-dimensional, and therefore does not apply to the algebras $\mathfrak{b}_k\!\sim\!\mathfrak{so}_{2k{+}1}$. 

Further below saturation, for $k\!<\!n$, there are even more possibilities. Indeed, such Pfaffian-like tensors can be constructed using subsets of generators according to the above. Consider the case of $\r{n}$-adjoint scattering in $\mathfrak{d}_{\r{n}{-}1}$ theory. In this case, we could construct Pfaffians using $\r{n}{-}1$ of the generators, but there are no single-particle invariants available from which we may build an $\r{n}$-adjoint tensor. There is, however, an invariant 2-adjoint tensor (namely the Cartan-Killing form), and we could construct an $\r{n}$-adjoint tensor using the Pfaffian built from any choice of $\r{n}{-}2$ of them wedged with the commutator. 

Let us try to be a bit more concrete. To describe the generators, let us denote the colour-tensor in the fundamental representation for the particle $a$ by $\underline{T}_a$. We can think of these $\underline{T}_a$ as two-forms on the $2(\r{n}{-}1)$-dimensional space on which the representation acts. In addition to these two-forms, there is a two-form associated with the commutator $[\underline{T}_a,\underline{T}_b]$. We can form Pfaffian-like invariants using the commutator as in $[\underline{T}_a,\underline{T}_b]\wedge\underline{T}_1\!\wedge\!\cdots\!\wedge\hat{\underline{T}}_a\!\wedge\!\cdots\!\wedge\hat{\underline{T}}_b\!\wedge\!\cdots\!\wedge\!\underline{T}_\r{n}$, where hats denote generators not appearing in the wedge-product. These tensors are not all independent, however: for every $a\!\in\![\r{n}]$, we have an identity $\sum_{b}[\underline{T}_a,\underline{T}_b]\wedge \dots  \hat{\underline{T}}_a \dots \hat{\underline{T}}_b \dots \wedge \underline{T}_n{=}0$, which is simply the invariance condition of $\underline{T}_1 \wedge \dots \wedge \hat{\underline{T}}_a\wedge \dots \wedge \underline{T}_n$ under a gauge transformation; but these relations are not themselves independent: they have one syzygy (their sum vanishes identically). 

Thus, for every $\binom{\r{n}}{2}$ pair of indices we can construct a new invariant tensor; these satisfy $\r{n}$ identities of which only $\r{n}{-}1$ are independent. Thus, from this construction we expect to be able to build $\binom{\r{n}}{2}$ new tensors, spanning $\binom{\r{n}}{2}{-}\r{n}{+}1{=}\binom{\r{n}{-}1}{2}$ independent objects. Moreover,  since these new invariants involve the Levi-Civita tensor they are invariants for the \emph{special} orthogonal group; the other invariants which do not involve the Levi-Civita tensor are invariants for the \emph{orthogonal} group and therefore they should be independent. 

Indeed, this agrees with our explicit calculations. Consulting Table~\ref{pure_glue_counting_table_d}, one can check that $\r{8}$ adjoints, $\mathcal{C}^{\r{8}}_{\mathfrak{d}_9}{=}6202{=}!\r{8}$, $\mathcal{C}^{\r{8}}_{\mathfrak{d}_8}{=}6203{=}!!\r{8}{+}1$ and for $\mathcal{C}^{\r{8}}_{\mathfrak{d}_7}{=}6223{=}!!\r{8}{+}\binom{7}{2}.$

\subsection{Counting in the Limit of Large Multiplicity (for a Fixed Lie Algebra)}\label{subsection_large_multiplicities_of_adjoints}

For any Lie algebra of fixed rank $\b{k}$, however, there will be an unbounded range of \emph{multiplicities} which are \emph{outside}/above the saturation condition for $\b{k}$; and for the exceptional algebras, no notion of a `large rank limit' exists. For any algebra at sufficiently low rank, even relatively low-multiplicity scattering will be outside saturation, and it is interesting to see how $\mathcal{C}^{\r{n}}_{\mathfrak{g}}$ compares with saturation in these instances. We have tabulated many of these numbers in \mbox{Appendix~\ref{appendix_enumerating_ranks}} and provided them also as an ancillary file attached to this work.

On general grounds, as we review in \mbox{Appendix~\ref{asymptotic_appendix}}, it turns out that for any fixed Lie algebra $\mathfrak{g}$,
\eq{\mathcal{C}^{\r{n}}_{\mathfrak{g}}\asympt{\r{n}\to\infty}\#\,\,\b{d}^{\r{n}}\r{n}^{{-}\b{d}/2}\exp\!\left\{{-}\frac{\b{d}^2}{48\,\r{n}}\right\}\,\Big(1{+}\mathcal{O}(1/\r{n})\Big)\,}
where $\b{d}{=}\mathrm{dim}(\mathfrak{g})$ and `$\#$' is some fixed constant that can be computed for any algebra. (These constants have been tabulated for all algebras of rank $k\!\leq\!8$ in \mbox{Appendix~\ref{asymptotic_appendix}}.) This is steep growth indeed, but dramatically slower than the \emph{factorial} growth of $!\r{n}$ or $!!\r{n}$. 

\begin{figure}[t!]\vspace{-20pt}$$\fwbox{0pt}{\fig{-100pt}{1}{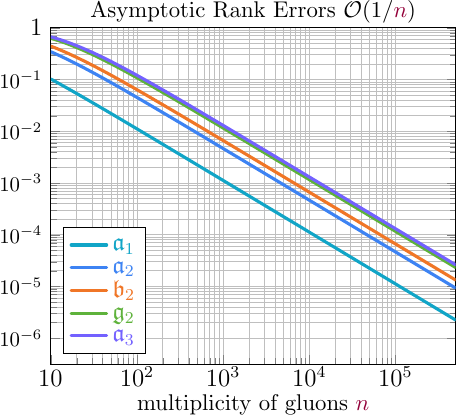}\hspace{20pt}\fig{-100pt}{1}{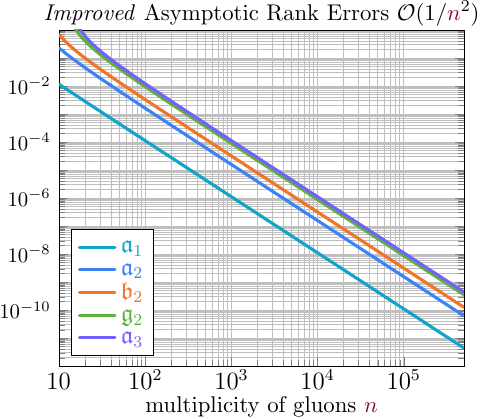}}$$\vspace{-20pt}\caption{Comparison of $\mathcal{C}^{\r{n}}_{\mathfrak{g}}$ with the asymptotic formulae (\ref{improved_asymptotics}) with and without the $\mathcal{O}(1/\r{n})$ correction terms through reasonably high multiplicity.\label{asymptotic_number_error_plots}}\vspace{-15pt}\end{figure}

In some cases, the $\mathcal{O}(1/\r{n})$ corrections are known or can be deduced from the data directly. For example, 
\eq{\begin{split}
\mathfrak{a}_1\!\!:\;\;\mathcal{C}^{\r{n}}_{\mathfrak{a}_1}\asympt{\r{n}\to\infty}&= \frac{1}{8\sqrt{3\,\pi}}\b{3}^{\r{n}}\exp\!\left\{\hspace{-0.5pt}\text{--}\frac{\b{3}^2\hspace{-3pt}}{\hspace{1pt}48\hspace{1pt}\r{n}}\,\right\}\Big(1{-}\frac{9}{8\,\r{n}}{+}\mathcal{O}\!\big(1/\r{n}^2\big)\Big)\;;\\
\mathfrak{a}_2\!\!:\;\;\mathcal{C}^{\r{n}}_{\mathfrak{a}_2}\asympt{\r{n}\to\infty}&=\frac{4}{81\sqrt{3}\,\pi}\hspace{0pt}\b{8}^{\hspace{-0.25pt}\r{n}\text{+}2}\r{n}^{\hspace{-0.5pt}\text{--}\frac{\b{8}}{2}}\exp\!\left\{\hspace{-0.5pt}\text{--}\frac{\b{8}^2\hspace{-3pt}}{\hspace{1pt}48\hspace{1pt}\r{n}}\,\right\}\Big(\!1{-}\frac{14}{3\,\r{n}}{+}\mathcal{O}\!\left(1/\r{n}^2\right)\!\!\Big)\;;\\
\mathfrak{b}_2\!\!:\;\;\mathcal{C}^{\r{n}}_{\mathfrak{b}_2}\asympt{\r{n}\to\infty}&=\frac{125}{1296\,\pi}\hspace{0pt}\b{10}^{\hspace{-0.25pt}\r{n}\text{+}2}\r{n}^{\hspace{-0.5pt}\text{--}\frac{\b{10}}{2}}\exp\!\left\{\hspace{-0.5pt}\text{--}\frac{\b{10}^2\hspace{-3pt}}{\hspace{1pt}48\hspace{1pt}\r{n}}\,\right\}\Big(\!1{-}\frac{20}{3\,\r{n}}{+}\mathcal{O}\!\left(1/\r{n}^2\right)\!\!\Big)\,;\\
\mathfrak{g}_2\!\!:\;\;\mathcal{C}^{\r{n}}_{\mathfrak{g}_2}\asympt{\r{n}\to\infty}&=\frac{84035}{442368\sqrt{3}\,\pi}\b{14}^{\hspace{-0.25pt}\r{n}\text{+}2}\r{n}^{\hspace{-0.5pt}\text{--}\frac{\b{14}}{2}}\exp\!\left\{\hspace{-0.5pt}\text{--}\frac{\b{14}^2\hspace{-3pt}}{\hspace{1pt}48\hspace{1pt}\r{n}}\,\right\}\Big(\!1{-}\frac{35}{3\,\r{n}}{+}\mathcal{O}\!\left(1/\r{n}^2\right)\!\!\Big)\,;\\
\mathfrak{a}_3\!\!:\;\;\mathcal{C}^{\r{n}}_{\mathfrak{a}_3}\asympt{\r{n}\to\infty}&=\frac{3}{2^{21}\pi^{3/2}}\b{15}^{\hspace{-0.25pt}\r{n}\text{+}15/2}\r{n}^{\hspace{-0.5pt}\text{--}\frac{\b{15}}{2}}\exp\!\left\{\hspace{-0.5pt}\text{--}\frac{\b{15}^2\hspace{-3pt}}{\hspace{1pt}48\hspace{1pt}\r{n}}\,\right\}\Big(\!1{-}\frac{105}{8\,\r{n}}{+}\mathcal{O}\!\left(1/\r{n}^2\right)\!\!\Big)\,.
\end{split}\label{improved_asymptotics}}
In \mbox{Figure~\ref{asymptotic_number_error_plots}}, we show how direct computation of $\mathcal{C}^{\r{n}}_{\mathfrak{g}}$ compares with the asymptotic formulae both with and without the $\mathcal{O}(1/\r{n})$ corrections.

\subsubsection{Surveys of Asymptotic Growth at Moderately Large Multiplicity}

\begin{figure}[ht!]\vspace{-0pt}$$\begin{array}{c}\fwbox{0pt}{\fig{-100pt}{1}{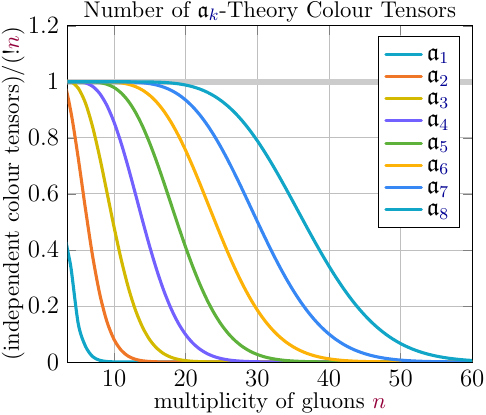}\hspace{15pt}\fig{-100pt}{1}{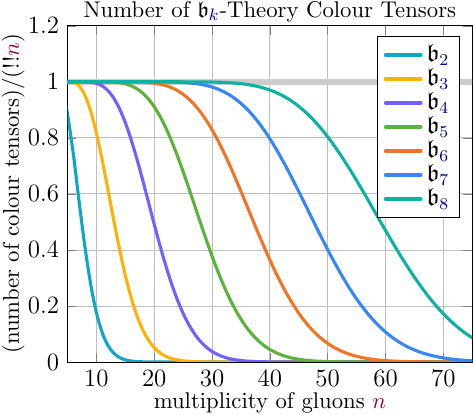}}\\[110pt]
\fwbox{0pt}{\fig{-100pt}{1}{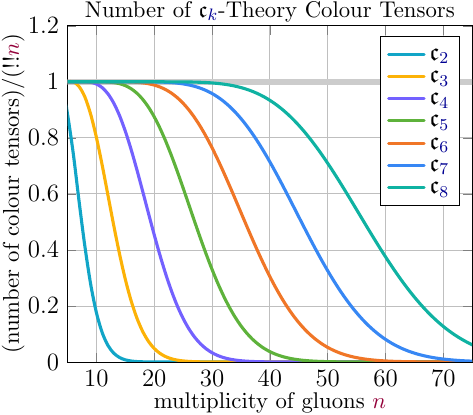}\hspace{15pt}\fig{-100pt}{1}{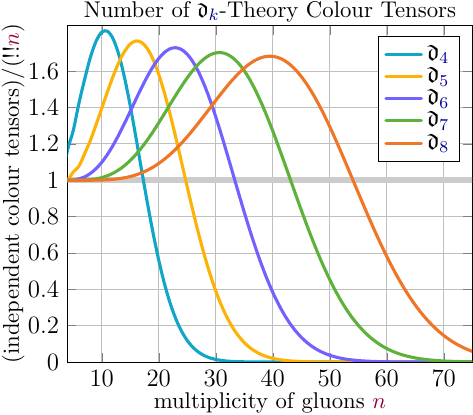}}\\[110pt]
\fwbox{0pt}{\fig{-100pt}{1}{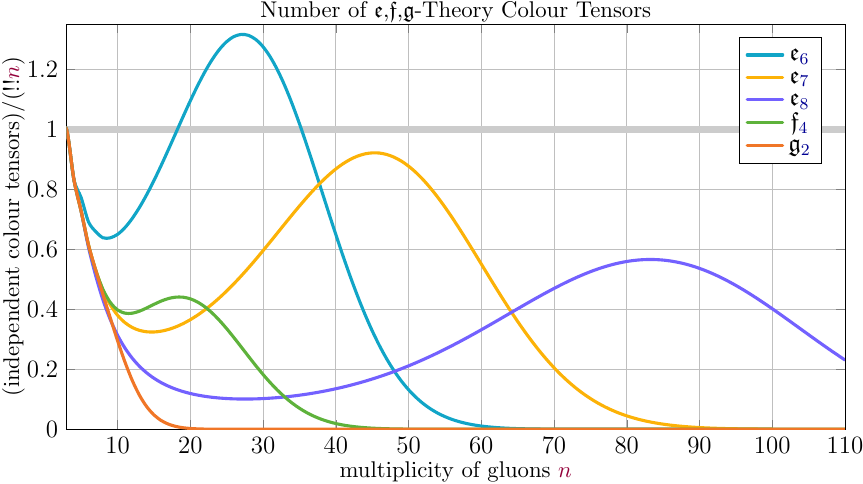}}
\end{array}$$\vspace{-20pt}\caption{$\mathcal{C}^{\r{n}}_{\mathfrak{g}}$ relative to saturation numbers for all Lie algebras through rank $8$.\label{adjn_plots}}\vspace{-90pt}\end{figure}

For any modest-rank Lie algebra $\mathcal{C}^{\r{n}}_{\mathfrak{g}}$ can be computed reasonably efficiently through $\r{n}\!\sim\!\mathcal{O}(100)$ (and in some instances, much higher) using readily available computer algebra packages (such as \cite{sage,magma} or the \textsc{Mathematica} package of \cite{Feger:2019tvk}). We have tabulated many of these results in \mbox{Appendix~\ref{appendix_enumerating_ranks}}, and included precise ranks to reasonably-high multiplicity for all rank $k\!\leq\!8$ algebras in the ancillary file attached to this work. We plot these numbers relative to their saturated values for a given multiplicity in \mbox{Figure~\ref{adjn_plots}}.

\begin{figure}[b!]\vspace{-10pt}$$\fig{-100pt}{1}{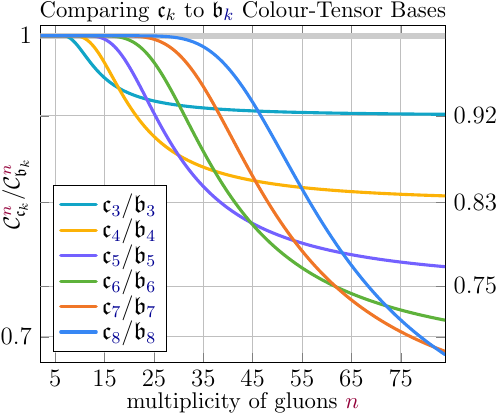}$$\vspace{-20pt}\caption{Relative numbers of independent colour-tensors for $\mathbf{ad}^{\r{n}}$ scattering in $\mathfrak{c}_k$ vs.~$\mathfrak{b}_k$.\label{ratioOfBCplot}}\vspace{-40pt}\end{figure}

The similarity between the counting for $\mathfrak{b}_k$ and $\mathfrak{c}_k$ gauge theories is striking. Although the counting in both cases saturate to $!!\r{n}$, below saturation (at higher $\r{n}$ for fixed rank) there are slightly fewer colour-tensors for $\mathfrak{c}_k$ gauge theories than $\mathfrak{b}_k$ theories. In fact, our analysis above suggests that the ratio $\mathcal{C}^{\r{n}}_{\mathfrak{c}_k}/\mathcal{C}^{\r{n}}_{\mathfrak{b}_k}$ goes to a constant as $\r{n}\!\to\!\infty$ for any fixed $k$. This is indeed seen in the explicit data, as indicated in \mbox{Figure~\ref{ratioOfBCplot}}, with the asymptotic ratios predicted by the asymptotic formulae given in \mbox{Appendix~\ref{asymptotic_appendix}} indicated on the right for $k{=}3,4,5$.\\

Although there is no large-rank limit for Lie algebras of type $\mathfrak{e},\mathfrak{f},\mathfrak{g}$, we have plotted the counting $\mathcal{C}^{\r{n}}_{\mathfrak{g}}$ against $!!\r{n}$ as only $\mathfrak{e}_6$ admits complex representations. The sequences that arise for $\mathcal{C}^{\r{n}}_{\mathfrak{e}_6}$, $\mathcal{C}^{\r{n}}_{\mathfrak{e}_7}$, $\mathcal{C}^{\r{n}}_{\mathfrak{e}_8}$, $\mathcal{C}^{\r{n}}_{\mathfrak{f}_4}$, and $\mathcal{C}^{\r{n}}_{\mathfrak{g}_2}$ are quite interesting---indeed, they are enshrined in the OEIS \cite{oeis} as \cite{oeis-e6,oeis-e7,oeis-e8,oeis-f4,oeis-g2}, respectively.

Because $\mathfrak{e}_6$'s fundamental representation is not self-conjugate, it should not be terribly surprising that $\mathcal{C}^{\r{n}}_{\mathfrak{e}_6}/(!!\r{n})$ exceeds 1 for a limited range of multiplicities (with a maximum occurring, suggestively, at $\r{n}{=}27$); to be clear $\mathcal{C}^{\r{n}}_{\mathfrak{g}}/(!\r{n})\!<\!1$ for all $\r{n}$, which is consistent with an always over-complete basis consisting of multi-traces over the fundamental representation's generators. However, we have not been able to construct explicit these for $\r{n}{=}19$, say, to directly verify that multi-traces over the fundamental representation $\mathbf{27}$ truly span the space of $\mathcal{C}^{\r{19}}_{\mathfrak{e}_6}{=}12\,702\,789\,216\,958\,134$ independent tensors. 

At low multiplicity, these algebras have among the fewest independent colour-tensors. In particular, $\mathfrak{e}_8$ gauge theory requires the fewest of any algebra other than $\mathfrak{a}_1$ for $\r{n}\!<\!10$. For $\r{n}{=}4$, this is true for all the exceptional algebras, each of which have only $5{=}\,(!!\r{4}){-}1$ independent tensors, which we can understand through the relations (which we have checked directly):
\vspace{-8pt}\begin{align}\fwboxL{200pt}{\mathfrak{e}_6\!:\;\;12\,\mathrm{tr}_\mathbf{F}(1\,\r{2\,3\,4}){=}\mathrm{tr}_\mathbf{F}(1\,\r{2})\mathrm{tr}_\mathbf{F}(\r{3\,4})}&\fwboxL{200pt}{\mathfrak{f}_4\!:\;\;12\,\mathrm{tr}_\mathbf{F}(1\,\r{2\,3\,4}){=}\mathrm{tr}_\mathbf{F}(1\,\r{2})\mathrm{tr}_\mathbf{F}(\r{3\,4})}\nonumber\\
\fwboxL{200pt}{\mathfrak{e}_7\!:\;\;24\,\mathrm{tr}_\mathbf{F}(1\,\r{2\,3\,4}){=}\mathrm{tr}_\mathbf{F}(1\,\r{2})\mathrm{tr}_\mathbf{F}(\r{3\,4})}&\fwboxL{200pt}{\mathfrak{g}_2\!:\;\;4\,\mathrm{tr}_\mathbf{F}(1\,\r{2\,3\,4}){=}\mathrm{tr}_\mathbf{F}(1\,\r{2})\mathrm{tr}_\mathbf{F}(\r{3\,4})}\nonumber\\
\fwboxL{200pt}{\mathfrak{e}_8\!:\;\;100\,\mathrm{tr}_\mathbf{F}(1\,\r{2\,3\,4}){=}\mathrm{tr}_\mathbf{F}(1\,\r{2})\mathrm{tr}_\mathbf{F}(\r{3\,4})\,;}\\[-28pt]\nonumber
\end{align}
here, indices coloured red should be cyclically-symmetrized. Notice that applying this logic to the case of $\mathfrak{e}_6$ only works because, although its fundamental representation is \emph{not} real, it is nevertheless true that its traces enjoy reflection symmetry through length 5 (and not beyond)---a fact which accounts for the absence of a $d^{\r{a\,b\,c}}$ tensor for $\mathfrak{e}_6$ \cite{birdtracks}.

~\newpage

\newpage
\section{Scattering of Gluons and Variously Charged Matter}\label{sec:adjoints_and_matter}

As soon as we turn our attention beyond the adjoint representation, there are never-ending combinations to be analyzed. It would be impossible to concisely describe all the combinations we have explored, and we do not even claim to have exhausted the `interesting' cases. Part of the purpose of what we describe here is to invite the reader to play these games themselves. Besides any particular relevance to physics these surveys may have, we find a wealth of connections to interesting combinatoric structure, and it is reasonable to expect some of the sequences we have not yet identified may have an interesting and rich structure.

\subsection{Counting the Colours of Gluons with \emph{Fundamental}-Charged Matter}\label{subsec:gluons_and_fundamentals}

Let us begin our studies with the case of a single type of matter particle charged under the representation $\mathbf{R}$ of $\mathfrak{g}$. For the scattering of $\b{q}$ fermion lines and $\r{n}$ gluons, the number of independent colour-tensors would be computed by 
\eq{\mathcal{C}\big[\mathbf{ad}^{\r{n}},(\mathbf{R}\bar{\mathbf{R}})^{\b{q}}\big]\equivL\,\,\mathcal{C}_{\mathfrak{g}}^{\r{n},\b{q}}(\mathbf{R})=m\big(\mathbf{ad}^{\otimes\r{n}}\!\otimes\!(\mathbf{R}\!\otimes\!\bar{\mathbf{R}})^{\otimes\b{q}}\!\!\to\!\mathbf{1}\big)\,.}

It is a general fact that for any (non-trivial) irreducible representation, the tensor product $\mathbf{R}\!\otimes\!\bar{\mathbf{R}}$ decomposes into irreducible representations according to
\eq{\mathbf{R}\!\otimes\!\bar{\mathbf{R}}=\mathbf{1}\!\oplus\!\mathbf{ad}\!\oplus\!(\cdots)\,.\label{general_rrbar_decomp}}
If we denote the combinations of irreducible representations (possibly with multiplicity) encoded by the `$(\cdots)$' above by `$\mathbf{Q}$', we see on general grounds  (by the associativity of the tensor product) that
\eq{\begin{split}\mathcal{C}^{\r{n}\,\b{q}}_{\mathfrak{g}}(\mathbf{R})&=\mathcal{C}^{\r{n}\,\b{q{-}1}}_{\mathfrak{g}}(\mathbf{R})\!+\mathcal{C}^{\r{n{+}1}\,\b{q{-}1}}_{\mathfrak{g}}(\mathbf{R}){+}m\big(\mathbf{ad}^{\otimes\r{n}}\!\!\otimes\!(\mathbf{R}\!\otimes\!\bar{\mathbf{R}})^{\otimes\b{q{-}1}}\!\!\otimes\!\mathbf{Q}\!\to\!\mathbf{1}\big)\,\\
&\geq\mathcal{C}^{\r{n}\,\b{q{-}1}}_{\mathfrak{g}}(\mathbf{R})\!+\mathcal{C}^{\r{n{+}1}\,\b{q{-}1}}_{\mathfrak{g}}(\mathbf{R})\,.
\end{split}\label{breakdown_for_Rs}}

In particular, for the fundamental representations (defined in \mbox{Table~\ref{classification_of_simple_lie_algebras_table}}), we have\footnote{Although one should generally avoid naming irreducible representations by their dimension alone, there are no ambiguities among those that appear here.}
\eq{\begin{split}
\fwboxL{20pt}{\mathfrak{a}_{\b{k}}\!\!:\;\;}&\mathbf{F}\!\otimes\!\bar{\mathbf{F}}
=\mathbf{1}\!\oplus\!\mathbf{ad}\\
\fwboxL{20pt}{\mathfrak{b}_{\b{k}}\!\!:\;\;}&\mathbf{F}\!\otimes\!\bar{\mathbf{F}}
=\mathbf{1}\!\oplus\!\mathbf{ad}\!\oplus\!\b{\big[}\mathbf{k(2k{+}3)}\b{\big]}\\
\fwboxL{20pt}{\mathfrak{c}_{\b{k}}\!\!:\;\;}&\mathbf{F}\!\otimes\!\bar{\mathbf{F}}
=\mathbf{1}\!\oplus\!\mathbf{ad}\!\oplus\!\b{\big[}\mathbf{(k{-}1)(2k{+}1)}\b{\big]}\\
\fwboxL{20pt}{\mathfrak{d}_{\b{k}}\!\!:\;\;}&\mathbf{F}\!\otimes\!\bar{\mathbf{F}}
=\mathbf{1}\!\oplus\!\mathbf{ad}\!\oplus\!\b{\big[}\mathbf{(k{+}1)(2k{-}1)}\b{\big]}\\
\fwboxL{20pt}{\mathfrak{e}_{\b{6}}\!\!:\;\;}&\mathbf{F}\!\otimes\!\bar{\mathbf{F}}
=\mathbf{1}\!\oplus\!\mathbf{ad}\!\oplus\!\b{\big[}\mathbf{650}\b{\big]}\\
\fwboxL{20pt}{\mathfrak{e}_{\b{7}}\!\!:\;\;}&\mathbf{F}\!\otimes\!\bar{\mathbf{F}}
=\mathbf{1}\!\oplus\!\mathbf{ad}\!\oplus\!\b{\big[}\mathbf{1463}\!\oplus\!\mathbf{1539}\b{\big]}\\
\fwboxL{20pt}{\mathfrak{e}_{\b{8}}\!\!:\;\;}&\mathbf{F}\!\otimes\!\bar{\mathbf{F}}
=\mathbf{1}\!\oplus\!\mathbf{ad}\!\oplus\!\b{\big[}\mathbf{3875}\!\oplus\!\mathbf{27000}\!\oplus\!\mathbf{30380}\b{\big]}\\
\fwboxL{20pt}{\mathfrak{f}_{\b{4}}\!\!:\;\;}&\mathbf{F}\!\otimes\!\bar{\mathbf{F}}
=\mathbf{1}\!\oplus\!\mathbf{ad}\!\oplus\!\b{\big[}\mathbf{F}\!\oplus\!\mathbf{273}\!\oplus\!\mathbf{324}\b{\big]}\\
\fwboxL{20pt}{\mathfrak{g}_{\b{2}}\!\!:\;\;}&\mathbf{F}\!\otimes\!\bar{\mathbf{F}}
=\mathbf{1}\!\oplus\!\mathbf{ad}\!\oplus\!\b{\big[}\mathbf{F}\!\oplus\!\mathbf{27}\b{\big]}
\end{split}}

\subsubsection{Scattering of Gluons and \emph{Fundamental} Matter in $\mathfrak{a}$-Theories}\label{subsec:anq_counting}

Notice in particular that for the fundamental representation of $\mathfrak{a}_k$, the remainder `$\mathbf{Q}$' does not exist. This is in fact the only irreducible representation of any Lie algebra which has this property, and leads to a very simple recurrence:
\eq{\mathcal{C}^{\r{n}\,\b{q}}_{\mathfrak{a}_k}(\mathbf{F})=\mathcal{C}^{\r{n}\,\b{q{-}1}}_{\mathfrak{a}_k}(\mathbf{F}){+}\mathcal{C}^{\r{n{+}1}\,\b{q{-}1}}_{\mathfrak{a}_k}(\mathbf{F})\,,}
which is easily solved:
\eq{\mathcal{C}^{\r{n}\,\b{q}}_{\mathfrak{a}_k}(\mathbf{F})=\sum_{m{=}0}^{\b{q}}\binom{\b{q}}{m}\mathcal{C}^{\r{n{+}m}}_{\mathfrak{a}_k}\,.}

Let us consider the simplest example of charged matter in any non-abelian gauge theory: the fundamental (`$\mathbf{2}$') of $\mathfrak{su}_2$. The resulting numbers of independent colour-tensors are given in \mbox{Table~\ref{a1f_table}}. 
\begin{table}[t!]\renewcommand{\arraystretch}{.125}$$\fwbox{0pt}{\begin{array}{|r@{$\,$}|@{$\,$}r|r|r|r|r|r|r|r|r@{$\,$}|}
\hline\multicolumn{1}{|c@{$\,$}|@{$\,$}}{\text{\backslashbox[14.05pt]{\raisebox{2pt}{$\fwboxL{0pt}{\hspace{-6pt}\b{q}}$\vspace{0pt}}}{\raisebox{-4.5pt}{\fwboxL{0pt}{$\r{\hspace{-1.5pt}$n$}$}}}}}
&\multicolumn{1}{c|}{\r{0}}&\multicolumn{1}{c|}{\r{1}}&\multicolumn{1}{c|}{\r{2}}&\multicolumn{1}{c|}{\r{3}}&\multicolumn{1}{c|}{\r{4}}&\multicolumn{1}{c|}{\r{5}}&\multicolumn{1}{c|}{\r{6}}&\multicolumn{1}{c|}{\r{7}}&\multicolumn{1}{c|}{\r{8}}\\\hline\hline
\fwbox{10pt}{\rule{0pt}{12pt}\b{0}}&\fwboxR{34pt}{1}&\fwboxR{34pt}{0}&\fwboxR{40pt}{1}&\fwboxR{40pt}{1}&\fwboxR{46pt}{3}&\fwboxR{52pt}{6}&\fwboxR{52pt}{15}&\fwboxR{54pt}{36}&\fwboxR{54pt}{91}\\
\fwbox{10pt}{\rule{0pt}{12pt}\b{1}}&\fwboxR{34pt}{1}&\fwboxR{34pt}{1}&\fwboxR{40pt}{2}&\fwboxR{40pt}{4}&\fwboxR{46pt}{9}&\fwboxR{52pt}{21}&\fwboxR{52pt}{51}&\fwboxR{54pt}{127}&\fwboxR{54pt}{323}\\
\fwbox{10pt}{\rule{0pt}{12pt}\b{2}}&\fwboxR{34pt}{2}&\fwboxR{34pt}{3}&\fwboxR{40pt}{6}&\fwboxR{40pt}{13}&\fwboxR{46pt}{30}&\fwboxR{52pt}{72}&\fwboxR{52pt}{178}&\fwboxR{54pt}{450}&\fwboxR{54pt}{1\hspace{-0.5pt},\hspace{-2.5pt}158}\\
\fwbox{10pt}{\rule{0pt}{12pt}\b{3}}&\fwboxR{34pt}{5}&\fwboxR{34pt}{9}&\fwboxR{40pt}{19}&\fwboxR{40pt}{43}&\fwboxR{46pt}{102}&\fwboxR{52pt}{250}&\fwboxR{52pt}{628}&\fwboxR{54pt}{1\hspace{-0.5pt},\hspace{-2.5pt}608}&\fwboxR{54pt}{4\hspace{-0.5pt},\hspace{-2.5pt}181}\\
\fwbox{10pt}{\rule{0pt}{12pt}\b{4}}&\fwboxR{34pt}{14}&\fwboxR{34pt}{28}&\fwboxR{40pt}{62}&\fwboxR{40pt}{145}&\fwboxR{46pt}{352}&\fwboxR{52pt}{878}&\fwboxR{52pt}{2\hspace{-0.5pt},\hspace{-2.5pt}236}&\fwboxR{54pt}{5\hspace{-0.5pt},\hspace{-2.5pt}789}&\fwboxR{54pt}{15\hspace{-0.5pt},\hspace{-2.5pt}190}\\
\fwbox{10pt}{\rule{0pt}{12pt}\b{5}}&\fwboxR{34pt}{42}&\fwboxR{34pt}{90}&\fwboxR{40pt}{207}&\fwboxR{40pt}{497}&\fwboxR{46pt}{1\hspace{-0.5pt},\hspace{-2.5pt}230}&\fwboxR{52pt}{3\hspace{-0.5pt},\hspace{-2.5pt}114}&\fwboxR{52pt}{8\hspace{-0.5pt},\hspace{-2.5pt}025}&\fwboxR{54pt}{20\hspace{-0.5pt},\hspace{-2.5pt}979}&\fwboxR{54pt}{55\hspace{-0.5pt},\hspace{-2.5pt}494}\\
\fwbox{10pt}{\rule{0pt}{12pt}\b{6}}&\fwboxR{34pt}{132}&\fwboxR{34pt}{297}&\fwboxR{40pt}{704}&\fwboxR{40pt}{1\hspace{-0.5pt},\hspace{-2.5pt}727}&\fwboxR{46pt}{4\hspace{-0.5pt},\hspace{-2.5pt}344}&\fwboxR{52pt}{11\hspace{-0.5pt},\hspace{-2.5pt}139}&\fwboxR{52pt}{29\hspace{-0.5pt},\hspace{-2.5pt}004}&\fwboxR{54pt}{76\hspace{-0.5pt},\hspace{-2.5pt}473}&\fwboxR{54pt}{203\hspace{-0.5pt},\hspace{-2.5pt}748}\\
\fwbox{10pt}{\rule{0pt}{12pt}\b{7}}&\fwboxR{34pt}{429}&\fwboxR{34pt}{1\hspace{-0.5pt},\hspace{-2.5pt}001}&\fwboxR{40pt}{2\hspace{-0.5pt},\hspace{-2.5pt}431}&\fwboxR{40pt}{6\hspace{-0.5pt},\hspace{-2.5pt}071}&\fwboxR{46pt}{15\hspace{-0.5pt},\hspace{-2.5pt}483}&\fwboxR{52pt}{40\hspace{-0.5pt},\hspace{-2.5pt}143}&\fwboxR{52pt}{105\hspace{-0.5pt},\hspace{-2.5pt}477}&\fwboxR{54pt}{280\hspace{-0.5pt},\hspace{-2.5pt}221}&\fwboxR{54pt}{751\hspace{-0.5pt},\hspace{-2.5pt}422}\\
\fwbox{10pt}{\rule{0pt}{12pt}\b{8}}&\fwboxR{34pt}{1\hspace{-0.5pt},\hspace{-2.5pt}430}&\fwboxR{34pt}{3\hspace{-0.5pt},\hspace{-2.5pt}432}&\fwboxR{40pt}{8\hspace{-0.5pt},\hspace{-2.5pt}502}&\fwboxR{40pt}{21\hspace{-0.5pt},\hspace{-2.5pt}554}&\fwboxR{46pt}{55\hspace{-0.5pt},\hspace{-2.5pt}626}&\fwboxR{52pt}{145\hspace{-0.5pt},\hspace{-2.5pt}620}&\fwboxR{52pt}{385\hspace{-0.5pt},\hspace{-2.5pt}698}&\fwboxR{54pt}{1\hspace{-0.5pt},\hspace{-2.5pt}031\hspace{-0.5pt},\hspace{-2.5pt}643}&\fwboxR{54pt}{2\hspace{-0.5pt},\hspace{-2.5pt}782\hspace{-0.5pt},\hspace{-2.5pt}476}\\
\fwbox{10pt}{\rule{0pt}{12pt}\b{9}}&\fwboxR{34pt}{4\hspace{-0.5pt},\hspace{-2.5pt}862}&\fwboxR{34pt}{11\hspace{-0.5pt},\hspace{-2.5pt}934}&\fwboxR{40pt}{30\hspace{-0.5pt},\hspace{-2.5pt}056}&\fwboxR{40pt}{77\hspace{-0.5pt},\hspace{-2.5pt}180}&\fwboxR{46pt}{201\hspace{-0.5pt},\hspace{-2.5pt}246}&\fwboxR{52pt}{531\hspace{-0.5pt},\hspace{-2.5pt}318}&\fwboxR{52pt}{1\hspace{-0.5pt},\hspace{-2.5pt}417\hspace{-0.5pt},\hspace{-2.5pt}341}&\fwboxR{54pt}{3\hspace{-0.5pt},\hspace{-2.5pt}814\hspace{-0.5pt},\hspace{-2.5pt}119}&\fwboxR{54pt}{10\hspace{-0.5pt},\hspace{-2.5pt}341\hspace{-0.5pt},\hspace{-2.5pt}280}\\
\fwbox{10pt}{\rule{0pt}{12pt}\b{10}}&\fwboxR{34pt}{16\hspace{-0.5pt},\hspace{-2.5pt}796}&\fwboxR{34pt}{41\hspace{-0.5pt},\hspace{-2.5pt}990}&\fwboxR{40pt}{107\hspace{-0.5pt},\hspace{-2.5pt}236}&\fwboxR{40pt}{278\hspace{-0.5pt},\hspace{-2.5pt}426}&\fwboxR{46pt}{732\hspace{-0.5pt},\hspace{-2.5pt}564}&\fwboxR{52pt}{1\hspace{-0.5pt},\hspace{-2.5pt}948\hspace{-0.5pt},\hspace{-2.5pt}659}&\fwboxR{52pt}{5\hspace{-0.5pt},\hspace{-2.5pt}231\hspace{-0.5pt},\hspace{-2.5pt}460}&\fwboxR{54pt}{14\hspace{-0.5pt},\hspace{-2.5pt}155\hspace{-0.5pt},\hspace{-2.5pt}399}&\fwboxR{54pt}{38\hspace{-0.5pt},\hspace{-2.5pt}563\hspace{-0.5pt},\hspace{-2.5pt}064}\\\hline
\end{array}}\vspace{-14pt}$$\renewcommand{\arraystretch}{1}\caption{$\mathcal{C}^{\r{n}\,\b{q}}_{\mathfrak{a}_1}(\mathbf{F})$: Scattering $\b{q}\!\times\!\mathbf{2}$-charged lines and $\r{n}\times$Adjoints($\mathbf{3}$s) in $\mathfrak{a}_1$ gauge theory.\label{a1f_table}}\end{table}

Notice that for $\r{n}{=}0$, $\mathcal{C}^{\r{0}\,\b{q}}_{\mathfrak{a}_1}(\mathbf{F})$ is simply the $\b{q}$th \emph{Catalan} number \cite{oeis-catalan}! The second row is easily identified as the Motzkin numbers \cite{oeis-motzkin}. Each column of \mbox{Table~\ref{a1f_table}} is obtained by subtracting successive entries of the previous column. As such, we may call this a `\emph{Catalan difference table}'.

The recursive structure obviously holds for all cases of fundamental matter in $\mathfrak{a}$-type gauge theories. In particular, it holds for the saturated case obtained in the limit of large rank. Defining 
\eq{!(\r{r},\b{q})\equivR\!\sum_{m{=}0}^{\b{q}}\,\binom{\b{q}}{m}!(\r{n{+}m})\,,\label{generalized_saturation_a_defined}}
we may tabulate the large-$k$ limit of $\mathcal{C}^{\r{n}\,\b{q}}_{\mathfrak{a}_k}(\mathbf{F})$ directly. These numbers are given in \mbox{Table~\ref{nq_derangements_table}}. We already knew that in the large-$k$ limit, $\mathcal{C}^{\r{n}\,\b{0}}_{\mathfrak{a}_{\infty}}(\mathbf{F})=!\r{n}$, but we can further read off from the table that $\mathcal{C}^{\r{0}\,\b{q}}_{\mathfrak{a}_{\infty}}(\mathbf{F})=\b{q}!$. Indeed, starting from either of these one may recursively fill in the rest of the table, leading to what is sometimes called \emph{Euler's difference table}, as given in \mbox{Table~\ref{nq_derangements_table}}.

\begin{table}[t!]\renewcommand{\arraystretch}{.125}$$\fwbox{0pt}{\begin{array}{|r@{$\,$}|@{$\,$}r|r|r|r|r|r|r|r|r@{$\,$}|}
\hline\multicolumn{1}{|c@{$\,$}|@{$\,$}}{\text{\backslashbox[14.05pt]{\raisebox{2pt}{$\fwboxL{0pt}{\hspace{-6pt}\b{q}}$\vspace{0pt}}}{\raisebox{-4.5pt}{\fwboxL{0pt}{$\r{\hspace{-1.5pt}$n$}$}}}}}
&\multicolumn{1}{c|}{\r{0}}&\multicolumn{1}{c|}{\r{1}}&\multicolumn{1}{c|}{\r{2}}&\multicolumn{1}{c|}{\r{3}}&\multicolumn{1}{c|}{\r{4}}&\multicolumn{1}{c|}{\r{5}}&\multicolumn{1}{c|}{\r{6}}\\\hline\hline
\fwbox{10pt}{\rule{0pt}{12pt}\b{0}}&\fwboxR{38pt}{1}&\fwboxR{46pt}{0}&\fwboxR{52pt}{1}&\fwboxR{57pt}{2}&\fwboxR{66pt}{9}&\fwboxR{72pt}{44}&\fwboxR{84pt}{265}\\
\fwbox{10pt}{\rule{0pt}{12pt}\b{1}}&\fwboxR{38pt}{1}&\fwboxR{46pt}{1}&\fwboxR{52pt}{3}&\fwboxR{57pt}{11}&\fwboxR{66pt}{53}&\fwboxR{72pt}{309}&\fwboxR{84pt}{2\hspace{-0.5pt},\hspace{-2.5pt}119}\\
\fwbox{10pt}{\rule{0pt}{12pt}\b{2}}&\fwboxR{38pt}{2}&\fwboxR{46pt}{4}&\fwboxR{52pt}{14}&\fwboxR{57pt}{64}&\fwboxR{66pt}{362}&\fwboxR{72pt}{2\hspace{-0.5pt},\hspace{-2.5pt}428}&\fwboxR{84pt}{18\hspace{-0.5pt},\hspace{-2.5pt}806}\\
\fwbox{10pt}{\rule{0pt}{12pt}\b{3}}&\fwboxR{38pt}{6}&\fwboxR{46pt}{18}&\fwboxR{52pt}{78}&\fwboxR{57pt}{426}&\fwboxR{66pt}{2\hspace{-0.5pt},\hspace{-2.5pt}790}&\fwboxR{72pt}{21\hspace{-0.5pt},\hspace{-2.5pt}234}&\fwboxR{84pt}{183\hspace{-0.5pt},\hspace{-2.5pt}822}\\
\fwbox{10pt}{\rule{0pt}{12pt}\b{4}}&\fwboxR{38pt}{24}&\fwboxR{46pt}{96}&\fwboxR{52pt}{504}&\fwboxR{57pt}{3\hspace{-0.5pt},\hspace{-2.5pt}216}&\fwboxR{66pt}{24\hspace{-0.5pt},\hspace{-2.5pt}024}&\fwboxR{72pt}{205\hspace{-0.5pt},\hspace{-2.5pt}056}&\fwboxR{84pt}{1\hspace{-0.5pt},\hspace{-2.5pt}965\hspace{-0.5pt},\hspace{-2.5pt}624}\\
\fwbox{10pt}{\rule{0pt}{12pt}\b{5}}&\fwboxR{38pt}{120}&\fwboxR{46pt}{600}&\fwboxR{52pt}{3\hspace{-0.5pt},\hspace{-2.5pt}720}&\fwboxR{57pt}{27\hspace{-0.5pt},\hspace{-2.5pt}240}&\fwboxR{66pt}{229\hspace{-0.5pt},\hspace{-2.5pt}080}&\fwboxR{72pt}{2\hspace{-0.5pt},\hspace{-2.5pt}170\hspace{-0.5pt},\hspace{-2.5pt}680}&\fwboxR{84pt}{22\hspace{-0.5pt},\hspace{-2.5pt}852\hspace{-0.5pt},\hspace{-2.5pt}200}\\
\fwbox{10pt}{\rule{0pt}{12pt}\b{6}}&\fwboxR{38pt}{720}&\fwboxR{46pt}{4\hspace{-0.5pt},\hspace{-2.5pt}320}&\fwboxR{52pt}{30\hspace{-0.5pt},\hspace{-2.5pt}960}&\fwboxR{57pt}{256\hspace{-0.5pt},\hspace{-2.5pt}320}&\fwboxR{66pt}{2\hspace{-0.5pt},\hspace{-2.5pt}399\hspace{-0.5pt},\hspace{-2.5pt}760}&\fwboxR{72pt}{25\hspace{-0.5pt},\hspace{-2.5pt}022\hspace{-0.5pt},\hspace{-2.5pt}880}&\fwboxR{84pt}{287\hspace{-0.5pt},\hspace{-2.5pt}250\hspace{-0.5pt},\hspace{-2.5pt}480}\\
\fwbox{10pt}{\rule{0pt}{12pt}\b{7}}&\fwboxR{38pt}{5\hspace{-0.5pt},\hspace{-2.5pt}040}&\fwboxR{46pt}{35\hspace{-0.5pt},\hspace{-2.5pt}280}&\fwboxR{52pt}{287\hspace{-0.5pt},\hspace{-2.5pt}280}&\fwboxR{57pt}{2\hspace{-0.5pt},\hspace{-2.5pt}656\hspace{-0.5pt},\hspace{-2.5pt}080}&\fwboxR{66pt}{27\hspace{-0.5pt},\hspace{-2.5pt}422\hspace{-0.5pt},\hspace{-2.5pt}640}&\fwboxR{72pt}{312\hspace{-0.5pt},\hspace{-2.5pt}273\hspace{-0.5pt},\hspace{-2.5pt}360}&\fwboxR{84pt}{3\hspace{-0.5pt},\hspace{-2.5pt}884\hspace{-0.5pt},\hspace{-2.5pt}393\hspace{-0.5pt},\hspace{-2.5pt}520}\\
\fwbox{10pt}{\rule{0pt}{12pt}\b{8}}&\fwboxR{38pt}{40\hspace{-0.5pt},\hspace{-2.5pt}320}&\fwboxR{46pt}{322\hspace{-0.5pt},\hspace{-2.5pt}560}&\fwboxR{52pt}{2\hspace{-0.5pt},\hspace{-2.5pt}943\hspace{-0.5pt},\hspace{-2.5pt}360}&\fwboxR{57pt}{30\hspace{-0.5pt},\hspace{-2.5pt}078\hspace{-0.5pt},\hspace{-2.5pt}720}&\fwboxR{66pt}{339\hspace{-0.5pt},\hspace{-2.5pt}696\hspace{-0.5pt},\hspace{-2.5pt}000}&\fwboxR{72pt}{4\hspace{-0.5pt},\hspace{-2.5pt}196\hspace{-0.5pt},\hspace{-2.5pt}666\hspace{-0.5pt},\hspace{-2.5pt}880}&\fwboxR{84pt}{56\hspace{-0.5pt},\hspace{-2.5pt}255\hspace{-0.5pt},\hspace{-2.5pt}149\hspace{-0.5pt},\hspace{-2.5pt}440}\\\hline
\end{array}}$$\vspace{-20pt}\renewcommand{\arraystretch}{1}\caption{$!(\r{n},\b{q})\text{: large-$k$ limit of scattering $\b{q}\!\times\!\mathbf{F}$-charged lines and $\r{n}\times$adjoints in $\mathfrak{a}_k$-theory.}$\label{nq_derangements_table}}\end{table}

These saturation numbers $!(\r{n},\b{q})$ provide a convenient reference-point in plotting the numbers of colour-tensors away from saturation (or for the exceptional algebras, where no notion of saturation exists). They are associated with the following exponential generating function in two variables:
\eq{!(\r{n},\b{q})\,\,\Leftrightarrow\,\,\underset{!(\r{n},\b{q})}{\text{EGF}}({x},y)\equivR\sum_{\r{n},\b{q}}\frac{(!(\r{n},\b{q}))}{(\r{n}!)(\b{q}!)}x^{\r{n}}y^{\b{q}}=\frac{e^{-{x}}}{1{-}{x}{-}y}.\,\label{generalized_derangements_egf}}
Also note that $!(\r{1},\b{q})\!=\!\b{q}(\b{q}!)$ and $!(\r{2},\b{q})\!=\!(\b{q}^2{+}\b{q}{+}1)(\b{q}!)$.

\subsubsection{Scattering of Gluons and \emph{Fundamental} Matter in $\mathfrak{b},\mathfrak{c},\mathfrak{d}$-Theories}

Outside of the case of the fundamental representation of $\mathfrak{a}_k$, there is always some contribution(s) to $\mathbf{R}\!\otimes\!\bar{\mathbf{R}}$ besides $\mathbf{1}$ and $\mathbf{ad}$. This prevents us from repeating the simple recursive construction of the $\mathfrak{a}_k$ case: generally speaking, knowledge of the number of independent adjoint tensors is not sufficient to recursively construct the numbers for some alternative particle content. Nevertheless, we may calculate these numbers directly in the large-rank limit (where $\mathfrak{b}_k,\mathfrak{c}_k,\mathfrak{d}_k$ are again found to saturate identically) and collect the results in Table~\ref{nq_dihedral_derangements_table}. We will again generalize the adjoint notation $!!\r{n}$ to $!!(\r{n},\b{q})$ in the obvious way (\emph{defined} as the entries in Table~\ref{nq_dihedral_derangements_table}). Amusingly, $\mathcal{C}^{\r{n}\,\b{0}}_{(\mathfrak{b},\mathfrak{c},\mathfrak{d})_{\infty}}(\mathbf{F})=!!\r{n}$ and $\mathcal{C}^{\r{0}\,\b{q}}_{(\mathfrak{b},\mathfrak{c},\mathfrak{d})_{\infty}}(\mathbf{F})=(2\b{q}-1)!!$, justifying the notation.\footnote{Also note that $({-}1)!!{=}1$ is well-defined due to the recurrence $n!!{=}n(n{-}2)!!$}

In spite of the lack of a recurrence and the diverse nature of the $\mathfrak{b}_k/\mathfrak{c}_k/\mathfrak{d}_k$ algebras, remarkably it turns out these numbers are \emph{again} captured by a simple exponential generating function:
\eq{!!(\r{n},\b{q})\,\,\Leftrightarrow\,\,\underset{!!(\r{n},\b{q})}{\text{EGF}}({x},y)\equivR\sum_{\r{n},\b{q}}\frac{(!!(\r{n},\b{q}))}{(\r{n}!)(\b{q}!)}x^{\r{n}}y^{\b{q}}=\frac{e^{{x}^2/4{-}x/2}}{\sqrt{1{-}{x}{-}2y}}.\,\label{generalized_dihedral_derangements_egf}}
Impressively, equations~(\ref{generalized_derangements_egf}) and~(\ref{generalized_dihedral_derangements_egf}) compactly encode the number of independent colour-tensors involving \emph{any} number of fundamental matter lines and adjoints at saturation for \emph{any} of the classical Lie algebras in the limit of large-rank. The specific criteria for saturation for each of the classical Lie algebras is given in Table~\ref{saturation_conditions_general} below. 

\begin{table}[t!]\renewcommand{\arraystretch}{.125}$$\fwbox{0pt}{\begin{array}{|r@{$\,$}|@{$\,$}r|r|r|r|r|r|r|r|r@{$\,$}|}
\hline\multicolumn{1}{|c@{$\,$}|@{$\,$}}{\text{\backslashbox[14.05pt]{\raisebox{2pt}{$\fwboxL{0pt}{\hspace{-6pt}\b{q}}$\vspace{0pt}}}{\raisebox{-4.5pt}{\fwboxL{0pt}{$\r{\hspace{-1.5pt}$n$}$}}}}}
&\multicolumn{1}{c|}{\r{0}}&\multicolumn{1}{c|}{\r{1}}&\multicolumn{1}{c|}{\r{2}}&\multicolumn{1}{c|}{\r{3}}&\multicolumn{1}{c|}{\r{4}}&\multicolumn{1}{c|}{\r{5}}&\multicolumn{1}{c|}{\r{6}}\\\hline\hline
\fwbox{10pt}{\rule{0pt}{12pt}\b{0}}&\fwboxR{38pt}{1}&\fwboxR{46pt}{0}&\fwboxR{52pt}{1}&\fwboxR{57pt}{1}&\fwboxR{66pt}{6}&\fwboxR{72pt}{22}&\fwboxR{84pt}{130}\\
\fwbox{10pt}{\rule{0pt}{12pt}\b{1}}&\fwboxR{38pt}{1}&\fwboxR{46pt}{1}&\fwboxR{52pt}{3}&\fwboxR{57pt}{10}&\fwboxR{66pt}{46}&\fwboxR{72pt}{252}&\fwboxR{84pt}{1\hspace{-0.5pt},\hspace{-2.5pt}642}\\
\fwbox{10pt}{\rule{0pt}{12pt}\b{2}}&\fwboxR{38pt}{3}&\fwboxR{46pt}{6}&\fwboxR{52pt}{21}&\fwboxR{57pt}{93}&\fwboxR{66pt}{510}&\fwboxR{72pt}{3\hspace{-0.5pt},\hspace{-2.5pt}306}&\fwboxR{84pt}{24\hspace{-0.5pt},\hspace{-2.5pt}762}\\
\fwbox{10pt}{\rule{0pt}{12pt}\b{3}}&\fwboxR{38pt}{15}&\fwboxR{46pt}{45}&\fwboxR{52pt}{195}&\fwboxR{57pt}{1\hspace{-0.5pt},\hspace{-2.5pt}050}&\fwboxR{66pt}{6\hspace{-0.5pt},\hspace{-2.5pt}750}&\fwboxR{72pt}{50\hspace{-0.5pt},\hspace{-2.5pt}280}&\fwboxR{84pt}{425\hspace{-0.5pt},\hspace{-2.5pt}490}\\
\fwbox{10pt}{\rule{0pt}{12pt}\b{4}}&\fwboxR{38pt}{105}&\fwboxR{46pt}{420}&\fwboxR{52pt}{2\hspace{-0.5pt},\hspace{-2.5pt}205}&\fwboxR{57pt}{13\hspace{-0.5pt},\hspace{-2.5pt}965}&\fwboxR{66pt}{103\hspace{-0.5pt},\hspace{-2.5pt}110}&\fwboxR{72pt}{867\hspace{-0.5pt},\hspace{-2.5pt}510}&\fwboxR{84pt}{8\hspace{-0.5pt},\hspace{-2.5pt}183\hspace{-0.5pt},\hspace{-2.5pt}490}\\
\fwbox{10pt}{\rule{0pt}{12pt}\b{5}}&\fwboxR{38pt}{945}&\fwboxR{46pt}{4\hspace{-0.5pt},\hspace{-2.5pt}725}&\fwboxR{52pt}{29\hspace{-0.5pt},\hspace{-2.5pt}295}&\fwboxR{57pt}{213\hspace{-0.5pt},\hspace{-2.5pt}570}&\fwboxR{66pt}{1\hspace{-0.5pt},\hspace{-2.5pt}782\hspace{-0.5pt},\hspace{-2.5pt}270}&\fwboxR{72pt}{16\hspace{-0.5pt},\hspace{-2.5pt}718\hspace{-0.5pt},\hspace{-2.5pt}940}&\fwboxR{84pt}{173\hspace{-0.5pt},\hspace{-2.5pt}965\hspace{-0.5pt},\hspace{-2.5pt}050}\\
\fwbox{10pt}{\rule{0pt}{12pt}\b{6}}&\fwboxR{38pt}{10\hspace{-0.5pt},\hspace{-2.5pt}395}&\fwboxR{46pt}{62\hspace{-0.5pt},\hspace{-2.5pt}370}&\fwboxR{52pt}{446\hspace{-0.5pt},\hspace{-2.5pt}985}&\fwboxR{57pt}{3\hspace{-0.5pt},\hspace{-2.5pt}690\hspace{-0.5pt},\hspace{-2.5pt}225}&\fwboxR{66pt}{34\hspace{-0.5pt},\hspace{-2.5pt}365\hspace{-0.5pt},\hspace{-2.5pt}870}&\fwboxR{72pt}{355\hspace{-0.5pt},\hspace{-2.5pt}737\hspace{-0.5pt},\hspace{-2.5pt}690}&\fwboxR{84pt}{4\hspace{-0.5pt},\hspace{-2.5pt}048\hspace{-0.5pt},\hspace{-2.5pt}041\hspace{-0.5pt},\hspace{-2.5pt}690}\\
\fwbox{10pt}{\rule{0pt}{12pt}\b{7}}&\fwboxR{38pt}{135\hspace{-0.5pt},\hspace{-2.5pt}135}&\fwboxR{46pt}{945\hspace{-0.5pt},\hspace{-2.5pt}945}&\fwboxR{52pt}{7\hspace{-0.5pt},\hspace{-2.5pt}702\hspace{-0.5pt},\hspace{-2.5pt}695}&\fwboxR{57pt}{71\hspace{-0.5pt},\hspace{-2.5pt}081\hspace{-0.5pt},\hspace{-2.5pt}010}&\fwboxR{66pt}{731\hspace{-0.5pt},\hspace{-2.5pt}080\hspace{-0.5pt},\hspace{-2.5pt}350}&\fwboxR{72pt}{8\hspace{-0.5pt},\hspace{-2.5pt}279\hspace{-0.5pt},\hspace{-2.5pt}991\hspace{-0.5pt},\hspace{-2.5pt}720}&\fwboxR{84pt}{102\hspace{-0.5pt},\hspace{-2.5pt}304\hspace{-0.5pt},\hspace{-2.5pt}492\hspace{-0.5pt},\hspace{-2.5pt}290}\\\hline\end{array}}$$\vspace{-24pt}\renewcommand{\arraystretch}{1}\caption{$!!(\r{n},\b{q})\text{: large-$k$ limit of scattering $\b{q}\!\times\!\mathbf{F}$-lines and $\r{n}\times$adjoints in $\mathfrak{b},\mathfrak{c},\mathfrak{d}$-theories.}$\label{nq_dihedral_derangements_table}}\vspace{0pt}\end{table}

\begin{table}[t!]
\vspace{-10pt}\centering\begin{tabular}{clc}
algebra&saturates for:&$\displaystyle\lim_{k\!\to\infty}\mathcal{C}^{\r{n},\b{q}}(\mathbf{F})$\\
$\mathfrak{a}_k$&$k~\!>\!\r{n}{+}\b{q}~{-}2$&$\phantom{!}!(\r{n},\b{q})$\\
$\mathfrak{b}_k$&$k~\!>\!\frac{1}{2}(\r{n}{+}\b{q}){-}1$&$!!(\r{n},\b{q})$\\
$\mathfrak{c}_k$&$k~\!>\!\frac{1}{2}\r{n}{+}\b{q}~{-}1$&$!!(\r{n},\b{q})$\\
$\mathfrak{d}_k$&$k~\!>\!\r{n}{+}\b{q}$&$!!(\r{n},\b{q})$\\
\end{tabular}\vspace{-10pt}
\caption{Saturation points for scattering of $\mathbf{F}$-matter and adjoints for classical algebras.\label{saturation_conditions_general}}\vspace{-10pt}\end{table}

It is important to note that while $!!\r{n}<\,!\r{n}$, it is \emph{not} true that $!!(\r{n}, \b{q})\!<\,!(\r{n}, \b{q})$ for all $\r{n}$ and $\b{q}$. This is perhaps startling: in $\mathfrak{a}_k$-type gauge theory the expectation would be that multi-traces over fundamental generators (which are counted combinatorially) should suffice to describe colour-structures involving fundamental matter and adjoints, but \emph{this can't be the whole story for the other classical Lie algebras}. If there were a combinatorial construction in terms of fundamental generators we would expect the same to apply to the other classical algebras: too many tensors due to (\emph{e.g.}) dihedral symmetry amongst the traces is one thing, but not enough tensors is quite another. Since we may (for sufficiently large $\b{q}$) have an excess, $!!(\r{n},\b{q})\!>!(\r{n},\b{q})$, \emph{multi-traces over fundamental generators alone can not be sufficient}. Analogous to the sub-saturation adjoint case for $\mathfrak{d}_k$ discussed in section~\ref{subsec:dragons}, there are necessarily tensors which can not be expressed in terms of multi-traces of the type familiar from $\mathfrak{a}_k$. It is not clear whether or not these additional tensors are generated at the perturbative level in specific theories, nor how they fit into the particular-charge-representation-agnostic colour-tensors described by \cite{Melia:2015ika,Johansson:2015oia,Ochirov:2019mtf} (albeit, limited to tree-level). We must postpone such explorations to future work.\\

Below saturation, little can be said without wading into algebra-specific analyses. Broadly speaking though, the deviation from saturation mirrors the adjoint case: $\mathcal{C}^{\r{n}\,\b{q}}_{\mathfrak{d}_{k}}(\mathbf{F})$ \emph{exceeds} $!!(\r{n},\b{q})$ with the excess reducing (eventually to 0) as rank increases, and $\mathcal{C}^{\r{n}\,\b{q}}_{\mathfrak{c}_{k}}(\mathbf{F})<\mathcal{C}^{\r{n}\,\b{q}}_{\mathfrak{b}_{k}}(\mathbf{F})$, both of which then saturate monotonically as rank increases, as may be seen from Figures~\ref{cnq_figure_bc_series} and~\ref{cnq_figure_d_series} shown below. In \mbox{Figure~\ref{cnq_figure_d_series}}, we also show how increasing the number of fundamental particles has an effect very similar to shifting the counting in multiplicity. 
\pagebreak

\begin{figure}[t!]$$\fwbox{0pt}{\fig{0pt}{1}{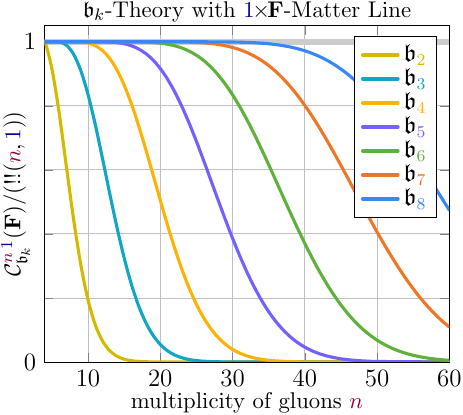}\hspace{10pt}\fig{0pt}{1}{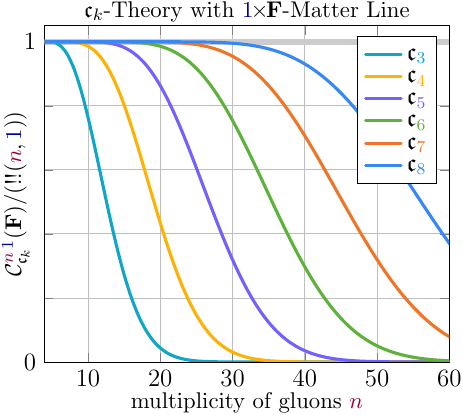}}$$\vspace{-35pt}\caption{$\mathcal{C}^{\r{n},\b{1}}_\mathfrak{g}(\mathbf{F})$ for each of the Lie algebras $\mathfrak{b},\mathfrak{c}$.\label{cnq_figure_bc_series}}\vspace{0pt}\end{figure}

\begin{figure}[t!]$$\fwbox{0pt}{\fig{0pt}{1}{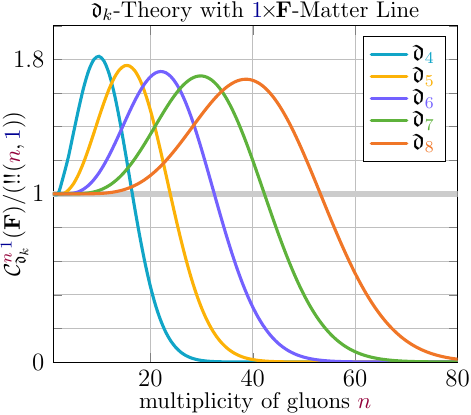}\hspace{10pt}\fig{0pt}{1}{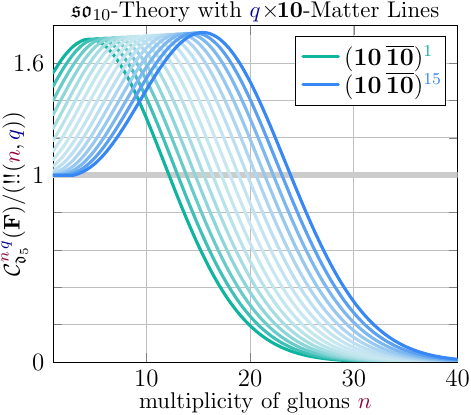}}$$\vspace{-35pt}\caption{$\mathcal{C}^{\r{n},\b{1}}_\mathfrak{d_k}(\mathbf{F})$ for Lie algebras $\mathfrak{d}_k$ through rank $k{=}8$, and for $\b{q}\!\leq\!15$ for $\mathfrak{d}_5$.\label{cnq_figure_d_series}}\vspace{0pt}\end{figure}

\subsubsection{Scattering of Gluons and \emph{Fundamental} Matter in $\mathfrak{e},\mathfrak{f},\mathfrak{g}$-Theories}

\begin{figure}[h!]$$\fwbox{0pt}{\fig{0pt}{1}{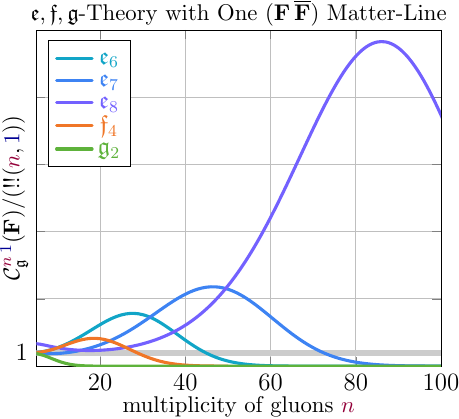}\hspace{10pt}\fig{0pt}{1}{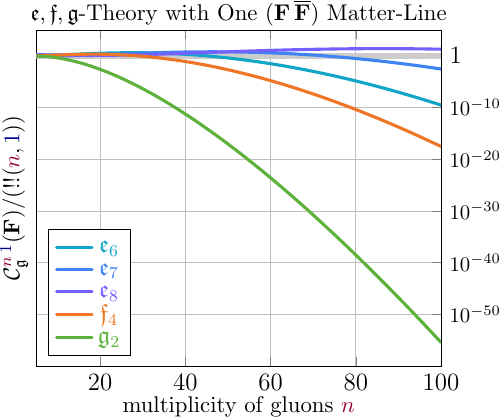}}$$\vspace{-35pt}\caption{$\mathcal{C}^{\r{n},\b{1}}_\mathfrak{g}(\mathbf{F})$ for each of the exceptional Lie algebras $\mathfrak{e},\mathfrak{f},\mathfrak{g}$.\label{cn1_eft_plots}}\vspace{0pt}\end{figure}

It is not \emph{a priori} clear that comparing the exceptional algebras to the classical ones at large rank is at all well motivated. In the case of scattering only adjoint matter, there was some expectation that an (over-)complete basis should exist of multi-trace tensors built from their fundamental representation's generators, which would be of size $!\r{n}$ for $\mathfrak{e}_6$ and $!!\r{n}$ for all the others. But when considering multiple numbers of charged matter, it is not at all clear there should exist a simple (over-)complete basis of symbols that make sense for all algebras. Nevertheless, in plotting these numbers, say, it is useful to compare to \emph{something} which grows (nearly) as rapidly. For this reason, in \mbox{Figure~\ref{cn1_eft_plots}} we have plotted $\mathcal{C}^{\r{n}\,\b{1}}_{\mathfrak{g}}(\mathbf{F})/(!!(\r{n},\b{1}))$ to get some sense of their relative sizes. 

One of the unique things about $\mathfrak{e}_8$ is that the fundamental \emph{is} the adjoint, $\mathbf{248}$. Nevertheless, if we are considering adjoint \emph{matter} this is different from the pure-adjoint analysis due to the breaking of Bose-symmetry. In Figure~\ref{cn1_eft_plots} we see familiar features: the exceptionals have an `excess' at low-multiplicities (they exceed the expectation from the classical algebras) and then fall off at higher multiplicity, just as in Figure~\ref{adjn_plots}. However, whereas in the adjoint case only $\mathfrak{e}_6$ exceeded $!!\r{n}$, all exceptional algebras with one fundamental-charged line exceed $!!(\r{n},\b{1})$ for some range of $\r{n}$ \emph{except} the case of $\mathfrak{g}_2$. On the subject of $\mathfrak{g}_2$, it is interesting to note that for $\r{n}{=}100$ gluons, scattering with one fundamental-charged line involves \emph{$56$ orders of magnitude fewer} colour tensors than the large-rank-saturated number $!!(\r{100},\b{1})$ (which is itself merely 4 times fewer than the case of $!(\r{100},\b{1})$). It is perhaps unfair to read too deeply into this, however; after all, it is not clear that this is even close to an apples-to-apples comparison; but it suggests that is a dearth of interesting and `exceptional' behaviour to be seen in the case of scattering matter in these gauge theories.

The above Figures are limited in that they only make the comparison for $\b{q}{=}1$. A more holistic comparison can be made from the data collected in Tables~\ref{e6nq_table}, \ref{e7nq_table}, \ref{e8nq_table}, \ref{f4nq_table}, \ref{g2nq_table}, which give the specific values of $\mathcal{C}^{\r{n}\,\b{q}}_{\mathfrak{g}}$ over modest ranges of adjoints and fundamental multiplicities. 

Interestingly, $\mathcal{C}^{\r{0},\b{q}}_{\mathfrak{f}_4}(\mathbf{F})$ has \emph{far} more colour-tensors than any other algebra (besides $\mathfrak{e}_8$). In particular, it is smaller than $\mathcal{C}^{\r{0},\b{q}}_{\mathfrak{e}_7}$ for all $\b{q}\!\leq\!22$. Despite this low-$\b{q}$ behavior, we expect on general grounds that all counting grows like a power of the dimension of the representation at large multiplicity; as such, we should not be surprised that the number of colours in $(\mathbf{56}\!\otimes\!\bar{\mathbf{56}})^{\b{q}}$ for $\mathfrak{e}_7$ eventually exceeds that of $(\mathbf{52}\!\otimes\!\bar{\mathbf{52}})^{\b{q}}$ for $\mathfrak{f}_4$.

\pagebreak

\begin{table}[h!]\renewcommand{\arraystretch}{.125}$$\fwbox{0pt}{\begin{array}{|r@{$\,$}|@{$\,$}r|r|r|r|r|r|r|r|r@{$\,$}|}
\hline\multicolumn{1}{|c@{$\,$}|@{$\,$}}{\text{\backslashbox[14.05pt]{\raisebox{2pt}{$\fwboxL{0pt}{\hspace{-6pt}\b{q}}$\vspace{0pt}}}{\raisebox{-4.5pt}{\fwboxL{0pt}{$\r{\hspace{-1.5pt}$n$}$}}}}}
&\multicolumn{1}{c|}{\r{0}}&\multicolumn{1}{c|}{\r{1}}&\multicolumn{1}{c|}{\r{2}}&\multicolumn{1}{c|}{\r{3}}&\multicolumn{1}{c|}{\r{4}}&\multicolumn{1}{c|}{\r{5}}&\multicolumn{1}{c|}{\r{6}}\\\hline\hline
\fwbox{10pt}{\rule{0pt}{12pt}\b{0}}&\fwboxR{38pt}{1}&\fwboxR{46pt}{0}&\fwboxR{52pt}{1}&\fwboxR{57pt}{1}&\fwboxR{66pt}{5}&\fwboxR{72pt}{17}&\fwboxR{84pt}{90}\\
\fwbox{10pt}{\rule{0pt}{12pt}\b{1}}&\fwboxR{38pt}{1}&\fwboxR{46pt}{1}&\fwboxR{52pt}{3}&\fwboxR{57pt}{10}&\fwboxR{66pt}{46}&\fwboxR{72pt}{257}&\fwboxR{84pt}{1\hspace{-0.5pt},\hspace{-2.5pt}732}\\
\fwbox{10pt}{\rule{0pt}{12pt}\b{2}}&\fwboxR{38pt}{3}&\fwboxR{46pt}{7}&\fwboxR{52pt}{27}&\fwboxR{57pt}{133}&\fwboxR{66pt}{810}&\fwboxR{72pt}{5\hspace{-0.5pt},\hspace{-2.5pt}865}&\fwboxR{84pt}{49\hspace{-0.5pt},\hspace{-2.5pt}295}\\
\fwbox{10pt}{\rule{0pt}{12pt}\b{3}}&\fwboxR{38pt}{20}&\fwboxR{46pt}{78}&\fwboxR{52pt}{413}&\fwboxR{57pt}{2\hspace{-0.5pt},\hspace{-2.5pt}687}&\fwboxR{66pt}{20\hspace{-0.5pt},\hspace{-2.5pt}691}&\fwboxR{72pt}{183\hspace{-0.5pt},\hspace{-2.5pt}760}&\fwboxR{84pt}{1\hspace{-0.5pt},\hspace{-2.5pt}847\hspace{-0.5pt},\hspace{-2.5pt}238}\\
\fwbox{10pt}{\rule{0pt}{12pt}\b{4}}&\fwboxR{38pt}{241}&\fwboxR{46pt}{1\hspace{-0.5pt},\hspace{-2.5pt}342}&\fwboxR{52pt}{9\hspace{-0.5pt},\hspace{-2.5pt}259}&\fwboxR{57pt}{75\hspace{-0.5pt},\hspace{-2.5pt}319}&\fwboxR{66pt}{703\hspace{-0.5pt},\hspace{-2.5pt}303}&\fwboxR{72pt}{7\hspace{-0.5pt},\hspace{-2.5pt}398\hspace{-0.5pt},\hspace{-2.5pt}154}&\fwboxR{84pt}{86\hspace{-0.5pt},\hspace{-2.5pt}425\hspace{-0.5pt},\hspace{-2.5pt}818}\\
\fwbox{10pt}{\rule{0pt}{12pt}\b{5}}&\fwboxR{38pt}{4\hspace{-0.5pt},\hspace{-2.5pt}523}&\fwboxR{46pt}{32\hspace{-0.5pt},\hspace{-2.5pt}885}&\fwboxR{52pt}{281\hspace{-0.5pt},\hspace{-2.5pt}433}&\fwboxR{57pt}{2\hspace{-0.5pt},\hspace{-2.5pt}753\hspace{-0.5pt},\hspace{-2.5pt}618}&\fwboxR{66pt}{30\hspace{-0.5pt},\hspace{-2.5pt}226\hspace{-0.5pt},\hspace{-2.5pt}701}&\fwboxR{72pt}{367\hspace{-0.5pt},\hspace{-2.5pt}078\hspace{-0.5pt},\hspace{-2.5pt}180}&\fwboxR{84pt}{4\hspace{-0.5pt},\hspace{-2.5pt}877\hspace{-0.5pt},\hspace{-2.5pt}668\hspace{-0.5pt},\hspace{-2.5pt}495}\\
\fwbox{10pt}{\rule{0pt}{12pt}\b{6}}&\fwboxR{38pt}{119\hspace{-0.5pt},\hspace{-2.5pt}858}&\fwboxR{46pt}{1\hspace{-0.5pt},\hspace{-2.5pt}075\hspace{-0.5pt},\hspace{-2.5pt}893}&\fwboxR{52pt}{11\hspace{-0.5pt},\hspace{-2.5pt}002\hspace{-0.5pt},\hspace{-2.5pt}721}&\fwboxR{57pt}{125\hspace{-0.5pt},\hspace{-2.5pt}764\hspace{-0.5pt},\hspace{-2.5pt}739}&\fwboxR{66pt}{1\hspace{-0.5pt},\hspace{-2.5pt}584\hspace{-0.5pt},\hspace{-2.5pt}806\hspace{-0.5pt},\hspace{-2.5pt}313}&\fwboxR{72pt}{21\hspace{-0.5pt},\hspace{-2.5pt}782\hspace{-0.5pt},\hspace{-2.5pt}379\hspace{-0.5pt},\hspace{-2.5pt}160}&\fwboxR{84pt}{323\hspace{-0.5pt},\hspace{-2.5pt}719\hspace{-0.5pt},\hspace{-2.5pt}876\hspace{-0.5pt},\hspace{-2.5pt}855}\\\hline
\end{array}}$$\renewcommand{\arraystretch}{1}\vspace{-24pt}\caption{$\mathcal{C}^{\r{n}\,\b{q}}_{\mathfrak{e}_6}(\mathbf{F})$: scattering $\b{q}\!\times\!\mathbf{F}$-lines and $\r{n}\times$adjoints in $\mathfrak{e}_6$ gauge theory.\label{e6nq_table}}\vspace{-10pt}\end{table}

\begin{table}[h!]\renewcommand{\arraystretch}{.125}\vspace{-10pt}$$\fwbox{0pt}{\begin{array}{|r@{$\,$}|@{$\,$}r|r|r|r|r|r|r|r|r@{$\,$}|}
\hline\multicolumn{1}{|c@{$\,$}|@{$\,$}}{\text{\backslashbox[14.05pt]{\raisebox{2pt}{$\fwboxL{0pt}{\hspace{-6pt}\b{q}}$\vspace{0pt}}}{\raisebox{-4.5pt}{\fwboxL{0pt}{$\r{\hspace{-1.5pt}$n$}$}}}}}
&\multicolumn{1}{c|}{\r{0}}&\multicolumn{1}{c|}{\r{1}}&\multicolumn{1}{c|}{\r{2}}&\multicolumn{1}{c|}{\r{3}}&\multicolumn{1}{c|}{\r{4}}&\multicolumn{1}{c|}{\r{5}}&\multicolumn{1}{c|}{\r{6}}\\\hline\hline
\fwbox{10pt}{\rule{0pt}{12pt}\b{0}}&\fwboxR{38pt}{1}&\fwboxR{46pt}{0}&\fwboxR{52pt}{1}&\fwboxR{57pt}{1}&\fwboxR{66pt}{5}&\fwboxR{72pt}{16}&\fwboxR{84pt}{80}\\
\fwbox{10pt}{\rule{0pt}{12pt}\b{1}}&\fwboxR{38pt}{1}&\fwboxR{46pt}{1}&\fwboxR{52pt}{3}&\fwboxR{57pt}{10}&\fwboxR{66pt}{45}&\fwboxR{72pt}{242}&\fwboxR{84pt}{1\hspace{-0.5pt},\hspace{-2.5pt}547}\\
\fwbox{10pt}{\rule{0pt}{12pt}\b{2}}&\fwboxR{38pt}{4}&\fwboxR{46pt}{9}&\fwboxR{52pt}{34}&\fwboxR{57pt}{163}&\fwboxR{66pt}{958}&\fwboxR{72pt}{6\hspace{-0.5pt},\hspace{-2.5pt}645}&\fwboxR{84pt}{53\hspace{-0.5pt},\hspace{-2.5pt}236}\\
\fwbox{10pt}{\rule{0pt}{12pt}\b{3}}&\fwboxR{38pt}{35}&\fwboxR{46pt}{130}&\fwboxR{52pt}{665}&\fwboxR{57pt}{4\hspace{-0.5pt},\hspace{-2.5pt}180}&\fwboxR{66pt}{31\hspace{-0.5pt},\hspace{-2.5pt}045}&\fwboxR{72pt}{265\hspace{-0.5pt},\hspace{-2.5pt}590}&\fwboxR{84pt}{2\hspace{-0.5pt},\hspace{-2.5pt}571\hspace{-0.5pt},\hspace{-2.5pt}732}\\
\fwbox{10pt}{\rule{0pt}{12pt}\b{4}}&\fwboxR{38pt}{546}&\fwboxR{46pt}{2\hspace{-0.5pt},\hspace{-2.5pt}912}&\fwboxR{52pt}{19\hspace{-0.5pt},\hspace{-2.5pt}446}&\fwboxR{57pt}{153\hspace{-0.5pt},\hspace{-2.5pt}572}&\fwboxR{66pt}{1\hspace{-0.5pt},\hspace{-2.5pt}394\hspace{-0.5pt},\hspace{-2.5pt}422}&\fwboxR{72pt}{14\hspace{-0.5pt},\hspace{-2.5pt}288\hspace{-0.5pt},\hspace{-2.5pt}220}&\fwboxR{84pt}{162\hspace{-0.5pt},\hspace{-2.5pt}987\hspace{-0.5pt},\hspace{-2.5pt}390}\\
\fwbox{10pt}{\rule{0pt}{12pt}\b{5}}&\fwboxR{38pt}{13\hspace{-0.5pt},\hspace{-2.5pt}467}&\fwboxR{46pt}{94\hspace{-0.5pt},\hspace{-2.5pt}995}&\fwboxR{52pt}{794\hspace{-0.5pt},\hspace{-2.5pt}682}&\fwboxR{57pt}{7\hspace{-0.5pt},\hspace{-2.5pt}630\hspace{-0.5pt},\hspace{-2.5pt}080}&\fwboxR{66pt}{82\hspace{-0.5pt},\hspace{-2.5pt}445\hspace{-0.5pt},\hspace{-2.5pt}844}&\fwboxR{72pt}{988\hspace{-0.5pt},\hspace{-2.5pt}740\hspace{-0.5pt},\hspace{-2.5pt}710}&\fwboxR{84pt}{13\hspace{-0.5pt},\hspace{-2.5pt}021\hspace{-0.5pt},\hspace{-2.5pt}697\hspace{-0.5pt},\hspace{-2.5pt}870}\\
\fwbox{10pt}{\rule{0pt}{12pt}\b{6}}&\fwboxR{38pt}{483\hspace{-0.5pt},\hspace{-2.5pt}340}&\fwboxR{46pt}{4\hspace{-0.5pt},\hspace{-2.5pt}272\hspace{-0.5pt},\hspace{-2.5pt}070}&\fwboxR{52pt}{43\hspace{-0.5pt},\hspace{-2.5pt}265\hspace{-0.5pt},\hspace{-2.5pt}320}&\fwboxR{57pt}{491\hspace{-0.5pt},\hspace{-2.5pt}758\hspace{-0.5pt},\hspace{-2.5pt}550}&\fwboxR{66pt}{6\hspace{-0.5pt},\hspace{-2.5pt}185\hspace{-0.5pt},\hspace{-2.5pt}350\hspace{-0.5pt},\hspace{-2.5pt}940}&\fwboxR{72pt}{85\hspace{-0.5pt},\hspace{-2.5pt}192\hspace{-0.5pt},\hspace{-2.5pt}166\hspace{-0.5pt},\hspace{-2.5pt}432}&\fwboxR{84pt}{1\hspace{-0.5pt},\hspace{-2.5pt}274\hspace{-0.5pt},\hspace{-2.5pt}173\hspace{-0.5pt},\hspace{-2.5pt}103\hspace{-0.5pt},\hspace{-2.5pt}012}\\\hline
\end{array}}$$\renewcommand{\arraystretch}{1}\vspace{-24pt}\caption{$\mathcal{C}^{\r{n}\,\b{q}}_{\mathfrak{e}_7}(\mathbf{F})$: scattering $\b{q}\!\times\!\mathbf{F}$-lines and $\r{n}\times$adjoints in $\mathfrak{e}_7$ gauge theory.\label{e7nq_table}}\vspace{-10pt}\end{table}

\begin{table}[h!]\renewcommand{\arraystretch}{.125}\vspace{-10pt}$$\fwbox{0pt}{\begin{array}{|r@{$\,$}|@{$\,$}r|r|r|r|r|r|r|r|r@{$\,$}|}
\hline\multicolumn{1}{|c@{$\,$}|@{$\,$}}{\text{\backslashbox[14.05pt]{\raisebox{2pt}{$\fwboxL{0pt}{\hspace{-6pt}\b{q}}$\vspace{0pt}}}{\raisebox{-4.5pt}{\fwboxL{0pt}{$\r{\hspace{-1.5pt}$n$}$}}}}}
&\multicolumn{1}{c|}{\r{0}}&\multicolumn{1}{c|}{\r{1}}&\multicolumn{1}{c|}{\r{2}}&\multicolumn{1}{c|}{\r{3}}&\multicolumn{1}{c|}{\r{4}}&\multicolumn{1}{c|}{\r{5}}&\multicolumn{1}{c|}{\r{6}}\\\hline\hline
\fwbox{10pt}{\rule{0pt}{12pt}\b{0}}&\fwboxR{38pt}{1}&\fwboxR{46pt}{0}&\fwboxR{52pt}{1}&\fwboxR{57pt}{1}&\fwboxR{66pt}{5}&\fwboxR{72pt}{16}&\fwboxR{84pt}{79}\\
\fwbox{10pt}{\rule{0pt}{12pt}\b{1}}&\fwboxR{38pt}{1}&\fwboxR{46pt}{1}&\fwboxR{52pt}{5}&\fwboxR{57pt}{16}&\fwboxR{66pt}{79}&\fwboxR{72pt}{421}&\fwboxR{84pt}{2\hspace{-0.5pt},\hspace{-2.5pt}674}\\
\fwbox{10pt}{\rule{0pt}{12pt}\b{2}}&\fwboxR{38pt}{5}&\fwboxR{46pt}{16}&\fwboxR{52pt}{79}&\fwboxR{57pt}{421}&\fwboxR{66pt}{2\hspace{-0.5pt},\hspace{-2.5pt}674}&\fwboxR{72pt}{19\hspace{-0.5pt},\hspace{-2.5pt}244}&\fwboxR{84pt}{156\hspace{-0.5pt},\hspace{-2.5pt}612}\\
\fwbox{10pt}{\rule{0pt}{12pt}\b{3}}&\fwboxR{38pt}{79}&\fwboxR{46pt}{421}&\fwboxR{52pt}{2\hspace{-0.5pt},\hspace{-2.5pt}674}&\fwboxR{57pt}{19\hspace{-0.5pt},\hspace{-2.5pt}244}&\fwboxR{66pt}{156\hspace{-0.5pt},\hspace{-2.5pt}612}&\fwboxR{72pt}{1\hspace{-0.5pt},\hspace{-2.5pt}423\hspace{-0.5pt},\hspace{-2.5pt}028}&\fwboxR{84pt}{14\hspace{-0.5pt},\hspace{-2.5pt}320\hspace{-0.5pt},\hspace{-2.5pt}350}\\
\fwbox{10pt}{\rule{0pt}{12pt}\b{4}}&\fwboxR{38pt}{2\hspace{-0.5pt},\hspace{-2.5pt}674}&\fwboxR{46pt}{19\hspace{-0.5pt},\hspace{-2.5pt}244}&\fwboxR{52pt}{156\hspace{-0.5pt},\hspace{-2.5pt}612}&\fwboxR{57pt}{1\hspace{-0.5pt},\hspace{-2.5pt}423\hspace{-0.5pt},\hspace{-2.5pt}028}&\fwboxR{66pt}{14\hspace{-0.5pt},\hspace{-2.5pt}320\hspace{-0.5pt},\hspace{-2.5pt}350}&\fwboxR{72pt}{158\hspace{-0.5pt},\hspace{-2.5pt}390\hspace{-0.5pt},\hspace{-2.5pt}872}&\fwboxR{84pt}{1\hspace{-0.5pt},\hspace{-2.5pt}912\hspace{-0.5pt},\hspace{-2.5pt}977\hspace{-0.5pt},\hspace{-2.5pt}222}\\
\fwbox{10pt}{\rule{0pt}{12pt}\b{5}}&\fwboxR{38pt}{156\hspace{-0.5pt},\hspace{-2.5pt}612}&\fwboxR{46pt}{1\hspace{-0.5pt},\hspace{-2.5pt}423\hspace{-0.5pt},\hspace{-2.5pt}028}&\fwboxR{52pt}{14\hspace{-0.5pt},\hspace{-2.5pt}320\hspace{-0.5pt},\hspace{-2.5pt}350}&\fwboxR{57pt}{158\hspace{-0.5pt},\hspace{-2.5pt}390\hspace{-0.5pt},\hspace{-2.5pt}872}&\fwboxR{66pt}{1\hspace{-0.5pt},\hspace{-2.5pt}912\hspace{-0.5pt},\hspace{-2.5pt}977\hspace{-0.5pt},\hspace{-2.5pt}222}&\fwboxR{72pt}{25\hspace{-0.5pt},\hspace{-2.5pt}083\hspace{-0.5pt},\hspace{-2.5pt}283\hspace{-0.5pt},\hspace{-2.5pt}995}&\fwboxR{84pt}{355\hspace{-0.5pt},\hspace{-2.5pt}246\hspace{-0.5pt},\hspace{-2.5pt}037\hspace{-0.5pt},\hspace{-2.5pt}162}\\\hline
\end{array}}$$\renewcommand{\arraystretch}{1}\vspace{-24pt}\caption{$\mathcal{C}^{\r{n}\,\b{q}}_{\mathfrak{e}_8}(\mathbf{F})$: scattering $\b{q}\!\times\!\mathbf{F}$-lines and $\r{n}\times$adjoints in $\mathfrak{e}_8$ gauge theory.\label{e8nq_table}}\vspace{-10pt}\end{table}

\begin{table}[h!]\renewcommand{\arraystretch}{.125}\vspace{-10pt}$$\fwbox{0pt}{\begin{array}{|r@{$\,$}|@{$\,$}r|r|r|r|r|r|r|r|r@{$\,$}|}\hline\multicolumn{1}{|c@{$\,$}|@{$\,$}}{\text{\backslashbox[14.05pt]{\raisebox{2pt}{$\fwboxL{0pt}{\hspace{-6pt}\b{q}}$\vspace{0pt}}}{\raisebox{-4.5pt}{\fwboxL{0pt}{$\r{\hspace{-1.5pt}$n$}$}}}}}
&\multicolumn{1}{c|}{\r{0}}&\multicolumn{1}{c|}{\r{1}}&\multicolumn{1}{c|}{\r{2}}&\multicolumn{1}{c|}{\r{3}}&\multicolumn{1}{c|}{\r{4}}&\multicolumn{1}{c|}{\r{5}}&\multicolumn{1}{c|}{\r{6}}\\\hline\hline
\fwbox{10pt}{\rule{0pt}{12pt}\b{0}}&\fwboxR{38pt}{1}&\fwboxR{46pt}{0}&\fwboxR{52pt}{1}&\fwboxR{57pt}{1}&\fwboxR{66pt}{5}&\fwboxR{72pt}{16}&\fwboxR{84pt}{80}\\
\fwbox{10pt}{\rule{0pt}{12pt}\b{1}}&\fwboxR{38pt}{1}&\fwboxR{46pt}{1}&\fwboxR{52pt}{3}&\fwboxR{57pt}{10}&\fwboxR{66pt}{46}&\fwboxR{72pt}{256}&\fwboxR{84pt}{1\hspace{-0.5pt},\hspace{-2.5pt}721}\\
\fwbox{10pt}{\rule{0pt}{12pt}\b{2}}&\fwboxR{38pt}{5}&\fwboxR{46pt}{12}&\fwboxR{52pt}{47}&\fwboxR{57pt}{236}&\fwboxR{66pt}{1\hspace{-0.5pt},\hspace{-2.5pt}466}&\fwboxR{72pt}{10\hspace{-0.5pt},\hspace{-2.5pt}816}&\fwboxR{84pt}{92\hspace{-0.5pt},\hspace{-2.5pt}358}\\
\fwbox{10pt}{\rule{0pt}{12pt}\b{3}}&\fwboxR{38pt}{70}&\fwboxR{46pt}{280}&\fwboxR{52pt}{1\hspace{-0.5pt},\hspace{-2.5pt}505}&\fwboxR{57pt}{9\hspace{-0.5pt},\hspace{-2.5pt}920}&\fwboxR{66pt}{77\hspace{-0.5pt},\hspace{-2.5pt}163}&\fwboxR{72pt}{689\hspace{-0.5pt},\hspace{-2.5pt}300}&\fwboxR{84pt}{6\hspace{-0.5pt},\hspace{-2.5pt}930\hspace{-0.5pt},\hspace{-2.5pt}735}\\
\fwbox{10pt}{\rule{0pt}{12pt}\b{4}}&\fwboxR{38pt}{1\hspace{-0.5pt},\hspace{-2.5pt}820}&\fwboxR{46pt}{10\hspace{-0.5pt},\hspace{-2.5pt}192}&\fwboxR{52pt}{70\hspace{-0.5pt},\hspace{-2.5pt}301}&\fwboxR{57pt}{569\hspace{-0.5pt},\hspace{-2.5pt}681}&\fwboxR{66pt}{5\hspace{-0.5pt},\hspace{-2.5pt}276\hspace{-0.5pt},\hspace{-2.5pt}607}&\fwboxR{72pt}{54\hspace{-0.5pt},\hspace{-2.5pt}772\hspace{-0.5pt},\hspace{-2.5pt}326}&\fwboxR{84pt}{627\hspace{-0.5pt},\hspace{-2.5pt}593\hspace{-0.5pt},\hspace{-2.5pt}687}\\
\fwbox{10pt}{\rule{0pt}{12pt}\b{5}}&\fwboxR{38pt}{70\hspace{-0.5pt},\hspace{-2.5pt}875}&\fwboxR{46pt}{508\hspace{-0.5pt},\hspace{-2.5pt}305}&\fwboxR{52pt}{4\hspace{-0.5pt},\hspace{-2.5pt}273\hspace{-0.5pt},\hspace{-2.5pt}773}&\fwboxR{57pt}{40\hspace{-0.5pt},\hspace{-2.5pt}920\hspace{-0.5pt},\hspace{-2.5pt}243}&\fwboxR{66pt}{437\hspace{-0.5pt},\hspace{-2.5pt}523\hspace{-0.5pt},\hspace{-2.5pt}898}&\fwboxR{72pt}{5\hspace{-0.5pt},\hspace{-2.5pt}147\hspace{-0.5pt},\hspace{-2.5pt}588\hspace{-0.5pt},\hspace{-2.5pt}746}&\fwboxR{84pt}{65\hspace{-0.5pt},\hspace{-2.5pt}870\hspace{-0.5pt},\hspace{-2.5pt}361\hspace{-0.5pt},\hspace{-2.5pt}350}\\\hline
\end{array}}$$\renewcommand{\arraystretch}{1}\vspace{-24pt}\caption{$\mathcal{C}^{\r{n}\,\b{q}}_{\mathfrak{f}_4}(\mathbf{F})$: scattering $\b{q}\!\times\!\mathbf{F}$-lines and $\r{n}\times$adjoints in $\mathfrak{f}_4$ gauge theory.\label{f4nq_table}}\vspace{-10pt}\end{table}

\begin{table}[h!]\renewcommand{\arraystretch}{.125}\vspace{-10pt}$$\fwbox{0pt}{\begin{array}{|r@{$\,$}|@{$\,$}r|r|r|r|r|r|r|r|r@{$\,$}|}
\hline\multicolumn{1}{|c@{$\,$}|@{$\,$}}{\text{\backslashbox[14.05pt]{\raisebox{2pt}{$\fwboxL{0pt}{\hspace{-6pt}\b{q}}$\vspace{0pt}}}{\raisebox{-4.5pt}{\fwboxL{0pt}{$\r{\hspace{-1.5pt}$n$}$}}}}}
&\multicolumn{1}{c|}{\r{0}}&\multicolumn{1}{c|}{\r{1}}&\multicolumn{1}{c|}{\r{2}}&\multicolumn{1}{c|}{\r{3}}&\multicolumn{1}{c|}{\r{4}}&\multicolumn{1}{c|}{\r{5}}&\multicolumn{1}{c|}{\r{6}}\\\hline\hline
\fwbox{10pt}{\rule{0pt}{12pt}\b{0}}&\fwboxR{38pt}{1}&\fwboxR{46pt}{0}&\fwboxR{52pt}{1}&\fwboxR{57pt}{1}&\fwboxR{66pt}{5}&\fwboxR{72pt}{16}&\fwboxR{84pt}{80}\\
\fwbox{10pt}{\rule{0pt}{12pt}\b{1}}&\fwboxR{38pt}{1}&\fwboxR{46pt}{1}&\fwboxR{52pt}{3}&\fwboxR{57pt}{10}&\fwboxR{66pt}{45}&\fwboxR{72pt}{236}&\fwboxR{84pt}{1\hspace{-0.5pt},\hspace{-2.5pt}421}\\
\fwbox{10pt}{\rule{0pt}{12pt}\b{2}}&\fwboxR{38pt}{4}&\fwboxR{46pt}{9}&\fwboxR{52pt}{33}&\fwboxR{57pt}{151}&\fwboxR{66pt}{817}&\fwboxR{72pt}{4\hspace{-0.5pt},\hspace{-2.5pt}984}&\fwboxR{84pt}{33\hspace{-0.5pt},\hspace{-2.5pt}357}\\
\fwbox{10pt}{\rule{0pt}{12pt}\b{3}}&\fwboxR{38pt}{35}&\fwboxR{46pt}{120}&\fwboxR{52pt}{545}&\fwboxR{57pt}{2\hspace{-0.5pt},\hspace{-2.5pt}932}&\fwboxR{66pt}{17\hspace{-0.5pt},\hspace{-2.5pt}827}&\fwboxR{72pt}{118\hspace{-0.5pt},\hspace{-2.5pt}945}&\fwboxR{84pt}{854\hspace{-0.5pt},\hspace{-2.5pt}135}\\
\fwbox{10pt}{\rule{0pt}{12pt}\b{4}}&\fwboxR{38pt}{455}&\fwboxR{46pt}{2\hspace{-0.5pt},\hspace{-2.5pt}002}&\fwboxR{52pt}{10\hspace{-0.5pt},\hspace{-2.5pt}626}&\fwboxR{57pt}{64\hspace{-0.5pt},\hspace{-2.5pt}078}&\fwboxR{66pt}{425\hspace{-0.5pt},\hspace{-2.5pt}194}&\fwboxR{72pt}{3\hspace{-0.5pt},\hspace{-2.5pt}041\hspace{-0.5pt},\hspace{-2.5pt}241}&\fwboxR{84pt}{23\hspace{-0.5pt},\hspace{-2.5pt}115\hspace{-0.5pt},\hspace{-2.5pt}050}\\
\fwbox{10pt}{\rule{0pt}{12pt}\b{5}}&\fwboxR{38pt}{7\hspace{-0.5pt},\hspace{-2.5pt}413}&\fwboxR{46pt}{38\hspace{-0.5pt},\hspace{-2.5pt}640}&\fwboxR{52pt}{230\hspace{-0.5pt},\hspace{-2.5pt}720}&\fwboxR{57pt}{1\hspace{-0.5pt},\hspace{-2.5pt}521\hspace{-0.5pt},\hspace{-2.5pt}300}&\fwboxR{66pt}{10\hspace{-0.5pt},\hspace{-2.5pt}833\hspace{-0.5pt},\hspace{-2.5pt}879}&\fwboxR{72pt}{82\hspace{-0.5pt},\hspace{-2.5pt}083\hspace{-0.5pt},\hspace{-2.5pt}517}&\fwboxR{84pt}{654\hspace{-0.5pt},\hspace{-2.5pt}527\hspace{-0.5pt},\hspace{-2.5pt}565}\\
\fwbox{10pt}{\rule{0pt}{12pt}\b{6}}&\fwboxR{38pt}{140\hspace{-0.5pt},\hspace{-2.5pt}833}&\fwboxR{46pt}{831\hspace{-0.5pt},\hspace{-2.5pt}600}&\fwboxR{52pt}{5\hspace{-0.5pt},\hspace{-2.5pt}446\hspace{-0.5pt},\hspace{-2.5pt}078}&\fwboxR{57pt}{38\hspace{-0.5pt},\hspace{-2.5pt}606\hspace{-0.5pt},\hspace{-2.5pt}096}&\fwboxR{66pt}{291\hspace{-0.5pt},\hspace{-2.5pt}542\hspace{-0.5pt},\hspace{-2.5pt}694}&\fwboxR{72pt}{2\hspace{-0.5pt},\hspace{-2.5pt}319\hspace{-0.5pt},\hspace{-2.5pt}076\hspace{-0.5pt},\hspace{-2.5pt}187}&\fwboxR{84pt}{19\hspace{-0.5pt},\hspace{-2.5pt}268\hspace{-0.5pt},\hspace{-2.5pt}675\hspace{-0.5pt},\hspace{-2.5pt}895}\\\hline
\end{array}}$$\renewcommand{\arraystretch}{1}\vspace{-24pt}\caption{$\mathcal{C}^{\r{n}\,\b{q}}_{\mathfrak{g}_2}(\mathbf{F})$: scattering $\b{q}\!\times\!\mathbf{F}$-lines and $\r{n}\times$adjoints in $\mathfrak{g}_2$ gauge theory.\label{g2nq_table}}\vspace{-40pt}\end{table}

\pagebreak

\subsection{The Breakdown of the Large-$N_c$ Limit for Some Charges of Matter}\label{subsec:dtensors}
From the discussion thus far one may conclude that saturation is an inevitable feature for the large-rank limit of the classical algebras. This is however \emph{not} the case. In the $\mathfrak{a}_k$ case the saturation of gluons and fundamental matter follows simply from the recursive construction of the difference tables and the fact that the pure-adjoint case (empirically) saturates, but already this is somewhat mysterious and a perhaps spurious explanation, as for the other classical algebras saturation \emph{does} occur despite the \emph{lack} of a recursive construction. 

A very simple prerequisite for `saturation' of colour tensors to make sense in a theory with charged matter would be that
\eq{\lim_{\r{k}\to\infty}m_{\mathfrak{g}_\b{\r{k}}}\big((\mathbf{R}\!\otimes\!\bar{\mathbf{R}})^{\b{q}}\!\!\to\!\mathbf{1}\big)<\infty}
for all $\b{q}$. If this limit did not exist, then taking the rank to be arbitrarily-large would require an unbounded number of \emph{new} independent colour-tensors for amplitudes involving $\b{q}$ matter lines charged in the representation $\mathbf{R}$. 

From this point of view, it should be surprising that saturation exists for the fundamental representations of the classical Lie algebras. But as we have seen, these limits do empirically exist. 

We suspect that a meaningful large-rank limit exists for all classical Lie algebras with matter charged under representations constructed from tensor-products of fundamental representations. But what about those representations \emph{not} obtainable through tensor products of (what we have defined to be) the fundamental representations? There is only one kind of exception: the spinor representations of $\mathfrak{so}(n)$. It turns out that these \emph{do not} saturate: we will see that
\eq{\begin{split}
\fwboxL{40pt}{\mathcal{C}^{\r{0},\b{2}}_{\mathfrak{b}_k}(\mathbf{S})}&=k{+}1\,\\
\fwboxL{40pt}{\mathcal{C}^{\r{0},\b{2}}_{\mathfrak{d}_k}(\mathbf{S}_{\pm})}&=\lceil(k{+}1)/2\rceil\,;\end{split}}
as such, no `large-${N_c}$' limit should exist for $\mathfrak{so}(k)$ gauge theories for scattering amplitudes involving matter charged under spin-representations.

\subsubsection*{Amplitudes Involving Spinor Representations of $\mathfrak{b}_k$ or $\mathfrak{d}_k$}

In $\mathfrak{d}_k$ there are two $\mathbf{2^{k-1}}$-dimensional representations which are the two spinor representations $\mathbf{S}_\pm$. We can consider spinors of similar or different chirality coupled to adjoints. Specifically, we will consider $\mathcal{C}_{\mathfrak{d}_k}[\mathbf{ad}^\r{n},(\mathbf{S}_{\pm}\bar{\mathbf{S}_{\pm}})^\b{q} ]$. There is, however, a distinction for even and odd ranks we must address first. For $\b{q}{=}1$ and $k$ even there is one $\mathbf{1}$ in $\mathbf{S}_\pm^2$ and none in $\mathbf{S}_{+}\otimes\mathbf{S}_{-}$, while for $k$ odd there is one $\mathbf{1}$ in $\mathbf{S}_{+}\otimes\mathbf{S}_{-}$ and none in $\mathbf{S}_\pm^2$. This is reflecting the fact that $\bar{\mathbf{S}_\pm} \simeq \mathbf{S}_\pm$ for $k$ even and $\bar{\mathbf{S}_\pm} \simeq \mathbf{S}_\mp$ for $k$ odd: spinor representations are real for $k$ even and complex for $k$ odd. For two spinors and $\r{n}$ gluons one saturates to \href{https://oeis.org/A000985}{[A000985]}, \emph{regardless} of rank. 

Next, consider $\b{q}{=}2$. Empirically, we find $\mathcal{C}_{\mathfrak{d}_k}[(\mathbf{S}_{\pm}\bar{\mathbf{S}_\pm})^2]=\lceil(k{+}1)/2\rceil$, which evidently grows \emph{linearly} with rank and will never saturate. We can understand this behaviour as follows (see \emph{e.g.}~\cite{Polchinski_1998} for the required background).  We denote $\psi_\alpha \in \mathbf{S}_+$, $\psi_{\alpha'} \in \mathbf{S}_-$, $\psi^\alpha \in \bar{\mathbf{S}_+}$ and $\psi^{\alpha'} \in \bar{\mathbf{S}_-}$.  For $k$ even $\bar{\mathbf{S}_\pm} \simeq \mathbf{S}_\pm$ and therefore $C_{\alpha \beta} \colon \bar{\mathbf{S}_+} \to \mathbf{S}_+$, which is sometimes called a \emph{charge conjugation matrix}.  We also have $C_{\alpha' \beta'} \colon \bar{\mathbf{S}_-} \to \mathbf{S}_-$.  The Pauli matrices act as $(\sigma^\mu)_\alpha{}^{\alpha'} \colon \mathbf{S}_- \to \mathbf{S}_+$ and $(\bar{\sigma}^\mu)_{\alpha'}{}^\alpha \colon \mathbf{S}_+ \to \mathbf{S}_-$.  A tensor product in $\mathbf{S}_+ \otimes \mathbf{S}_+$ has the tensor structure $M_{\alpha \beta}$ and can be decomposed in a basis consisting of
\eq{C_{\alpha \beta}, (\sigma^{[\mu_1} \bar{\sigma}^{\mu_2]} C)_{\alpha \beta}, \cdots,
  (\sigma^{[\mu_1} \cdots \bar{\sigma}^{\mu_k]} C)_{\alpha \beta}.}
These representations have dimensions $1, \binom{2 k}{2}, \dotsc, \frac 1 2 \binom{2 k}{k}$.  The last representation above, the rank $k$ form, has dimension $\frac 1 2 \binom{2 k}{k}$ since a rank $k$ form in a $2 k$-dimensional space is reducible and can be decomposed in a self-dual and anti-self-dual irreducibles.

Since $\operatorname{dim} (\mathbf{S}_{\pm}) = 2^{k - 1}$, we have
$$(2^{k - 1})^2 = 1 + \binom{2 k}{2} + \cdots + \binom{2 k}{2 k - 2} + \frac 1 2 \binom{2 k}{k}.$$
Indeed, this identity follows from $(1{+}1)^{2 k}{=}\sum_{n = 0}^{2 k} \binom{2 k}{n}$, $(1 - 1)^{2 k} = \sum_{n = 0}^{2 k} ({-}1)^n \binom{2 k}{n}$ and $\binom{2 k}{n} = \binom{2 k}{2 k{-}n}$.  Note that there are $\frac k 2{+}1$ terms in the decomposition of $\mathbf{S}_+ \otimes \mathbf{S}_+$. We can obtain all the invariants (number of independent tensors) of $\mathbf{S}_+^4$ by taking the decomposition of $\mathbf{S}_+^2$ and contracting it with the decomposition of $\mathbf{S}_+^2$ (there is only one way to do so).  There are $\frac k 2{+}1$ terms in the decomposition which yield the same number of tensors.\\

If $k$ is odd then we have $C_{\alpha \alpha'} \colon \bar{\mathbf{S}_-} \to \mathbf{S}_+$ and $C_{\alpha' \alpha} \colon \bar{\mathbf{S}_+} \to \mathbf{S}_-$.  Then, a tensor product $\mathbf{S}_+ \otimes \mathbf{S}_+$ with index structure $M_{\alpha \beta}$ can be decomposed in a basis
\eq{  (\sigma^{\mu_1} C)_{\alpha \beta},
  (\sigma^{[\mu_1} \bar{\sigma}^{\mu_2} \sigma^{\mu_3]} C)_{\alpha \beta}, \dotsc,
  (\sigma^{[\mu_1} \bar{\sigma}^{\mu_2} \cdots \sigma^{\mu_k]} C)_{\alpha \beta}.}
There are $\frac {k + 1} 2$ terms in the decomposition.  A similar counting holds for $\mathbf{S}_- \otimes \mathbf{S}_-$ with the tensor structure $M'_{\alpha' \beta'}$.  It follows analogously that there are $\frac {k + 1} 2$ invariants in $\mathbf{S}_+^2 \mathbf{S}_-^2$. In this case we can also do the calculation by decomposing $(\mathbf{S}_+ \otimes \mathbf{S}_-)^2$ by using the associativity of the tensor product.  The tensor product $\mathbf{S}_+ \otimes \mathbf{S}_-$ can be decomposed in a basis of
\eq{C_{\alpha \alpha'},
  (\sigma^{[\mu_1} \bar{\sigma}^{\mu_2]} C)_{\alpha \alpha'}, \dotsc,
  (\sigma^{[\mu_1} \cdots \bar{\sigma}^{\mu_{k - 1}]} C)_{\alpha \alpha'}.}
We again have $\frac {k{+}1} 2$ independent tensors.

The above analysis was for $\mathfrak{d}_k$ spinor representations, but a similar story holds for the $\mathbf{2^{k}}$ (real) spinor representation of $\mathfrak{b}_k$. In this case we find $\mathcal{C}_{\mathfrak{b}_k}[\mathbf{S}^\b{2}]=k{+}1$, which again grows linearly with rank.

~\pagebreak

\vspace{0pt}%
\section{Open Problems for Future Work}\label{sec:open_problems}\vspace{0pt}
In this work we have shown how representation theory algorithmically answers the question of how many colour-tensors appear in generic $\mathfrak{g}$-type gauge theory to all orders in perturbation theory. There are a number of refinements to the question which are deserving of follow-up.

\vspace{0pt}%
\subsection{Colour-Tensors Required in Perturbation Theory}\label{subsec:perturbation_theory}\vspace{0pt}

We've already seen that for pure $\mathfrak{su}_N$ gauge theory (and any other non-anomalous $\mathfrak{a}$-type gauge theory), this paper's methodology over counts the number of independent colour-tensors. The kinematic coefficient to the $d^{abc}$ tensor in pure Yang-Mills theory is zero to all loop-orders, and so despite there being two independent colour-tensors for $\mathfrak{a}_k$ $\r{n}{=}3$ only one appears in the theory. This mirrors the discussion of the deranged traces vs.\ dihedrally-deranged traces at 4-points in section~\ref{subsec:large_rank_limit}. One may wonder how many colour-tensors \emph{actually appear} in a particular theory with specific matter content, where the counting herein should serve as an upper bound. 

In the same vein, it would be extremely interesting to examine the \emph{loop-dependence} on the number of colour-factors in a theory. At tree-level the DDM-basis is widely used and of size $(n{-}2)!$ but because the asymptotics of our counting are \emph{exponential} and not factorial, at some point the number of colour-tensors \emph{to all-orders} will be fewer than the number in the DDM basis at \emph{tree-level}! We expect that low loop-orders will carve out subsets of the number of possible tensors and at a certain loop-order one will have generated all of the allowed colour-tensors, matching the counting in this work. We have already confirmed that this is the case in $\mathfrak{su}_2$-type gauge theories, but more systematic study is needed. Studying that loop-dependence for different algebras, theories, and particle content is extremely interesting and will be a major part of our follow-up work.

\vspace{0pt}%
\subsection{Construction of Explicit Bases of Colour-Tensors}\label{subsec:building_bases_of_colour_tensors}\vspace{0pt}

Bases of colour-tensors (perhaps over-complete) may be constructed in any number of ways: from traces of some single representation's generators, from traces of multiple representations' generators, from structure constants, etc. Each approach has its own advantages and disadvantages, but some of the weaknesses of the existing approaches motivate us to seek an alternative. First, the over-completeness of the bases should be remedied. The fact that the true size of the basis is unknown seems a remarkable gap in our understanding. Second, an orthogonalization of the tensors would prove immensely (and practically) useful. This would allow one to `project out' gauge-invariant pieces of the amplitude trivially, amongst other desirable features. From a computational complexity perspective it reduces the calculation of colour-summed cross-sections substantially since there is no `colour-screening', which for large multiplicity becomes significant; one need only consider the diagonal terms. Many of our methods are amenable to very high-multiplicity calculations, and it would be interesting to pursue this. 

So far we have simply stated the benefits of an orthogonal and minimal basis, but a more representation theoretic critique is that it is not always obvious what objects should be used to construct the tensors. This is especially true when considering exotic matter. In the case of a trace basis, there is the perennial question of which representation's generators we are using. And when is it possible to write all tensors in terms of traces of a \emph{single} representation? These are not straightforward questions to answer and we wish to sidestep them entirely by having some more canonical way of generating tensors, something more inherent to the algebras and representations involved in a process.

A new basis is only as good as it is usable, so there will be a few challenges to address: we will have to detail how to construct the tensors in a way that is adaptable to any algebra and representation, we will have to describe how to project between the new and existing bases, and we will have to describe any interesting structure or relations satisfied by the tensors so-constructed (of which there will be an abundance!). That story goes well beyond the bounds of this present work, and so we delay further discussion of these interesting topics to~\cite{paper2}.

\vspace{\fill}\vspace{-4pt}
\section*{Acknowledgments}%
\vspace{-4pt}
\noindent The authors gratefully acknowledge fruitful conversations with JJ Carrasco and Mark Spradlin. This work was supported in part by a grant from the US Department of Energy \mbox{(No.\ DE-SC00019066)}.

\appendix

\section{Multiplicities of Tensor Products' Decompositions}\label{computing_multiplicities_appendix}
Consider any simple Lie algebra $\mathfrak{g}$ of rank $k$ with irreducible representations labelled by Dynkin indices $\vec{w}\!\in\!\mathbb{Z}_{\geq0}^{k}$. For any three irreducible representations labelled by $\{\vec{w}_a,\vec{w}_b,\vec{w}_c\}$, let $m^{a\,b}_{\phantom{a\,b}\,c}$ denote the \emph{multiplicity} of the irreducible representation $\mathbf{c}$ in the decomposition of the tensor product $\mathbf{a}\!\otimes\!\mathbf{b}$. That is, 
\eq{\mathbf{a}\!\otimes\!\mathbf{b}\equivL\mathbf{c}^{\oplus{m}^{a\,b}_{\phantom{a\,b}\,c}}\!\oplus\ldots\,\equivL \big({m}^{a\,b}_{\phantom{a\,b}\,c}\big)\mathbf{c}\oplus\ldots\,.}
(Note that we have allowed ourselves the (very slight) abuse of notation: $r\,\mathbf{c}\equivR\mathbf{c}^{\oplus r}$.) In terms of this, the complete tensor product decomposition of $\mathbf{a}\!\otimes\!\mathbf{b}$ could be written
\eq{\mathbf{a}\!\otimes\!\mathbf{b}=\bigoplus_{c}m^{a\,b}_{\phantom{a\,b}\,c}\,\mathbf{c}\,.}
It is worth noting that this sum involves at most a finite number of terms.

Associativity of the tensor product allows us to write 
\eq{\mathbf{a}\!\otimes\!\mathbf{b}\!\otimes\!\mathbf{c}=\big(\mathbf{a}\!\otimes\!\mathbf{b}\big)\!\otimes\!\mathbf{c}=\bigoplus_{\mathbf{d}}\Big[\sum_{\ast}m^{a\,b}_{\phantom{a\,b}\,\ast}m^{\ast\,c}_{\phantom{\ast\,c}\,d}\Big]\mathbf{d}\equivL\bigoplus_{\mathbf{d}}m^{a\,b\,c}_{\phantom{a\,b\,c}\,d}\mathbf{d}\label{abc_tensor_product}}
with the obvious generalization to an arbitrary number of indices. 

The numbers $m^{a\cdots b}_{\phantom{a\cdots b}\,c}$ is fully symmetric (by the definition of the tensor product) in $\{a,\ldots,b\}$ but not generally symmetric under permutations involving the final index $c$. Let us define fully symmetric tensors 
\eq{m^{a\cdots b}\equivR m^{a\cdots b}_{\phantom{a\cdots b}\,1}\,,}
where we have used $\vec{w}_1\!=\![0,\ldots,0]$ to denote the trivial representation. The two-index tensor is especially interesting:
\eq{m^{a\,b}\equivR m^{a\,b}_{\phantom{a\,b}\,1}=\left\{\begin{array}{l@{$\;\;\;$}l}1&\mathbf{a}\!=\!\bar{\mathbf{b}}\\0&\text{else}.\end{array}\right.}
as it defines a non-degenerate bilinear form (a `\emph{metric}') on the space of irreducible representations, and can be used (in the physicists' vernacular) to `raise/lower' indices. Indeed, using (\ref{abc_tensor_product}) we see that 
\eq{m^{a\,b\,c}=m^{a\,b}_{\phantom{a\,b}\,\ast}m^{\ast\,c}\equivR \sum_{\ast}m^{a\,b}_{\phantom{a\,b}\,\ast}m^{\ast\,c}\,.}
where we have now made use of the summation convention: repeated indices should always be summed (unless stated otherwise).

In this work, we have primarily been interested in the multiplicity of $\mathbf{1}$ in the tensor products of representations relevant to coloured scattering amplitudes---namely, $\mathbf{ad}(\mathfrak{g})^{\otimes\r{n}}\bigotimes_{i}(\mathbf{R}_i\!\otimes\!\bar{\mathbf{R}_i})^{\otimes\b{m_i}}$ which would encode the colour-dependence of amplitudes involving $\r{n}$ gluons and $\b{m_i}$ lines of Fermions transforming in the representation $\mathbf{R}_i$. The analysis above naturally suggests that the ways in which these numbers can be recursively computed in terms of just the fundamental three index objects $m^{a\,b}_{\phantom{a\,b}\,c}$---and in many different ways. However one chooses to recursively compute them, we end up with terms being counted by products of $m$'s, and summing over all the possible irreducible representations that connect them. 

Such a sum is extremely reminiscent of the colour tensors generated by products of structure constants arising from the scattering of gluons. However, in our counting game, we always end with a trivalent \emph{tree} graph---with the edges labeling irreducible representations of the algebra.

Associating these with tensors constructed from the actual representations involved turns out to immediately translate this counting exercise into an \emph{explicit construction of independent tensors}. These tensors turn out to have many remarkable properties (such as orthogonality), as we describe in the forthcoming work \cite{paper2}.

\newpage
\section[Decomposition of Tensor Products into Irreducible Representations]{Decomposing Tensor Products into Irreducible Reps}%
\label{racahspeiser-Appendix}

To decompose the tensor products into irreducible representations we can use the Racah-Speiser algorithm\footnote{Sometimes also attributed to Brauer, Klimyk and Steinberg (see ref.~\cite{MR126508}).} (see ref.~\cite{MR170975, MR171552}), which works as follows.  Consider two representations $R_{\Lambda_1}$ and $R_{\Lambda_2}$ with highest weights $\Lambda_1$, $\Lambda_2$.  Then we have
\eq{R_{\Lambda_1} \otimes R_{\Lambda_2} = \bigoplus_{\lambda\in\operatorname{wt} R_{\Lambda_2}} R_{\Lambda_1 + \lambda}}
and the sum is over the weights of the representation $R_{\Lambda_2}$ including multiplicities; the same construction can be done by swapping $R_{\Lambda_1}$ and $R_{\Lambda_2}$.

Sometimes when adding $\Lambda_1 + \lambda$ we do not obtain a valid highest weight since some of the entries of the weight vector become negative.  In that case the algorithm instructs us to reflect by an element of the Weyl group such that the entries are positive.  For such representations we have
\eq{R_{\Lambda} = \operatorname{sign}(w) R_{w \cdot \Lambda}, }
for $w$ in the Weyl group of the Lie algebra.  The dot action is defined by
\eq{w \cdot \Lambda = w (\Lambda{+}\rho){-}\rho,}
where $\rho$ is the Weyl vector
\eq{\rho = \frac 1 2 \sum_{\alpha > 0} \alpha,}
which is the sum of positive roots.

We illustrate the procedure on a few examples.  $\mathfrak{su}_3$ has Cartan matrix
\eq{C\equivR\, \begin{pmatrix}
2 & -1 \\
-1 & 2
\end{pmatrix}}
and a reflection $w_i$ acts via
\eq{w_i[\dotsc, \lambda_j, \dotsc] =  [\dotsc, \lambda_j{-}\lambda_i C_{i j}, \dotsc].}
For the adjoint we have
\eq{\operatorname{wt}(\mathbf{ad}) = \{[1, 1], [-1, 2], [2, -1], [0, 0]^2, [1, -2], [-2, 1], [-1, -1]\},}
where the weight $[0, 0]$ occurs with multiplicity 2.

So the calculation we need to do is to start with some dots in the positive Weyl chamber (which are the vectors with positive Dynkin labels) and start with initial condition that the weight at $[1, 1]$ has multiplicity one and the rest zero.  Then add all the weights of the adjoint to it and reflect (or fold over) in the hyperplanes orthogonal to $\alpha_1$ and $\alpha_2$ and passing through $-\rho$ where $\rho$ is the Weyl vector.  If we do an odd number of reflections, subtract the multiplicities otherwise add.  Repeat as many times as necessary.

We have $\mathbf{ad} = R_{[1, 1]}$.  Let us apply the Racah-Speiser algorithm to decomposing $\mathbf{ad} \otimes \mathbf{ad}$.  We need to compute $R_{[1, 1]} \otimes R_{[1, 1]}$.  Adding all the weights of $R_{[1,1]}$ we find
\eq{R_{[1, 1]} \otimes R_{[1, 1]} = R_{[2, 2]} \oplus R_{[0, 3]} \oplus R_{[3, 0]} \oplus R_{[1, 1]}^{\oplus 2} \oplus R_{[-1, 2]} \oplus R_{[2, -1]} \oplus R_{[0, 0]}.}
The weights $[-1, 2]$ and $[2, -1]$ are outside the positive Weyl chamber so we need to reflect them.  Since $s_1 \cdot [-1, 2] = [-1, 2]$ so we have $R_{[-1, 2]} = -R_{[-1, 2]}$ which implies that $R_{[-1, 2]} = 0$.  Similarly, $R_{[2, -1]} = 0$.  In the end, we have
\eq{R_{[1, 1]} \otimes R_{[1, 1]} = R_{[2, 2]} \oplus R_{[0, 3]} \oplus R_{[3, 0]} \oplus R_{[1, 1]}^{\oplus 2} \oplus R_{[0, 0]}.
}

The root system of $\mathfrak{g}_2$ can be conveniently described inside $\mathbb{R}^3$.  We have the following roots $\pm (e_i - e_j)$ for $i < j$ (short roots) and $\pm (2 e_i - e_j - e_k)$ for $i, j, k$ all different (long roots).

The simple roots are $\alpha_1 = e_1 - e_2$ and $\alpha_2 = -2 e_1 + e_2 + e_3$.  We have $\alpha_1^2 = 2$, $\alpha_2^2 = 6$ and $\alpha_1 \cdot \alpha_2 = -3$.  The Cartan matrix is
\eq{C_{i j} = \alpha_i \cdot \alpha_j^\vee =
\begin{pmatrix}
2 & -1\\ -3 & 2
\end{pmatrix}.}

The six positive roots can be written in terms of the simple roots as follows
\begin{gather}
e_1 - e_2 = \alpha_1, \\
e_3 - e_1 = \alpha_1 + \alpha_2, \\
e_3 - e_2 = 2 \alpha_1 + \alpha_2, \\
2 e_1 - e_2 - e_3 = \alpha_2, \\
-(2 e_2 - e_1 - e_3) = 3 \alpha_1 + \alpha_2, \\
2 e_3 - e_1 - e_2 = 3 \alpha_1 + 2 \alpha_2.
\end{gather}
In terms of fundamental weights we have
\eq{\alpha_1 = 2 \omega_1{-}\omega_2, \\
\alpha_2 = {-}3 \omega_1{+}2 \omega_2.}
The highest root is
\eq{\theta = 3 \alpha_1{+}2 \alpha_2 = \omega_2.}
In this basis the positive roots above read
\begin{gather}
e_1 - e_2 = \alpha_1 = 2 \omega_1 - \omega_2, \\
e_3 - e_1 = \alpha_1 + \alpha_2 = -\omega_1 + \omega_2, \\
e_3 - e_2 = 2 \alpha_1 + \alpha_2 = \omega_1, \\
2 e_1 - e_2 - e_3 = \alpha_2 = -3 \omega_1 + 2 \omega_2, \\
-(2 e_2 - e_1 - e_3) = 3 \alpha_1 + \alpha_2 = 3 \omega_1 - \omega_2, \\
2 e_3 - e_1 - e_2 = 3 \alpha_1 + 2 \alpha_2 = \omega_2.
\end{gather}
The weights in the adjoint representation are therefore $[2, -1]$, $[-1, 1]$, $[1, 0]$, $[-3, 2]$, $[3, -1]$, $[0, 1]$, $[0, 0]$ twice and then the negatives of the above.

Then, the Racah-Speiser theorem implies that
\begin{multline}
[0, 1] \otimes [0, 1] = [2, 0] \oplus [-1, 2] \oplus [1, 1] \oplus [-3, 3] \oplus
[3, 0] \oplus [0, 2] \oplus [0, 1]^{\oplus 2} \oplus [-2, 2] \oplus \\
[1, 0] \oplus [-1, 1] \oplus [3, -1] \oplus [-3, 2] \oplus [0, 0].
\end{multline}
We can immediately drop the terms $ [-1, 2]$, $ [-1, 1]$ and $[3, -1]$.  Next, we need to reflect $[-3, 3]$, $[-2, 2]$ and $[-3, 2]$. We have
\begin{gather}
s_1 \cdot [-3, 3] = s_1([-2, 4]) - [1, 1] = [2, 2] - [1, 1] = [1, 1], \\
s_1 \cdot [-2, 2] = s_1([-1, 3]) - [1, 1] = [1, 2] - [1, 1] = [0, 1], \\
s_1 \cdot [-3, 2] = s_1([-2, 3]) - [1, 1] = [2, 1] - [1, 1] = [1, 0].
\end{gather}
Here we have used the fact that
\begin{gather}
s_1([\lambda_1, \lambda_2]) = [-\lambda_1, \lambda_1 + \lambda_2], \\
s_2([\lambda_1, \lambda_2]) = [\lambda_1 + 3 \lambda_2, -\lambda_2].
\end{gather}

In then end we have
\eq{[0, 1] \otimes [0, 1] = [2, 0]  \oplus [1, 1] \ominus [1, 1] \oplus
[3, 0] \oplus [0, 2] \oplus [0, 1]^{\oplus 2} \ominus [0, 1] \oplus
[1, 0] \ominus [1, 0] \oplus [0, 0]}
which simplifies to
\eq{[0, 1] \otimes [0, 1] = [2, 0]  \oplus [3, 0] \oplus [0, 2] \oplus [0, 1] \oplus [0, 0].}

\section{Numbers of Colour-Factors for Low Multiplicity/Ranks}\label{appendix_enumerating_ranks}

In the tables below we list the number of colour-structures for pure glue, for a given algebra and a given number of external lines (see Tables~\ref{pure_glue_counting_table_a}, \ref{pure_glue_counting_table_b}, \ref{pure_glue_counting_table_c}, \ref{pure_glue_counting_table_d}).  We list several numbers for each such combination.  The number of colour-structures is in black with the digits which coincide with the saturation value at high rank in red.  In parentheses and in light blue we list the difference between the saturation value (at high rank) and the actual number of colour-structures.  When saturation is reached we colour the number in grey and the difference is zero (denoted by ${\color{gray} (0)}$). For the $\mathfrak{d}$-type algebras (see Table~\ref{pure_glue_counting_table_d}), where these differences are negative for low ranks, we denote the excess using bared numbers. 

\begin{landscape}\begin{table}\caption{Number of independent colour-tensors for scattering $\r{n}$ adjoints for $\mathfrak{a}_k$ gauge theory.\\[-10pt]\label{pure_glue_counting_table_a}}\vspace{-10pt}$$\fwbox{0pt}{\fig{-20pt}{1}{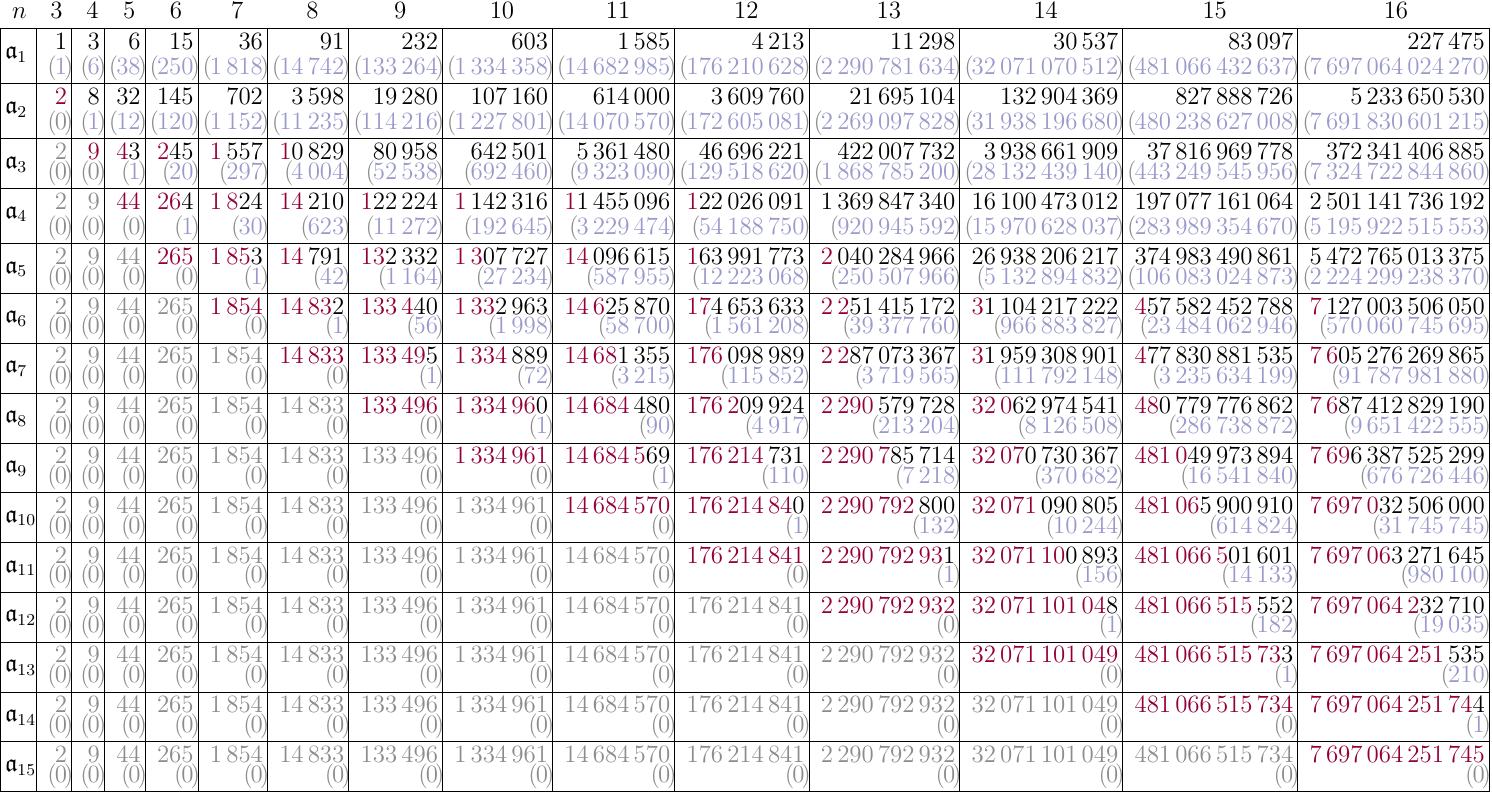}}$$\end{table}\end{landscape}

\begin{landscape}\begin{table}\vspace{-18pt}\caption{Number of independent colour-tensors for scattering $\r{n}$ adjoints for $\mathfrak{b}_k$ gauge theory.\\[-10pt]\label{pure_glue_counting_table_b}}\vspace{-10pt}$$\fwbox{0pt}{\fig{-20pt}{1}{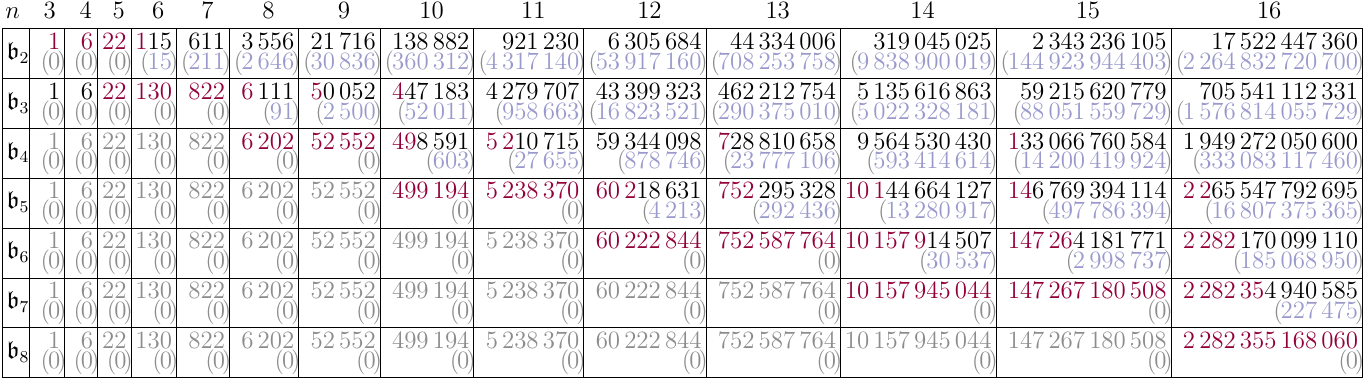}}$$\end{table}
\begin{table}\vspace{-10pt}\caption{Number of independent colour-tensors for scattering $\r{n}$ adjoints for $\mathfrak{c}_k$ gauge theory.\\[-10pt]\label{pure_glue_counting_table_c}}\vspace{-10pt}$$\fwbox{0pt}{\fig{00pt}{1}{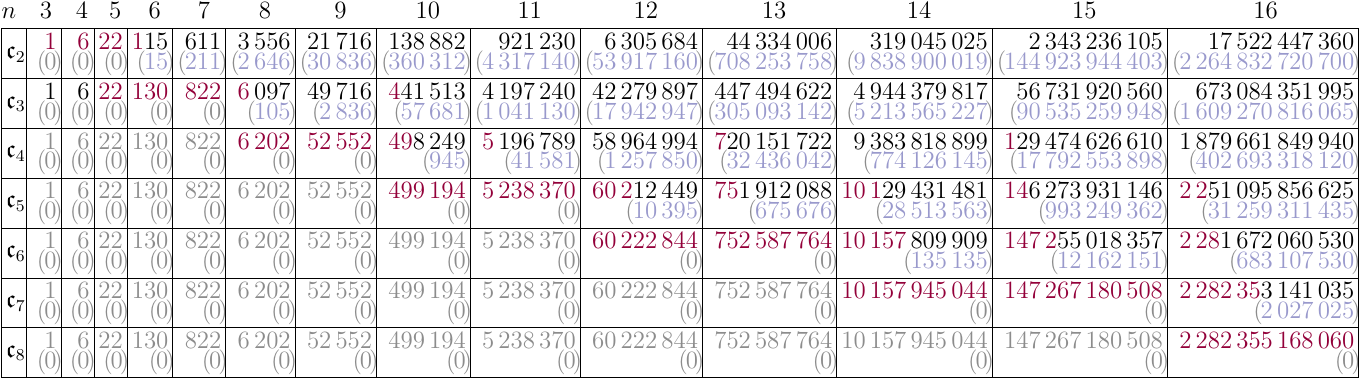}}$$\end{table}\end{landscape}

\begin{landscape}\begin{table}\vspace{-20pt}\caption{Number of independent colour-tensors for scattering $\r{n}$ adjoints for $\mathfrak{d}_k$ gauge theory.\\[-10pt]\label{pure_glue_counting_table_d}}\vspace{-10pt}$$\fwbox{0pt}{\hspace{-0pt}\fig{0pt}{1}{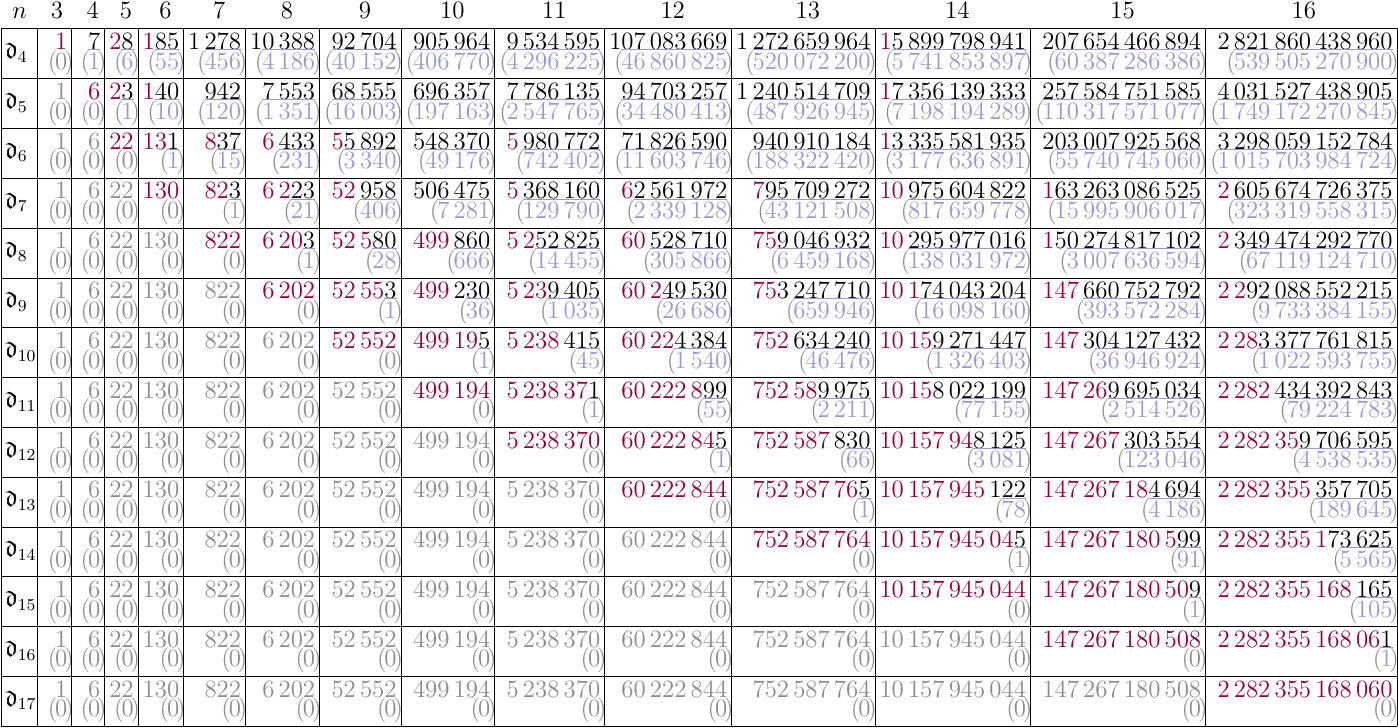}}$$\end{table}\end{landscape}

\begin{landscape}\begin{table}\caption{Number of independent colour-tensors for scattering $\r{n}$ adjoints for $\mathfrak{e}_k$, $\mathfrak{f}_4$ or $\mathfrak{g}_{2}$ gauge theory.\\[-10pt]\label{pure_glue_counting_table_efg}}\vspace{-10pt}$$\fwbox{0pt}{\hspace{-0pt}\fig{0pt}{1}{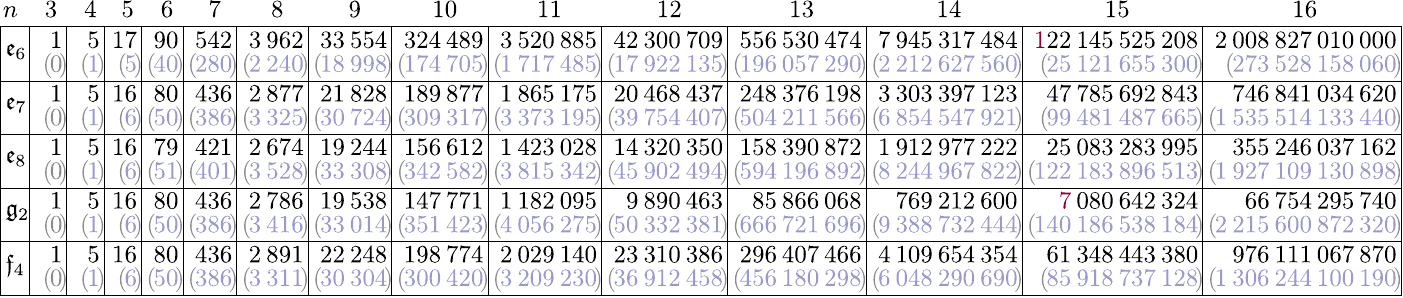}}$$\end{table}\end{landscape}

\newpage
\section[Asymptotic Growth of Colour-Factors for Large Multiplicity]{Asymptotic Growth of Colour-Factors for Large $\r{n}$}\label{asymptotic_appendix}

In ref.~\cite{MR1218274}, Biane has computed the asymptotic expansion in $n$ of the multiplicity of $R_\mu$ in the decomposition of $R_\lambda^{\otimes\r{n}}$.  In ref.~\cite{TATE2004402}, theorem 8, a more detailed explanation of the terms in Biane's formula is given.  Ref.~\cite{TATE2004402} uses the notation $A$ for the Biane's quadratic form $q$.  We adopt their notation below.  We have
\eq{A_\lambda = \frac 1 {\operatorname{dim} R_\lambda} \sum_{\mu\in\operatorname{wt}(R_\lambda)} m_{\mu} \mu \otimes \mu,}
where $R_\lambda$ is the irreducible representation with highest weight $\lambda$, $\operatorname{wt}(R_\lambda)$ is the set of weights of the representation $R_\lambda$, $m_\mu$ is the multiplicity of the weight $\mu$.  We denote by $P$ the weight lattice.  Since $\mu\!\in\!P\!\subset\!\mathfrak{h}^*$, we can think of it as a map $\mathfrak{h}^*\!\otimes\!\mathfrak{h}^*\!\to\!\mathbb{C}$ or, in matrix form, as a linear map $\mathfrak{h}^*\!\to\!\mathfrak{h}$.  We can express the weights in a basis of fundamental weights $\omega$ with integer coefficients:
\eq{A_\lambda = \sum_{i, j} (A_\lambda)_{i j} \omega_i \otimes \omega_j}
where
\eq{(A_\lambda)_{i j} = \frac 1 {\operatorname{dim} R_\lambda} \sum_{\mu \in \operatorname{wt}(R_\lambda)} m_\mu \mu(\alpha_i^\vee) \mu(\alpha_j^\vee).}
For the inverse we have
\eq{A^{-1}_\lambda = \sum_{k, l} (A^{-1}_\lambda)_{k l} \alpha_k^\vee \otimes \alpha_l^\vee.}

Then, we have
\begin{multline}
	\label{eq:asymptotic-mult2}
	\text{mult}(R_\mu, R_\lambda^{\otimes\r{n}}) = \frac {\lvert \Pi(G)\rvert (\operatorname{dim} R_\lambda)^n (\operatorname{dim} R_\mu) \prod_{\alpha > 0} \langle A_\lambda^{-1} \alpha, \rho\rangle}{\sqrt{\det A_\lambda} (2 \pi)^{\text{rank} G/2} \r{n}^{\operatorname{dim} G / 2}}\\
	\Bigl(\exp(-\langle A_\lambda^{-1} (\mu + \rho), \mu + \rho\rangle/(2\r{n})) + \mathcal{O}(\r{n}^{-1})\Bigr),
\end{multline}
where $\Pi(G)$ is the determinant of the Cartan matrix of the Lie algebra of $G$.

For $\lambda{=}\theta$, when we are dealing with the adjoint representation, we have
\eq{ (A_\theta)_{i j} = \frac 1 {\operatorname{dim} \mathfrak{g}} \sum_{\alpha \in \Delta} \alpha(\alpha_i^\vee) \alpha(\alpha_j^\vee).}

We know that (see ref.~\cite[eqs.~13.174, 13.184]{MR1424041})
\eq{(\lambda, \mu) = \frac 1 {2 g^\vee} \sum_{\alpha \in \Delta} (\lambda, \alpha) (\alpha, \mu),}
where $g^\vee$ is the dual Coxeter number.\footnote{The dual Coxeter number is defined as follows.  If $\theta$ is the highest root, then decompose it as $\theta = \sum_i a_i^\vee \alpha_i^\vee$ and $g^\vee := 1 + \sum_i a_i^\vee$.  The Coxeter number is $g: = 1 + \sum_i a_i$ where $\theta = \sum_i a_i \alpha_i$.}  Using this we find
\eq{(A_\theta)_{i j} = \frac {2 g^\vee}{\operatorname{dim} \mathfrak{g}} (\alpha_i^\vee, \alpha_j^\vee).}
Since $\omega_i$ is the dual basis to $\alpha^\vee_j$, we have that
\eq{(A_\theta^{-1})_{i j} = \frac {\operatorname{dim} \mathfrak{g}}{2 g^\vee} (\omega_i, \omega_j).}

If $\lambda{=}\theta$ and $\mu{=}0$ we have for the quantity in the exponent
\eq{\langle A_\lambda^{-1} (\mu{+}\rho), \mu{+}\rho\rangle =
  (A_\theta^{-1})_{i j} \rho_i \rho_j =
  \frac {\operatorname{dim} \mathfrak{g}}{2 g^\vee} \lvert \rho\rvert^2.}
Next, we can use the Freudenthal-de Vries strange formula (see refs.~\cite{MR260926},~\cite[eq.~13.179]{MR1424041})
\eq{\lvert \rho\rvert^2 = \frac{\hspace{-1pt}\phantom{{}^\vee\hspace{-1pt}}g^\vee}{12} \operatorname{dim} \mathfrak{g}.}

It follows that the quantity in the exponent is
\eq{\frac {(\operatorname{dim} \mathfrak{g})^2}{48 \r{n}}.}

From Biane's formula the asymptotic behavior have two terms, 
\eq{\exp\!\left(-\frac{\mathrm{dim}(\mathfrak{g})^2}{48\r{n}}\right) + \mathcal{O}(\r{n}^{-1})\,.}
The leading term (the exponential) will then dominate the sub-leading terms for 
\eq{\r{n}\log (\r{n}) \gg \frac{\mathrm{dim}(\mathfrak{g})^2}{48}\,.}
Algebras of smaller dimensions will evidently converge to the (leading term in the) asymptotic formulae for lower $\r{n}$ than for algebras of larger dimension.

\subsection{\texorpdfstring{$\mathfrak{a}$}{a}-Type Gauge Theory}
\vspace{-20pt}\eq{\begin{split}\fwboxR{40pt}{\mathfrak{a}_{1}\!\!:\,\,\mathcal{C}^{\r{n}}_{\mathfrak{a}_{1}}\asympt{\r{n}\to\infty}}&\hspace{4pt}\fwboxL{300pt}{\left(\hspace{-2pt}\frac{1\hspace{-4pt}}{2^{\hspace{-0.25pt}3}\!\,\,}\sqrt{\hspace{-2pt}\frac{3}{\pi}\hspace{-2pt}}\hspace{-4pt}\hspace{5pt}\right)\hspace{-2pt}\b{3}^{\hspace{-0.25pt}\r{n}\text{+}1}\r{n}^{\hspace{-0.5pt}\text{--}\frac{\b{3}}{2}}\exp\!\left\{\hspace{-0.5pt}\text{--}\frac{\b{3}^2\hspace{-3pt}}{\hspace{1pt}48\hspace{1pt}\r{n}}\,\right\}\Big(\!1{+}\mathcal{O}\!\left(1/\r{n}\right)\!\!\Big);}\\
\fwboxR{40pt}{\mathfrak{a}_{2}\!\!:\,\,\mathcal{C}^{\r{n}}_{\mathfrak{a}_{2}}\asympt{\r{n}\to\infty}}&\hspace{4pt}\fwboxL{300pt}{\left(\hspace{-2pt}\frac{2^{\hspace{-0.25pt}2}\!\hspace{-4pt}}{3^{\hspace{-0.25pt}4}\!\,\,\pi^{}}\sqrt{\hspace{-2pt}\frac{1}{3}\hspace{-2pt}}\hspace{-4pt}\hspace{5pt}\right)\hspace{-2pt}\b{8}^{\hspace{-0.25pt}\r{n}\text{+}2}\r{n}^{\hspace{-0.5pt}\text{--}\frac{\b{8}}{2}}\exp\!\left\{\hspace{-0.5pt}\text{--}\frac{\b{8}^2\hspace{-3pt}}{\hspace{1pt}48\hspace{1pt}\r{n}}\,\right\}\Big(\!1{+}\mathcal{O}\!\left(1/\r{n}\right)\!\!\Big);}\\
\fwboxR{40pt}{\mathfrak{a}_{3}\!\!:\,\,\mathcal{C}^{\r{n}}_{\mathfrak{a}_{3}}\asympt{\r{n}\to\infty}}&\hspace{4pt}\fwboxL{300pt}{\left(\hspace{-2pt}\frac{3^{\hspace{-0.25pt}5}\!\hspace{-2pt}\cdot\hspace{-3pt}5^{\hspace{-0.25pt}4}\!\hspace{-4pt}}{2^{\hspace{-0.25pt}21}\!\,\,\pi^{}}\sqrt{\hspace{-2pt}\frac{3\hspace{-2pt}\cdot\hspace{-3pt}5}{\pi}\hspace{-2pt}}\hspace{-4pt}\hspace{5pt}\right)\hspace{-2pt}\b{15}^{\hspace{-0.25pt}\r{n}\text{+}3}\r{n}^{\hspace{-0.5pt}\text{--}\frac{\b{15}}{2}}\exp\!\left\{\hspace{-0.5pt}\text{--}\frac{\b{15}^2\hspace{-3pt}}{\hspace{1pt}48\hspace{1pt}\r{n}}\,\right\}\Big(\!1{+}\mathcal{O}\!\left(1/\r{n}\right)\!\!\Big);}\\
\fwboxR{40pt}{\mathfrak{a}_{4}\!\!:\,\,\mathcal{C}^{\r{n}}_{\mathfrak{a}_{4}}\asympt{\r{n}\to\infty}}&\hspace{4pt}\fwboxL{300pt}{\left(\hspace{-2pt}\frac{2^{\hspace{-0.25pt}9}\!\hspace{-2pt}\cdot\hspace{-3pt}3^{\hspace{-0.25pt}8}\!\hspace{-4pt}}{5^{\hspace{-0.25pt}12}\!\,\,\pi^{2}}\sqrt{\hspace{-2pt}\frac{1}{5}\hspace{-2pt}}\hspace{-4pt}\hspace{5pt}\right)\hspace{-2pt}\b{24}^{\hspace{-0.25pt}\r{n}\text{+}6}\r{n}^{\hspace{-0.5pt}\text{--}\frac{\b{24}}{2}}\exp\!\left\{\hspace{-0.5pt}\text{--}\frac{\b{24}^2\hspace{-3pt}}{\hspace{1pt}48\hspace{1pt}\r{n}}\,\right\}\Big(\!1{+}\mathcal{O}\!\left(1/\r{n}\right)\!\!\Big);}\\
\fwboxR{40pt}{\mathfrak{a}_{5}\!\!:\,\,\mathcal{C}^{\r{n}}_{\mathfrak{a}_{5}}\asympt{\r{n}\to\infty}}&\hspace{4pt}\fwboxL{300pt}{\left(\hspace{-2pt}\frac{5^{\hspace{-0.25pt}9}\!\hspace{-2pt}\cdot\hspace{-3pt}7^{\hspace{-0.25pt}8}\!\hspace{-4pt}}{2^{\hspace{-0.25pt}29}\!\hspace{-2pt}\cdot\hspace{-3pt}3^{\hspace{-0.25pt}14}\!\,\,\pi^{2}}\sqrt{\hspace{-2pt}\frac{5\hspace{-2pt}\cdot\hspace{-3pt}7}{\pi}\hspace{-2pt}}\hspace{-4pt}\hspace{5pt}\right)\hspace{-2pt}\b{35}^{\hspace{-0.25pt}\r{n}\text{+}9}\r{n}^{\hspace{-0.5pt}\text{--}\frac{\b{35}}{2}}\exp\!\left\{\hspace{-0.5pt}\text{--}\frac{\b{35}^2\hspace{-3pt}}{\hspace{1pt}48\hspace{1pt}\r{n}}\,\right\}\Big(\!1{+}\mathcal{O}\!\left(1/\r{n}\right)\!\!\Big);}\\
\fwboxR{40pt}{\mathfrak{a}_{6}\!\!:\,\,\mathcal{C}^{\r{n}}_{\mathfrak{a}_{6}}\asympt{\r{n}\to\infty}}&\hspace{4pt}\fwboxL{300pt}{\left(\hspace{-2pt}\frac{2^{\hspace{-0.25pt}33}\!\hspace{-2pt}\cdot\hspace{-3pt}3^{\hspace{-0.25pt}17}\!\hspace{-2pt}\cdot\hspace{-3pt}5^{\hspace{-0.25pt}2}\!\hspace{-4pt}}{7^{\hspace{-0.25pt}24}\!\,\,\pi^{3}}\sqrt{\hspace{-2pt}\frac{1}{7}\hspace{-2pt}}\hspace{-4pt}\hspace{5pt}\right)\hspace{-2pt}\b{48}^{\hspace{-0.25pt}\r{n}\text{+}12}\r{n}^{\hspace{-0.5pt}\text{--}\frac{\b{48}}{2}}\exp\!\left\{\hspace{-0.5pt}\text{--}\frac{\b{48}^2\hspace{-3pt}}{\hspace{1pt}48\hspace{1pt}\r{n}}\,\right\}\Big(\!1{+}\mathcal{O}\!\left(1/\r{n}\right)\!\!\Big);}\\
\fwboxR{40pt}{\mathfrak{a}_{7}\!\!:\,\,\mathcal{C}^{\r{n}}_{\mathfrak{a}_{7}}\asympt{\r{n}\to\infty}}&\hspace{4pt}\fwboxL{300pt}{\left(\hspace{-2pt}\frac{3^{\hspace{-0.25pt}36}\!\hspace{-2pt}\cdot\hspace{-3pt}5^{\hspace{-0.25pt}3}\!\hspace{-2pt}\cdot\hspace{-3pt}7^{\hspace{-0.25pt}15}\!\hspace{-4pt}}{2^{\hspace{-0.25pt}112}\!\,\,\pi^{3}}\sqrt{\hspace{-2pt}\frac{7}{\pi}\hspace{-2pt}}\hspace{-4pt}\hspace{5pt}\right)\hspace{-2pt}\b{63}^{\hspace{-0.25pt}\r{n}\text{+}17}\r{n}^{\hspace{-0.5pt}\text{--}\frac{\b{63}}{2}}\exp\!\left\{\hspace{-0.5pt}\text{--}\frac{\b{63}^2\hspace{-3pt}}{\hspace{1pt}48\hspace{1pt}\r{n}}\,\right\}\Big(\!1{+}\mathcal{O}\!\left(1/\r{n}\right)\!\!\Big);}\\
\fwboxR{40pt}{\mathfrak{a}_{8}\!\!:\,\,\mathcal{C}^{\r{n}}_{\mathfrak{a}_{8}}\asympt{\r{n}\to\infty}}&\hspace{4pt}\fwboxL{300pt}{\left(\hspace{-2pt}\frac{2^{\hspace{-0.25pt}51}\!\hspace{-2pt}\cdot\hspace{-3pt}5^{\hspace{-0.25pt}22}\!\hspace{-2pt}\cdot\hspace{-3pt}7^{\hspace{-0.25pt}2}\!\hspace{-4pt}}{3^{\hspace{-0.25pt}70}\!\,\,\pi^{4}}\hspace{5pt}\right)\hspace{-2pt}\b{80}^{\hspace{-0.25pt}\r{n}\text{+}22}\r{n}^{\hspace{-0.5pt}\text{--}\frac{\b{80}}{2}}\exp\!\left\{\hspace{-0.5pt}\text{--}\frac{\b{80}^2\hspace{-3pt}}{\hspace{1pt}48\hspace{1pt}\r{n}}\,\right\}\Big(\!1{+}\mathcal{O}\!\left(1/\r{n}\right)\!\!\Big).}\end{split}\label{a_series_asymptotics}}

In the case of $\mathfrak{a}_1$ an extra sub-leading term is known if we also expand the exponential in the limit $n\!\to\!\infty$.  Indeed, we have (see refs.~\cite{oeis-riordan, 4977010})
\begin{equation}
  \label{eq:riordan-asympt-extra}
  \mathcal{C}\big[\mathbf{ad}(\mathfrak{a}_{1})^{\r{n}}\big]\asympt{\r{n}\to\infty} \frac {3^{\r{n} + 2}}{\sqrt{3 \r{n} \pi} 8 \r{n}} (1 - \frac {21}{16 \r{n}} + \mathcal{O}(\r{n}^{-2})).
\end{equation}
To the same order in $\r{n}^{-1}$ this can be rewritten as
\begin{equation}
  \label{eq:riordan-asympt-extra2}
  \mathcal{C}^{\r{n}}_{\mathfrak{a}_1}\asympt{\r{n}\to\infty} \frac{1}{8}\sqrt{\frac{3}{\pi}}\,\b{3}^{\r{n}{+}1}\exp\!\left\{\hspace{-0.5pt}\text{--}\frac{\b{3}^2\hspace{-3pt}}{\hspace{1pt}48\hspace{1pt}\r{n}}\,\right\}\Big(1{-}\frac{9}{8\,\r{n}}\Big)\Big(1{+}\mathcal{O}\!\big(1/\r{n}^2\big)\Big)\,.
\end{equation}
While the leading order formula matches the data well for $\mathfrak{a}_1$ for other algebras (of larger dimension) the convergence of data to the asymptotic formula will happen at some (potentially large) multiplicity. Other examples of $\mathcal{O}(1/\r{n})$ obtained by direct comparison with `data' at large multiplicity were given in (\ref{improved_asymptotics}).

\subsection{\texorpdfstring{$\mathfrak{b}$}{b}-Type Gauge Theory}
\vspace{-20pt}\eq{\begin{split}\fwboxR{40pt}{\mathfrak{b}_{2}\!\!:\,\,\mathcal{C}^{\r{n}}_{\mathfrak{b}_{2}}\asympt{\r{n}\to\infty}}&\hspace{4pt}\fwboxL{300pt}{\left(\hspace{-2pt}\frac{5^{\hspace{-0.25pt}3}\!\hspace{-4pt}}{2^{\hspace{-0.25pt}4}\!\hspace{-2pt}\cdot\hspace{-3pt}3^{\hspace{-0.25pt}4}\!\,\,\pi^{}}\hspace{5pt}\right)\hspace{-2pt}\b{10}^{\hspace{-0.25pt}\r{n}\text{+}2}\r{n}^{\hspace{-0.5pt}\text{--}\frac{\b{10}}{2}}\exp\!\left\{\hspace{-0.5pt}\text{--}\frac{\b{10}^2\hspace{-3pt}}{\hspace{1pt}48\hspace{1pt}\r{n}}\,\right\}\Big(\!1{+}\mathcal{O}\!\left(1/\r{n}\right)\!\!\Big);}\\
\fwboxR{40pt}{\mathfrak{b}_{3}\!\!:\,\,\mathcal{C}^{\r{n}}_{\mathfrak{b}_{3}}\asympt{\r{n}\to\infty}}&\hspace{4pt}\fwboxL{300pt}{\left(\hspace{-2pt}\frac{3^{\hspace{-0.25pt}8}\!\hspace{-2pt}\cdot\hspace{-3pt}7^{\hspace{-0.25pt}6}\!\hspace{-4pt}}{2^{\hspace{-0.25pt}11}\!\hspace{-2pt}\cdot\hspace{-3pt}5^{\hspace{-0.25pt}10}\!\,\,\pi^{}}\sqrt{\hspace{-2pt}\frac{3\hspace{-2pt}\cdot\hspace{-3pt}7}{5\,\pi}\hspace{-2pt}}\hspace{-4pt}\hspace{5pt}\right)\hspace{-2pt}\b{21}^{\hspace{-0.25pt}\r{n}\text{+}4}\r{n}^{\hspace{-0.5pt}\text{--}\frac{\b{21}}{2}}\exp\!\left\{\hspace{-0.5pt}\text{--}\frac{\b{21}^2\hspace{-3pt}}{\hspace{1pt}48\hspace{1pt}\r{n}}\,\right\}\Big(\!1{+}\mathcal{O}\!\left(1/\r{n}\right)\!\!\Big);}\\
\fwboxR{40pt}{\mathfrak{b}_{4}\!\!:\,\,\mathcal{C}^{\r{n}}_{\mathfrak{b}_{4}}\asympt{\r{n}\to\infty}}&\hspace{4pt}\fwboxL{300pt}{\left(\hspace{-2pt}\frac{2^{\hspace{-0.25pt}4}\!\hspace{-2pt}\cdot\hspace{-3pt}3^{\hspace{-0.25pt}24}\!\hspace{-2pt}\cdot\hspace{-3pt}5^{\hspace{-0.25pt}2}\!\hspace{-4pt}}{7^{\hspace{-0.25pt}17}\!\,\,\pi^{2}}\hspace{5pt}\right)\hspace{-2pt}\b{36}^{\hspace{-0.25pt}\r{n}\text{+}8}\r{n}^{\hspace{-0.5pt}\text{--}\frac{\b{36}}{2}}\exp\!\left\{\hspace{-0.5pt}\text{--}\frac{\b{36}^2\hspace{-3pt}}{\hspace{1pt}48\hspace{1pt}\r{n}}\,\right\}\Big(\!1{+}\mathcal{O}\!\left(1/\r{n}\right)\!\!\Big);}\\
\fwboxR{40pt}{\mathfrak{b}_{5}\!\!:\,\,\mathcal{C}^{\r{n}}_{\mathfrak{b}_{5}}\asympt{\r{n}\to\infty}}&\hspace{4pt}\fwboxL{300pt}{\left(\hspace{-2pt}\frac{5^{\hspace{-0.25pt}16}\!\hspace{-2pt}\cdot\hspace{-3pt}7^{\hspace{-0.25pt}2}\!\hspace{-2pt}\cdot\hspace{-3pt}11^{\hspace{-0.25pt}13}\!\hspace{-4pt}}{2^{\hspace{-0.25pt}20}\!\hspace{-2pt}\cdot\hspace{-3pt}3^{\hspace{-0.25pt}47}\!\,\,\pi^{2}}\sqrt{\hspace{-2pt}\frac{5\hspace{-2pt}\cdot\hspace{-3pt}11}{\pi}\hspace{-2pt}}\hspace{-4pt}\hspace{5pt}\right)\hspace{-2pt}\b{55}^{\hspace{-0.25pt}\r{n}\text{+}14}\r{n}^{\hspace{-0.5pt}\text{--}\frac{\b{55}}{2}}\exp\!\left\{\hspace{-0.5pt}\text{--}\frac{\b{55}^2\hspace{-3pt}}{\hspace{1pt}48\hspace{1pt}\r{n}}\,\right\}\Big(\!1{+}\mathcal{O}\!\left(1/\r{n}\right)\!\!\Big);}\\
\fwboxR{40pt}{\mathfrak{b}_{6}\!\!:\,\,\mathcal{C}^{\r{n}}_{\mathfrak{b}_{6}}\asympt{\r{n}\to\infty}}&\hspace{4pt}\fwboxL{300pt}{\left(\hspace{-2pt}\frac{3^{\hspace{-0.25pt}30}\!\hspace{-2pt}\cdot\hspace{-3pt}5^{\hspace{-0.25pt}5}\!\hspace{-2pt}\cdot\hspace{-3pt}7^{\hspace{-0.25pt}3}\!\hspace{-2pt}\cdot\hspace{-3pt}13^{\hspace{-0.25pt}18}\!\hspace{-4pt}}{2^{\hspace{-0.25pt}7}\!\hspace{-2pt}\cdot\hspace{-3pt}11^{\hspace{-0.25pt}38}\!\,\,\pi^{3}}\hspace{5pt}\right)\hspace{-2pt}\b{78}^{\hspace{-0.25pt}\r{n}\text{+}21}\r{n}^{\hspace{-0.5pt}\text{--}\frac{\b{78}}{2}}\exp\!\left\{\hspace{-0.5pt}\text{--}\frac{\b{78}^2\hspace{-3pt}}{\hspace{1pt}48\hspace{1pt}\r{n}}\,\right\}\Big(\!1{+}\mathcal{O}\!\left(1/\r{n}\right)\!\!\Big);}\\
\fwboxR{40pt}{\mathfrak{b}_{7}\!\!:\,\,\mathcal{C}^{\r{n}}_{\mathfrak{b}_{7}}\asympt{\r{n}\to\infty}}&\hspace{4pt}\fwboxL{300pt}{\left(\hspace{-2pt}\frac{3^{\hspace{-0.25pt}40}\!\hspace{-2pt}\cdot\hspace{-3pt}5^{\hspace{-0.25pt}30}\!\hspace{-2pt}\cdot\hspace{-3pt}7^{\hspace{-0.25pt}27}\!\hspace{-2pt}\cdot\hspace{-3pt}11^{\hspace{-0.25pt}2}\!\hspace{-4pt}}{2^{\hspace{-0.25pt}30}\!\hspace{-2pt}\cdot\hspace{-3pt}13^{\hspace{-0.25pt}52}\!\,\,\pi^{3}}\sqrt{\hspace{-2pt}\frac{3\hspace{-2pt}\cdot\hspace{-3pt}5\hspace{-2pt}\cdot\hspace{-3pt}7}{13\,\pi}\hspace{-2pt}}\hspace{-4pt}\hspace{5pt}\right)\hspace{-2pt}\b{105}^{\hspace{-0.25pt}\r{n}\text{+}29}\r{n}^{\hspace{-0.5pt}\text{--}\frac{\b{105}}{2}}\exp\!\left\{\hspace{-0.5pt}\text{--}\frac{\b{105}^2\hspace{-3pt}}{\hspace{1pt}48\hspace{1pt}\r{n}}\,\right\}\Big(\!1{+}\mathcal{O}\!\left(1/\r{n}\right)\!\!\Big);}\\
\fwboxR{40pt}{\mathfrak{b}_{8}\!\!:\,\,\mathcal{C}^{\r{n}}_{\mathfrak{b}_{8}}\asympt{\r{n}\to\infty}}&\hspace{4pt}\fwboxL{300pt}{\left(\hspace{-2pt}\frac{2^{\hspace{-0.25pt}51}\!\hspace{-2pt}\cdot\hspace{-3pt}7^{\hspace{-0.25pt}6}\!\hspace{-2pt}\cdot\hspace{-3pt}11^{\hspace{-0.25pt}3}\!\hspace{-2pt}\cdot\hspace{-3pt}13^{\hspace{-0.25pt}2}\!\hspace{-2pt}\cdot\hspace{-3pt}17^{\hspace{-0.25pt}29}\!\hspace{-4pt}}{3^{\hspace{-0.25pt}45}\!\hspace{-2pt}\cdot\hspace{-3pt}5^{\hspace{-0.25pt}58}\!\,\,\pi^{4}}\hspace{5pt}\right)\hspace{-2pt}\b{136}^{\hspace{-0.25pt}\r{n}\text{+}39}\r{n}^{\hspace{-0.5pt}\text{--}\frac{\b{136}}{2}}\exp\!\left\{\hspace{-0.5pt}\text{--}\frac{\b{136}^2\hspace{-3pt}}{\hspace{1pt}48\hspace{1pt}\r{n}}\,\right\}\Big(\!1{+}\mathcal{O}\!\left(1/\r{n}\right)\!\!\Big).}\end{split}\label{b_series_asymptotics}}

\subsection{\texorpdfstring{$\mathfrak{c}$}{c}-Type Gauge Theory}
\vspace{-20pt}\eq{\begin{split}\fwboxR{40pt}{\mathfrak{c}_{3}\!\!:\,\,\mathcal{C}^{\r{n}}_{\mathfrak{c}_{3}}\asympt{\r{n}\to\infty}}&\hspace{4pt}\fwboxL{300pt}{\left(\hspace{-2pt}\frac{3^{\hspace{-0.25pt}8}\!\hspace{-2pt}\cdot\hspace{-3pt}5\hspace{-2pt}\cdot\hspace{-3pt}7^{\hspace{-0.25pt}6}\!\hspace{-4pt}}{2^{\hspace{-0.25pt}36}\!\,\,\pi^{}}\sqrt{\hspace{-2pt}\frac{3\hspace{-2pt}\cdot\hspace{-3pt}7}{2\,\pi}\hspace{-2pt}}\hspace{-4pt}\hspace{5pt}\right)\hspace{-2pt}\b{21}^{\hspace{-0.25pt}\r{n}\text{+}4}\r{n}^{\hspace{-0.5pt}\text{--}\frac{\b{21}}{2}}\exp\!\left\{\hspace{-0.5pt}\text{--}\frac{\b{21}^2\hspace{-3pt}}{\hspace{1pt}48\hspace{1pt}\r{n}}\,\right\}\Big(\!1{+}\mathcal{O}\!\left(1/\r{n}\right)\!\!\Big);}\\
\fwboxR{40pt}{\mathfrak{c}_{4}\!\!:\,\,\mathcal{C}^{\r{n}}_{\mathfrak{c}_{4}}\asympt{\r{n}\to\infty}}&\hspace{4pt}\fwboxL{300pt}{\left(\hspace{-2pt}\frac{3^{\hspace{-0.25pt}24}\!\hspace{-2pt}\cdot\hspace{-3pt}7\hspace{-4pt}}{2^{\hspace{-0.25pt}5}\!\hspace{-2pt}\cdot\hspace{-3pt}5^{\hspace{-0.25pt}16}\!\,\,\pi^{2}}\hspace{5pt}\right)\hspace{-2pt}\b{36}^{\hspace{-0.25pt}\r{n}\text{+}8}\r{n}^{\hspace{-0.5pt}\text{--}\frac{\b{36}}{2}}\exp\!\left\{\hspace{-0.5pt}\text{--}\frac{\b{36}^2\hspace{-3pt}}{\hspace{1pt}48\hspace{1pt}\r{n}}\,\right\}\Big(\!1{+}\mathcal{O}\!\left(1/\r{n}\right)\!\!\Big);}\\
\fwboxR{40pt}{\mathfrak{c}_{5}\!\!:\,\,\mathcal{C}^{\r{n}}_{\mathfrak{c}_{5}}\asympt{\r{n}\to\infty}}&\hspace{4pt}\fwboxL{300pt}{\left(\hspace{-2pt}\frac{5^{\hspace{-0.25pt}16}\!\hspace{-2pt}\cdot\hspace{-3pt}7^{\hspace{-0.25pt}2}\!\hspace{-2pt}\cdot\hspace{-3pt}11^{\hspace{-0.25pt}13}\!\hspace{-4pt}}{2^{\hspace{-0.25pt}64}\!\hspace{-2pt}\cdot\hspace{-3pt}3^{\hspace{-0.25pt}20}\!\,\,\pi^{2}}\sqrt{\hspace{-2pt}\frac{5\hspace{-2pt}\cdot\hspace{-3pt}11}{3\,\pi}\hspace{-2pt}}\hspace{-4pt}\hspace{5pt}\right)\hspace{-2pt}\b{55}^{\hspace{-0.25pt}\r{n}\text{+}14}\r{n}^{\hspace{-0.5pt}\text{--}\frac{\b{55}}{2}}\exp\!\left\{\hspace{-0.5pt}\text{--}\frac{\b{55}^2\hspace{-3pt}}{\hspace{1pt}48\hspace{1pt}\r{n}}\,\right\}\Big(\!1{+}\mathcal{O}\!\left(1/\r{n}\right)\!\!\Big);}\\
\fwboxR{40pt}{\mathfrak{c}_{6}\!\!:\,\,\mathcal{C}^{\r{n}}_{\mathfrak{c}_{6}}\asympt{\r{n}\to\infty}}&\hspace{4pt}\fwboxL{300pt}{\left(\hspace{-2pt}\frac{3^{\hspace{-0.25pt}30}\!\hspace{-2pt}\cdot\hspace{-3pt}5^{\hspace{-0.25pt}5}\!\hspace{-2pt}\cdot\hspace{-3pt}11\hspace{-2pt}\cdot\hspace{-3pt}13^{\hspace{-0.25pt}18}\!\hspace{-4pt}}{2^{\hspace{-0.25pt}33}\!\hspace{-2pt}\cdot\hspace{-3pt}7^{\hspace{-0.25pt}36}\!\,\,\pi^{3}}\hspace{5pt}\right)\hspace{-2pt}\b{78}^{\hspace{-0.25pt}\r{n}\text{+}21}\r{n}^{\hspace{-0.5pt}\text{--}\frac{\b{78}}{2}}\exp\!\left\{\hspace{-0.5pt}\text{--}\frac{\b{78}^2\hspace{-3pt}}{\hspace{1pt}48\hspace{1pt}\r{n}}\,\right\}\Big(\!1{+}\mathcal{O}\!\left(1/\r{n}\right)\!\!\Big);}\\
\fwboxR{40pt}{\mathfrak{c}_{7}\!\!:\,\,\mathcal{C}^{\r{n}}_{\mathfrak{c}_{7}}\asympt{\r{n}\to\infty}}&\hspace{4pt}\fwboxL{300pt}{\left(\hspace{-2pt}\frac{3^{\hspace{-0.25pt}40}\!\hspace{-2pt}\cdot\hspace{-3pt}5^{\hspace{-0.25pt}30}\!\hspace{-2pt}\cdot\hspace{-3pt}7^{\hspace{-0.25pt}27}\!\hspace{-2pt}\cdot\hspace{-3pt}11^{\hspace{-0.25pt}2}\!\hspace{-2pt}\cdot\hspace{-3pt}13\hspace{-4pt}}{2^{\hspace{-0.25pt}225}\!\,\,\pi^{3}}\sqrt{\hspace{-2pt}\frac{3\hspace{-2pt}\cdot\hspace{-3pt}5\hspace{-2pt}\cdot\hspace{-3pt}7}{\pi}\hspace{-2pt}}\hspace{-4pt}\hspace{5pt}\right)\hspace{-2pt}\b{105}^{\hspace{-0.25pt}\r{n}\text{+}29}\r{n}^{\hspace{-0.5pt}\text{--}\frac{\b{105}}{2}}\exp\!\left\{\hspace{-0.5pt}\text{--}\frac{\b{105}^2\hspace{-3pt}}{\hspace{1pt}48\hspace{1pt}\r{n}}\,\right\}\Big(\!1{+}\mathcal{O}\!\left(1/\r{n}\right)\!\!\Big);}\\
\fwboxR{40pt}{\mathfrak{c}_{8}\!\!:\,\,\mathcal{C}^{\r{n}}_{\mathfrak{c}_{8}}\asympt{\r{n}\to\infty}}&\hspace{4pt}\fwboxL{300pt}{\left(\hspace{-2pt}\frac{5^{\hspace{-0.25pt}10}\!\hspace{-2pt}\cdot\hspace{-3pt}7^{\hspace{-0.25pt}6}\!\hspace{-2pt}\cdot\hspace{-3pt}11^{\hspace{-0.25pt}3}\!\hspace{-2pt}\cdot\hspace{-3pt}13^{\hspace{-0.25pt}2}\!\hspace{-2pt}\cdot\hspace{-3pt}17^{\hspace{-0.25pt}29}\!\hspace{-4pt}}{3^{\hspace{-0.25pt}113}\!\,\,\pi^{4}}\hspace{5pt}\right)\hspace{-2pt}\b{136}^{\hspace{-0.25pt}\r{n}\text{+}39}\r{n}^{\hspace{-0.5pt}\text{--}\frac{\b{136}}{2}}\exp\!\left\{\hspace{-0.5pt}\text{--}\frac{\b{136}^2\hspace{-3pt}}{\hspace{1pt}48\hspace{1pt}\r{n}}\,\right\}\Big(\!1{+}\mathcal{O}\!\left(1/\r{n}\right)\!\!\Big).}\\[-20pt]\end{split}\label{c_series_asymptotics}}

\subsection{\texorpdfstring{$\mathfrak{d}$}{d}-Type Gauge Theory}
\vspace{-22pt}\eq{\begin{split}\fwboxR{40pt}{\mathfrak{d}_{4}\!\!:\,\,\mathcal{C}^{\r{n}}_{\mathfrak{d}_{4}}\asympt{\r{n}\to\infty}}&\hspace{4pt}\fwboxL{300pt}{\left(\hspace{-2pt}\frac{5\hspace{-2pt}\cdot\hspace{-3pt}7^{\hspace{-0.25pt}7}\!\hspace{-4pt}}{2^{\hspace{-0.25pt}10}\!\hspace{-2pt}\cdot\hspace{-3pt}3^{\hspace{-0.25pt}11}\!\,\,\pi^{2}}\hspace{5pt}\right)\hspace{-2pt}\b{28}^{\hspace{-0.25pt}\r{n}\text{+}7}\r{n}^{\hspace{-0.5pt}\text{--}\frac{\b{28}}{2}}\exp\!\left\{\hspace{-0.5pt}\text{--}\frac{\b{28}^2\hspace{-3pt}}{\hspace{1pt}48\hspace{1pt}\r{n}}\,\right\}\Big(\!1{+}\mathcal{O}\!\left(1/\r{n}\right)\!\!\Big);}\\
\fwboxR{40pt}{\mathfrak{d}_{5}\!\!:\,\,\mathcal{C}^{\r{n}}_{\mathfrak{d}_{5}}\asympt{\r{n}\to\infty}}&\hspace{4pt}\fwboxL{300pt}{\left(\hspace{-2pt}\frac{3^{\hspace{-0.25pt}28}\!\hspace{-2pt}\cdot\hspace{-3pt}5^{\hspace{-0.25pt}13}\!\hspace{-2pt}\cdot\hspace{-3pt}7\hspace{-4pt}}{2^{\hspace{-0.25pt}81}\!\,\,\pi^{2}}\sqrt{\hspace{-2pt}\frac{5}{2\,\pi}\hspace{-2pt}}\hspace{-4pt}\hspace{5pt}\right)\hspace{-2pt}\b{45}^{\hspace{-0.25pt}\r{n}\text{+}11}\r{n}^{\hspace{-0.5pt}\text{--}\frac{\b{45}}{2}}\exp\!\left\{\hspace{-0.5pt}\text{--}\frac{\b{45}^2\hspace{-3pt}}{\hspace{1pt}48\hspace{1pt}\r{n}}\,\right\}\Big(\!1{+}\mathcal{O}\!\left(1/\r{n}\right)\!\!\Big);}\\
\fwboxR{40pt}{\mathfrak{d}_{6}\!\!:\,\,\mathcal{C}^{\r{n}}_{\mathfrak{d}_{6}}\asympt{\r{n}\to\infty}}&\hspace{4pt}\fwboxL{300pt}{\left(\hspace{-2pt}\frac{3^{\hspace{-0.25pt}24}\!\hspace{-2pt}\cdot\hspace{-3pt}7^{\hspace{-0.25pt}2}\!\hspace{-2pt}\cdot\hspace{-3pt}11^{\hspace{-0.25pt}15}\!\hspace{-4pt}}{2^{\hspace{-0.25pt}35}\!\hspace{-2pt}\cdot\hspace{-3pt}5^{\hspace{-0.25pt}29}\!\,\,\pi^{3}}\hspace{5pt}\right)\hspace{-2pt}\b{66}^{\hspace{-0.25pt}\r{n}\text{+}18}\r{n}^{\hspace{-0.5pt}\text{--}\frac{\b{66}}{2}}\exp\!\left\{\hspace{-0.5pt}\text{--}\frac{\b{66}^2\hspace{-3pt}}{\hspace{1pt}48\hspace{1pt}\r{n}}\,\right\}\Big(\!1{+}\mathcal{O}\!\left(1/\r{n}\right)\!\!\Big);}\\
\fwboxR{40pt}{\mathfrak{d}_{7}\!\!:\,\,\mathcal{C}^{\r{n}}_{\mathfrak{d}_{7}}\asympt{\r{n}\to\infty}}&\hspace{4pt}\fwboxL{300pt}{\left(\hspace{-2pt}\frac{5^{\hspace{-0.25pt}6}\!\hspace{-2pt}\cdot\hspace{-3pt}7^{\hspace{-0.25pt}23}\!\hspace{-2pt}\cdot\hspace{-3pt}11\hspace{-2pt}\cdot\hspace{-3pt}13^{\hspace{-0.25pt}20}\!\hspace{-4pt}}{2^{\hspace{-0.25pt}112}\!\hspace{-2pt}\cdot\hspace{-3pt}3^{\hspace{-0.25pt}32}\!\,\,\pi^{3}}\sqrt{\hspace{-2pt}\frac{7\hspace{-2pt}\cdot\hspace{-3pt}13}{3\,\pi}\hspace{-2pt}}\hspace{-4pt}\hspace{5pt}\right)\hspace{-2pt}\b{91}^{\hspace{-0.25pt}\r{n}\text{+}25}\r{n}^{\hspace{-0.5pt}\text{--}\frac{\b{91}}{2}}\exp\!\left\{\hspace{-0.5pt}\text{--}\frac{\b{91}^2\hspace{-3pt}}{\hspace{1pt}48\hspace{1pt}\r{n}}\,\right\}\Big(\!1{+}\mathcal{O}\!\left(1/\r{n}\right)\!\!\Big);}\\
\fwboxR{40pt}{\mathfrak{d}_{8}\!\!:\,\,\mathcal{C}^{\r{n}}_{\mathfrak{d}_{8}}\asympt{\r{n}\to\infty}}&\hspace{4pt}\fwboxL{300pt}{\left(\hspace{-2pt}\frac{3^{\hspace{-0.25pt}45}\!\hspace{-2pt}\cdot\hspace{-3pt}5^{\hspace{-0.25pt}34}\!\hspace{-2pt}\cdot\hspace{-3pt}11^{\hspace{-0.25pt}2}\!\hspace{-2pt}\cdot\hspace{-3pt}13\hspace{-4pt}}{2^{\hspace{-0.25pt}8}\!\hspace{-2pt}\cdot\hspace{-3pt}7^{\hspace{-0.25pt}55}\!\,\,\pi^{4}}\hspace{5pt}\right)\hspace{-2pt}\b{120}^{\hspace{-0.25pt}\r{n}\text{+}34}\r{n}^{\hspace{-0.5pt}\text{--}\frac{\b{120}}{2}}\exp\!\left\{\hspace{-0.5pt}\text{--}\frac{\b{120}^2\hspace{-3pt}}{\hspace{1pt}48\hspace{1pt}\r{n}}\,\right\}\Big(\!1{+}\mathcal{O}\!\left(1/\r{n}\right)\!\!\Big).}\\[-10pt]\end{split}\label{d_series_asymptotics}}

\subsection{Exceptional Gauge Theories}
\vspace{-26pt}\begin{align}\fwboxR{40pt}{\mathfrak{e}_{6}\!\!:\,\,\mathcal{C}^{\r{n}}_{\mathfrak{e}_{6}}\asympt{\r{n}\to\infty}}&\hspace{4pt}\fwboxL{300pt}{\left(\hspace{-2pt}\frac{5^{\hspace{-0.25pt}5}\!\hspace{-2pt}\cdot\hspace{-3pt}7^{\hspace{-0.25pt}3}\!\hspace{-2pt}\cdot\hspace{-3pt}11\hspace{-2pt}\cdot\hspace{-3pt}13^{\hspace{-0.25pt}18}\!\hspace{-4pt}}{2^{\hspace{-0.25pt}77}\!\hspace{-2pt}\cdot\hspace{-3pt}3^{\hspace{-0.25pt}11}\!\,\,\pi^{3}}\sqrt{\hspace{-2pt}\frac{1}{3}\hspace{-2pt}}\hspace{-4pt}\hspace{5pt}\right)\hspace{-2pt}\b{78}^{\hspace{-0.25pt}\r{n}\text{+}21}\r{n}^{\hspace{-0.5pt}\text{--}\frac{\b{78}}{2}}\exp\!\left\{\hspace{-0.5pt}\text{--}\frac{\b{78}^2\hspace{-3pt}}{\hspace{1pt}48\hspace{1pt}\r{n}}\,\right\}\Big(\!1{+}\mathcal{O}\!\left(1/\r{n}\right)\!\!\Big);}\nonumber\\
\fwboxR{40pt}{\mathfrak{e}_{7}\!\!:\,\,\mathcal{C}^{\r{n}}_{\mathfrak{e}_{7}}\asympt{\r{n}\to\infty}}&\hspace{4pt}\fwboxL{300pt}{\left(\hspace{-2pt}\frac{5^{\hspace{-0.25pt}10}\!\hspace{-2pt}\cdot\hspace{-3pt}7^{\hspace{-0.25pt}34}\!\hspace{-2pt}\cdot\hspace{-3pt}11^{\hspace{-0.25pt}3}\!\hspace{-2pt}\cdot\hspace{-3pt}13^{\hspace{-0.25pt}2}\!\hspace{-2pt}\cdot\hspace{-3pt}17\hspace{-2pt}\cdot\hspace{-3pt}19^{\hspace{-0.25pt}28}\!\hspace{-4pt}}{2^{\hspace{-0.25pt}89}\!\hspace{-2pt}\cdot\hspace{-3pt}3^{\hspace{-0.25pt}111}\!\,\,\pi^{3}}\frac{1}{\sqrt{\pi}}\hspace{-4pt}\hspace{5pt}\right)\hspace{-2pt}\b{133}^{\hspace{-0.25pt}\r{n}\text{+}38}\r{n}^{\hspace{-0.5pt}\text{--}\frac{\b{133}}{2}}\exp\!\left\{\hspace{-0.5pt}\text{--}\frac{\b{133}^2\hspace{-3pt}}{\hspace{1pt}48\hspace{1pt}\r{n}}\,\right\}\Big(\!1{+}\mathcal{O}\!\left(1/\r{n}\right)\!\!\Big);}\nonumber\\
\fwboxR{40pt}{\mathfrak{e}_{8}\!\!:\,\,\mathcal{C}^{\r{n}}_{\mathfrak{e}_{8}}\asympt{\r{n}\to\infty}}&\hspace{4pt}\fwboxL{300pt}{\left(\hspace{-2pt}\frac{7^{\hspace{-0.25pt}14}\!\hspace{-2pt}\cdot\hspace{-3pt}11^{\hspace{-0.25pt}8}\!\hspace{-2pt}\cdot\hspace{-3pt}13^{\hspace{-0.25pt}6}\!\hspace{-2pt}\cdot\hspace{-3pt}17^{\hspace{-0.25pt}4}\!\hspace{-2pt}\cdot\hspace{-3pt}19^{\hspace{-0.25pt}3}\!\hspace{-2pt}\cdot\hspace{-3pt}23^{\hspace{-0.25pt}2}\!\hspace{-2pt}\cdot\hspace{-3pt}29\hspace{-2pt}\cdot\hspace{-3pt}31^{\hspace{-0.25pt}48}\!\hspace{-4pt}}{2^{\hspace{-0.25pt}11}\!\hspace{-2pt}\cdot\hspace{-3pt}3^{\hspace{-0.25pt}77}\!\hspace{-2pt}\cdot\hspace{-3pt}5^{\hspace{-0.25pt}103}\!\,\,\pi^{4}}\hspace{5pt}\right)\hspace{-2pt}\b{248}^{\hspace{-0.25pt}\r{n}\text{+}76}\r{n}^{\hspace{-0.5pt}\text{--}\frac{\b{248}}{2}}\exp\!\left\{\hspace{-0.5pt}\text{--}\frac{\b{248}^2\hspace{-3pt}}{\hspace{1pt}48\hspace{1pt}\r{n}}\,\right\}\Big(\!1{+}\mathcal{O}\!\left(1/\r{n}\right)\!\!\Big);}\nonumber\\
\fwboxR{40pt}{\mathfrak{f}_{4}\!\!:\,\,\mathcal{C}^{\r{n}}_{\mathfrak{f}_{4}}\asympt{\r{n}\to\infty}}&\hspace{4pt}\fwboxL{300pt}{\left(\hspace{-2pt}\frac{5^{\hspace{-0.25pt}4}\!\hspace{-2pt}\cdot\hspace{-3pt}7^{\hspace{-0.25pt}2}\!\hspace{-2pt}\cdot\hspace{-3pt}11\hspace{-2pt}\cdot\hspace{-3pt}13^{\hspace{-0.25pt}13}\!\hspace{-4pt}}{3^{\hspace{-0.25pt}45}\!\,\,\pi^{2}}\hspace{5pt}\right)\hspace{-2pt}\b{52}^{\hspace{-0.25pt}\r{n}\text{+}13}\r{n}^{\hspace{-0.5pt}\text{--}\frac{\b{52}}{2}}\exp\!\left\{\hspace{-0.5pt}\text{--}\frac{\b{52}^2\hspace{-3pt}}{\hspace{1pt}48\hspace{1pt}\r{n}}\,\right\}\Big(\!1{+}\mathcal{O}\!\left(1/\r{n}\right)\!\!\Big);}\label{efg_series_asymptotics}\\
\fwboxR{40pt}{\mathfrak{g}_{2}\!\!:\,\,\mathcal{C}^{\r{n}}_{\mathfrak{g}_{2}}\asympt{\r{n}\to\infty}}&\hspace{4pt}\fwboxL{300pt}{\left(\hspace{-2pt}\frac{5\hspace{-2pt}\cdot\hspace{-3pt}7^{\hspace{-0.25pt}5}\!\hspace{-4pt}}{2^{\hspace{-0.25pt}14}\!\hspace{-2pt}\cdot\hspace{-3pt}3^{\hspace{-0.25pt}3}\!\,\,\pi^{}}\sqrt{\hspace{-2pt}\frac{1}{3}\hspace{-2pt}}\hspace{-4pt}\hspace{5pt}\right)\hspace{-2pt}\b{14}^{\hspace{-0.25pt}\r{n}\text{+}2}\r{n}^{\hspace{-0.5pt}\text{--}\frac{\b{14}}{2}}\exp\!\left\{\hspace{-0.5pt}\text{--}\frac{\b{14}^2\hspace{-3pt}}{\hspace{1pt}48\hspace{1pt}\r{n}}\,\right\}\Big(\!1{+}\mathcal{O}\!\left(1/\r{n}\right)\!\!\Big).}\nonumber\\[-25pt]\nonumber\end{align}

The numerical prefactors are complicated, but their factorization has some interesting features.  For example, for $\mathfrak{e}_8$ it contains $1$, $7$, $11$, $13$, $17$, $19$, $23$, $29$ which are the exponents of $\mathfrak{e}_8$.  The exponents are the integers $n_i$ such that the Cartan matrix has $4 \cos^2 \frac {\pi n_i}{2 g^\vee}$ as eigenvalues, where $g^\vee$ is the dual Coxeter number.  For $\mathfrak{e}_8$ the Coxeter number is $30$ and we also have the prime number $g + 1 = 31$ in the prefactor (see Table~\ref{tab:liealg-data}).  For the non-simply-laced algebras the Coxeter number and the dual Coxeter numbers are different.

\begin{table}[b!]\vspace{-20pt}\renewcommand{\arraystretch}{1}
\centering
\begin{tabular}{c|c|c|c}
algebra & exponents & $g^\vee$ & $g$ \\
\hline
$\mathfrak{a}_k$ & $1, 2, \dotsc, k$ & $k + 1$ & $k + 1$ \\
$\mathfrak{b}_k$ & $1, 3, \dotsc, 2 k - 1$ & $2 k - 1$ & $2 k$ \\
$\mathfrak{c}_k$ & $1, 3, \dotsc, 2 k - 1$ & $k + 1$ & $2 k$ \\
$\mathfrak{d}_k$ & $1, 3, \dotsc, 2 k - 3, k - 1$ & $2 k - 2$ & $2 k - 2$ \\
$\mathfrak{e}_6$ & $1, 4, 5, 7, 8, 11$ & $12$ & $12$ \\
$\mathfrak{e}_7$ & $1, 5, 7, 9, 11, 13, 17$ & $18$ & $18$ \\
$\mathfrak{e}_8$ & $1, 7, 11, 13, 17, 19, 23, 29$ & $30$ & $30$ \\
$\mathfrak{f}_4$ & $1, 5, 7, 11$ & $9$ & $12$ \\
$\mathfrak{g}_2$ & $1, 5$ & $4$ & $6$
\end{tabular}\vspace{-10pt}
\caption{Indices, Coxeter number $g$ and dual Coxeter number $g^\vee$ for simple Lie algebras.}%
\label{tab:liealg-data}\vspace{-30pt}
\end{table}

\newpage
\providecommand{\href}[2]{#2}\begingroup\raggedright\endgroup

\end{document}

%% file: main_headers.tex
\let\olditemize\itemize\renewcommand{\itemize}{\vspace{-2pt}\olditemize\setlength{\itemsep}{1pt}\setlength{\parskip}{0pt}\setlength{\parsep}{-0pt}}
\let\oldenumerate\enumerate\renewcommand{\enumerate}{\vspace{-4pt}\oldenumerate\setlength{\itemsep}{1pt}\setlength{\parskip}{0pt}\setlength{\parsep}{0pt}}
\makeatletter\renewcommand\section{\addtocontents{toc}{\protect\addvspace{-2.25\p@}}\@startsection {section}{1}{\z@}{-0.0ex \@plus .2ex \@minus 0.2ex}{1ex \@plus.1ex\@minus .5ex}{\normalfont\large\bfseries}}
\renewcommand\subsection{\addtocontents{toc}{\protect\addvspace{-2.5\p@}}\@startsection {subsection}{1}{\z@}{0.5ex \@plus .2ex \@minus 0.2ex}{0.75ex \@plus.1ex\@minus .5ex}{\normalfont\bfseries}}
\renewcommand\subsubsection{\addtocontents{toc}{\protect\addvspace{-2.5\p@}}\@startsection {subsubsection}{1}{\z@}{0.5ex \@plus .2ex \@minus 0.2ex}{0.75ex \@plus.1ex\@minus .5ex}{\normalfont\bfseries}}
\newcommand{\eq}[1]{\vspace{-0.5pt}\begin{equation}#1\vspace{-0.5pt}\end{equation}}

\newcommand{\fwbox}[2]{\text{\makebox[#1][c]{$\hspace{-150pt}\displaystyle#2\hspace{-150pt}$}}}
\newcommand{\fwboxL}[2]{\text{\makebox[#1][l]{$#2$}}}
\newcommand{\fwboxR}[2]{\text{\makebox[#1][r]{$#2$}}}
\newcommand{\equivR}{\fwbox{14.5pt}{\hspace{-0pt}\fwboxR{0pt}{\raisebox{0.47pt}{\hspace{1.25pt}:\hspace{-4pt}}}=\fwboxL{0pt}{}}}
\newcommand{\equivL}{\fwbox{14.5pt}{\fwboxR{0pt}{}=\fwboxL{0pt}{\raisebox{0.47pt}{\hspace{-4pt}:\hspace{1.25pt}}}}}
\newcommand{\fig}[3]{\raisebox{#1}{\includegraphics[scale=#2]{#3}}}

\renewcommand{\phi}{\varphi}
\renewcommand{\bar}{\overline}
\renewcommand{\hat}{\widehat}

\newsavebox{\relsuperscript}
\newlength{\relsuperscriptsize}
\newcommand{\asympt}[1]{\sbox{\relsuperscript}{\ensuremath{\scriptstyle#1}}\setlength{\relsuperscriptsize}{\wd\relsuperscript}\ensuremath{\mathrel{\overset{\usebox{\relsuperscript}}{\resizebox{\relsuperscriptsize}{\height}{\ensuremath{\sim}}}}}}

\definecolor{rindou1}{rgb}{0.4431,0.2862,0.7960}
\definecolor{rindou2}{rgb}{0.0078,0.1215,0.4392}
\definecolor{lapis}{rgb}{0.0.0470,0.2941,0.5568}
\definecolor{emerald}{rgb}{0.31, 0.78, 0.47}
\definecolor{pinegreen}{rgb}{0.0, 0.47, 0.44}
\definecolor{jade}{rgb}{0.0, 0.66, 0.42}
\definecolor{teal}{rgb}{0.0, 0.5, 0.5}
\definecolor{totalCount}{rgb}{0,0,0.575}
\definecolor{topCount}{rgb}{0.575,0.0,0.225}
\definecolor{dim}{rgb}{0.55,0.55,0.55}
\definecolor{deemph}{rgb}{0.25,0.25,0.25}
\definecolor{hblue}{rgb}{0,0,0.575}
\definecolor{hred}{rgb}{0.575,0.0,0.225}
\definecolor{hgreen}{rgb}{0.0,0.4,0.2}
\definecolor{hteal}{rgb}{0.0,0.445,0.6451}

\renewcommand{\r}[1]{{\color{hred}#1}}
\renewcommand{\b}[1]{{\color{hblue}#1}}
\newcommand{\g}[1]{{\color{hteal}#1}}